%% file: hve-with-apps.tex
\documentclass[11pt,letterpaper]{report}
\usepackage{times,mathptmx}
\DeclareMathAlphabet{\mathcal}{OMS}{cmsy}{m}{n}
\usepackage{fullpage}
\usepackage{amsmath,amsfonts,amssymb,amsthm,amscd}
\usepackage{tabularx,booktabs}
\usepackage{hyperref,cite,url}
\hypersetup{pdfpagemode=UseNone,pdfstartview=}
\newcommand{\vect}[1]{\mathbf{#1}}
\newcommand{\svs}{\vspace{0.7mm}}
\newcommand{\vs}{\vspace{1.5mm}}
\theoremstyle{plain} % plain
\newtheorem{theorem}{Theorem}[section]
\newtheorem{lemma}[theorem]{Lemma}

\theoremstyle{definition} % definition
\newtheorem{definition}{Definition}[section]

\theoremstyle{remark} % remark
\newtheorem{remark}{Remark}
\newcommand{\G}{\mathbb{G}}
\newcommand{\Z}{\mathbb{Z}}

\newcommand{\lb}{\linebreak[0]}

\begin{document}

\title{
    {Efficient Hidden Vector Encryptions and Its Applications}\\
}
\author{
    A Thesis for the Degree of Doctor of Philosophy\\ \\
    Kwangsu Lee\\ \\
    Department of Information Security,\\
    Graduate School of Information Management and Security,\\
    Korea University.
}
\date{February 2011}
%
%\maketitle

\begin{titlepage}
\begin{center}
    \vspace*{2.0cm}
    \LARGE{Efficient Hidden Vector Encryptions and Its Applications}\footnote{
    Advisor: Dong Hoon Lee.}\\

    \vspace{0.8cm}
    \Large{Kwangsu Lee}

    \vspace{4.0cm}
    \large{A Thesis for the Degree of Doctor of Philosophy}

    \vfill

    \large{
    Department of Information Security,\\
    Graduate School of Information Management and Security,\\
    Korea University}\\

    \vspace*{0.5cm}
    \large{February 2011}
    \vspace*{2.0cm}
\end{center}
\end{titlepage}

% abstract

\chapter*{Abstract}
\input{"chap-abstract"}

%\vs \noindent {\bf Keywords:} Adaptive security, Bilinear pairing.

%\newpage
\tableofcontents
%\newpage

% contents of thesis

\chapter{Introduction}
\input{"chap-introduction"}

\chapter{Background} \label{ch:back}
\input{"chap-background"}

\chapter{Previous Work} \label{ch:prev-work}
\input{"chap-previous-work"}

\chapter{Efficient HVE with Short Tokens} \label{ch:hve-token}
\input{"chap-hve-short-token"}

\chapter{Convert HVE from Composite to Prime Order Groups} \label{ch:convert-hve}
\input{"chap-convert-hve-prime"}

\chapter{Fully Secure HVE with Short Tokens} \label{ch:full-hve}
\input{"chap-fully-secure-hve"}

\chapter{Applications} \label{ch:hve-apps}
\input{"chap-hve-applications"}

\chapter{Generic Group Model} \label{ch:generic-model}
\input{"chap-generic-group-model"}

\chapter{Conclusion}
\input{"chap-conclusion"}

% references

\bibliographystyle{plain}
\bibliography{hve-with-apps}

\end{document}

%% file: chap-abstract.tex
Predicate encryption is a new paradigm of public key encryption that enables
searches on encrypted data. Using the predicate encryption, we can search
keywords or attributes on encrypted data without decrypting the ciphertexts.
In predicate encryption, a ciphertext is associated with attributes and a
token corresponds to a predicate. The token that corresponds to a predicate
$f$ can decrypt the ciphertext associated with attributes $\vect{x}$ if and
only if $f(\vect{x})=1$.
Hidden vector encryption (HVE) is a special kind of predicate encryption. HVE
supports the evaluation of conjunctive equality, comparison, and subset
operations between attributes in ciphertexts and attributes in tokens.
Currently, several HVE schemes were proposed where the ciphertext size, the
token size, and the decryption cost are proportional to the number of
attributes in the ciphertext. In this thesis, we consider the efficiency, the
generality, and the security of HVE schemes. The results of this thesis are
described as follows.

The first results of this thesis are efficient HVE schemes where the token
consists of just four group elements and the decryption only requires four
bilinear map computations, independent of the number of attributes in the
ciphertext. The construction uses composite order bilinear groups and is
selectively secure under the well-known assumptions.
The second results are efficient HVE schemes that are secure under any kind
of pairing types. To achieve our goals, we proposed a general framework that
converts HVE schemes from composite order bilinear groups to prime order
bilinear groups. Using the framework, we convert the previous HVE schemes
from composite order bilinear groups to prime order bilinear groups.
The third results are fully secure HVE schemes with short tokens. Previous
HVE schemes were proven to be secure only in the selective security model
where the capabilities of the adversaries are severely restricted. Using the
dual system encryption techniques, we construct fully secure HVE schemes with
match revealing property in composite order groups.

%% file: chap-introduction.tex
\section{Overview}

Public-key encryption is one of the most fundamental primitives in modern
cryptography. In public-key encryption, a sender encrypts a message $M$ under
a public key $\textsf{PK}$, and the receiver who has a private key
$\textsf{SK}$ that corresponds to the public key $\textsf{PK}$ can only
decrypt the ciphertext. This simple \textit{``all-or-nothing''} semantics for
decryption is sufficient for traditional secure communication systems.
However, as the applications of public-key encryption come to be various, a
more complex semantics for decryption is necessary to specify the set of
receivers. For instance, suppose that the ciphertexts associated with
keywords are in a database server, and a user who has permission to read the
ciphertexts that are associated with some keywords may want to decrypt that
ciphertexts. \textit{Predicate encryption} provides this kind of complex
semantics in public-key encryption. In predicate encryption, a ciphertext is
associated with attributes and a token corresponds to a predicate. The token
$\textsf{TK}_{f}$ that corresponds to a predicate $f$ can decrypt the
ciphertext $\textsf{CT}$ that is associated with attributes $\vect{x}$ if and
only if $f(\vect{x})=1$. A ciphertext in predicate encryption hides not only
a message $M$ but also attributes $\vect{x}$. Currently, the expressiveness
of predicates in predicate encryption is limited. The most expressive
predicate encryption scheme is the one proposed by Katz, Sahai, and Waters in
\cite{KatzSW08}, and it supports inner product predicates.

Predicate encryption enables efficient data processing in the cloud computing
systems where users' data is stored in un-trusted remote servers. In the case
of traditional public-key encryption, a user encrypts messages and then
uploads the ciphertexts to the remote servers. If the user needs information
about the ciphertexts, then he should download all the ciphertexts from the
remote servers to decrypt them. Thus, this approach demands unnecessary data
transfers and data decryption. In the case of predicate encryption, a user
creates ciphertexts that are associated with related attributes $\vect{x}$
and then stores them in the remote servers. If the user wishes to acquire
information about the ciphertexts, then he generates a token $\textsf{TK}_f$
that matches a predicate $f$ and transfers the token to the remote server.
Next the remote server retrieves all the ciphertexts that satisfy
$f(\vect{x})=1$ using the token $\textsf{TK}_f$ by evaluating $f(\vect{x})$,
and then it returns the retrieved ciphertexts to the user. In this case, the
remote server cannot learn any information except the boolean value of
$f(\vect{x})$.

Hidden vector encryption (HVE) is a particular kind of predicate encryption
and it was introduced by Boneh and Waters \cite{BonehW07}. HVE supports
evaluations of conjunctive equality, comparison, and subset predicates on
encrypted data. For example, if a ciphertext is associated with a vector
$\vect{x}=(x_1, \ldots, x_l)$ of attributes and a token is associated with a
vector $\vect{\sigma} = (\sigma_1, \ldots, \sigma_l)$ of attributes where an
attribute is in a set $\Sigma$, then it can evaluate predicates like $(x_i =
\sigma_i)$, $(x_i \geq \sigma)$, and $(x_i \in A)$ where $A$ is a subset of
$\Sigma$. Additionally, it supports conjunctive combination of these
primitive predicates by extending the size of ciphertexts. After the
introduction of HVE based on composite order bilinear groups, several HVE
schemes have been proposed in \cite{KatzSW08,ShiW08,IovinoP08,OkamotoT09}.
Katz, Sahai, and Waters \cite{KatzSW08} proposed a predicate encryption
scheme that supports inner product predicates and they showed that it implies
an HVE scheme. Shi and Waters \cite{ShiW08} presented a delegatable HVE
scheme that enables the delegation of user's capabilities to others, and they
showed that it implies an anonymous hierarchical identity-based encryption
(HIBE) scheme. Iovino and Persiano \cite{IovinoP08} constructed an HVE scheme
based on prime order bilinear groups, but the number of attributes in
$\Sigma$ is restricted when it is compared to other HVE schemes. Okamoto and
Takashima \cite{OkamotoT09} proposed a hierarchical predicate encryption
scheme for inner products under prime order bilinear groups, and it also
implies an HVE scheme.

\section{Our Motivations}

When cryptographic schemes are applied to the real applications, they should
be efficient and secure against strong adversaries. To meet this
requirements, we should consider the three issues of efficiency, generality,
and security.

The first issue is the \textit{efficiency}. Generally efficiency is measured
in terms of the size of ciphertexts and the cost of search operations. If the
size of ciphertexts in previous HVE schemes is considered, the number of
ciphertext elements is proportional to the number of attributes in
ciphertexts, and the size of group elements of ciphertexts is proportional to
the size of group order. If the cost of search operations is considered, the
number of pairing operations is proportional to the number of token elements.
Therefore, it is important to shorten the number of token elements to reduce
the cost of search operations. Additionally, it is better to use prime order
bilinear groups than to use composite order bilinear groups in terms of
efficiency.

The second issue is the \textit{generality} such that the HVE schemes can be
based on any kind of pairing types. Pairing in bilinear groups is classified
as three types \cite{GalbraithPS08}. Bilinear groups with Type 1 pairing
corresponds to symmetric bilinear groups, and bilinear groups with Type 2 or
Type 3 pairing correspond to asymmetric bilinear groups. The previous HVE
schemes were constructed under composite order bilinear groups that exist in
Type 1 pairing, or under asymmetric bilinear groups of prime order. The
problem of cryptographic scheme's dependency to specific pairing types is
that the cryptographic scheme can be useless if a successful attack for a
specific pairing type is found. The solution for this problem is to design a
cryptographic scheme to be independent of a specific pairing type. That is,
the cryptographic scheme should have generality that it is secure under any
kind of pairing types. Thus, if a security weakness of some pairing types is
found, then the security of the scheme is guaranteed by just replacing the
underlying pairing type without re-designing the cryptographic scheme.
However, it is an open problem to construct an efficient HVE scheme that is
secure under any kind of pairing types.

The third issue is the \textit{security} such that the HVE schemes should be
secure against strong adversaries. The security models of HVE are categorized
as two kinds: the selective security model and the full security model. In
the selective security model, an adversary should commit  target vectors
before he receive a public key. Thus, this model severely restricts the
capability of the adversary. In the full security model, the adversary
commits target vectors at the challenge stage. The previous HVE schemes were
proved in the selective security model since it is easy to prove its security
in this model. However, the right security model for HVE is the full security
model that does not restrict the ability of the adversary. Therefore, it is
an important problem to design an HVE scheme that is secure in the full
security model.

\section{Our Contributions}

The results of this thesis are divided as three folds. The detailed results
are described as follows.

Our first results are efficient HVE scheme with short tokens. In composite
order bilinear groups, we constructed HVE schemes that have the constant size
of tokens and the constant cost of pairing operations, and we proved its
selective security under the decisional Bilinear Diffie-Hellman (BDH) and the
decisional Composite 3-Party Diffie-Hellman (C3DH) assumptions. The
ciphertext consists of $l+O(1)$ group elements, the token consists of four
group elements, and the decryption requires four pairing computations. Though
our construction in composite order bilinear groups is algebraically similar
to the one by Shi and Waters in \cite{ShiW08}, we achieved the constant size
of tokens and the constant cost of decryption, in contrast to the
construction of Shi and Waters. Additionally, we converted our construction
in composite order groups to asymmetric bilinear groups of prime order where
isomorphisms between two groups are not efficiently computable.

Our second results are efficient HVE schemes that are secure under any kind
of pairing types. To achieve our goals, we proposed a general framework that
converts HVE schemes from composite order bilinear groups to prime order
bilinear groups. The conversion method of this paper is similar to the
conversion method of Freeman in terms of using product groups and vector
orthogonality, but it has the following three differences. The first
difference is that Freeman's method is related to the subgroup decision (SGD)
assumption in prime order bilinear groups, whereas our method is not related
to the SGD assumption. The second difference is that Freeman's method only
works in asymmetric bilinear groups of prime order, whereas our method works
in any bilinear groups of prime order. The third difference is that the
cryptographic schemes from Freeman's method use complex assumptions that
depend on complex basis vectors, whereas the HVE schemes from our method use
simple assumptions that are independent of basis vectors. We first convert
the HVE scheme of Boneh and Waters, the delegatable HVE scheme of Shi and
Waters, and the efficient HVE scheme with constant cost of pairing of Lee and
Lee from composite order bilinear groups to prime order bilinear groups. Next
we prove that these converted HVE schemes are selectively secure under the
decisional Bilinear Diffie-Hellman (BDH) and the decisional Parallel 3-party
Diffie-Hellman (P3DH) assumptions. Through these conversion, we constructed
the first delegatable HVE scheme and efficient HVE scheme with constant cost
of pairing in any bilinear groups of prime order.

Our third results are fully secure HVE schemes with short tokens. To
construct fully secure HVE schemes, we adapt the dual system encryption
technique of Waters \cite{Waters09,LewkoW10}. In the dual system encryption,
the ciphertexts and tokens can be a normal-type or a semi-functional type.
The dual system encryption technique achieves the full security model by
using two properties such that the normal-type and the semi-functional type
are indistinguishable and the decryption of the semi-functional ciphertext
using the semi-functional token always fails. We propose a fully secure HVE
scheme with short tokens in composite order bilinear groups with four primes.

\section{Related Works}

Predicate encryption in public-key encryption was presented by Boneh et al.
\cite{BonehCOP04}. They proposed a public-key encryption scheme with keyword
search (PEKS) using Boneh and Franklin's identity-based encryption (IBE)
scheme \cite{BonehF01,BonehF03}, and their construction corresponds to the
implementation of an equality predicate. Abdalla et al. \cite{AbdallaBC+05}
proved that anonymous IBE implies predicate encryption of an equality query,
and they proposed the definition of anonymous HIBE by extending anonymous
IBE. Several anonymous HIBE constructions were proposed in \cite{BoyenW06,
ShiW08,SeoKOS09}. A predicate encryption scheme for a comparison query was
constructed by Boneh et al. in \cite{BonehSW06,BonehW06}, and it can be used
to construct a fully collusion resistant traitor tracing scheme. By extending
comparison predicates, Shi et al. \cite{ShiBC+07} considered
multi-dimensional range predicates on encrypted data under a weaker security
model.

Research on predicate encryption was dramatically advanced by the
introduction of HVE by Boneh and Waters \cite{BonehW07}. An HVE scheme is a
predicate encryption scheme of conjunctive equality, comparison, and subset
predicates. After that, Shi and Waters \cite{ShiW08} presented the definition
of the delegation in predicate encryption, and they proposed a delegatable
HVE scheme. Iovino and Persiano \cite{IovinoP08} constructed an HVE scheme
based on prime order bilinear groups with a restricted number of attributes.
Katz, Sahai, and Waters \cite{KatzSW08} proposed the most expressive
predicate encryption scheme of inner product predicates, and they showed that
it implies anonymous IBE, HVE, and predicate encryption for disjunctions,
polynomials, CNF \& DNF formulas, or threshold predicates. Okamoto and
Takashima \cite{OkamotoT09} constructed a hierarchical predicate encryption
scheme for inner products under prime order bilinear groups using the notion
of dual pairing vector spaces.

Predicate encryption in symmetric encryption was considered by Goldreich and
Ostrovsky \cite{GoldreichO96}. Song et al. \cite{SongWP00} proposed an
efficient scheme that supports an equality predicate. Shen, Shi, and Waters
\cite{ShenSW09} introduced the formal definition of predicate privacy, and
they presented a symmetric predicate encryption scheme with predicate privacy
of inner product predicates using composite order bilinear groups. Blundo et
al. \cite{BlundoIP09} proposed a symmetric HVE scheme that provides weaker
predicate privacy under prime order asymmetric bilinear groups.

Other research direction that is related with predicate encryption is
functional encryption. In functional encryption, a ciphertext is associated
with attributes $\vect{x}$, and a private key is associated with a function
$f$. If $f(\vect{x})=1$, then a receiver who has a private key that
corresponds to the function $f$ can decrypt the ciphertext that corresponds
to attributes $\vect{x}$. The main difference between predicate encryption
and functional encryption is that the attribute hiding property is not
provided in functional encryption, whereas the attribute hiding property was
the essential one in predicate encryption.
The identity-based encryption (IBE) is the most simple type of functional
encryption, and it provide an equality function for an identity in
ciphertexts \cite{BonehF01,BonehF03,BonehB04e,Waters05,Gentry06}. The
hierarchical IBE (HIBE) is an extension of IBE, and it provides a conjunctive
equality function for a hierarchical identity in ciphertexts \cite{GentryS02,
BonehB04e,BonehBG05,BoyenW06,SeoKOS09,Waters09,LewkoW10}. The attribute-based
encryption (ABE) is also an extension of IBE, and it provides the most
general function that consists of AND, OR, NOT, and threshold gates
\cite{SahaiW05,GoyalPSW06,BethencourtSW07,OstrovskySW07,LewkoOSTW10}.

%% file: chap-background.tex
In this chapter, we define HVE and give the formal definition of its security
model. Then we give the necessary background on bilinear groups and
complexity assumptions.

\section{Hidden Vector Encryption} \label{back-hve}

Let $\Sigma$ be a finite set of attributes and let $*$ be a special symbol
not in $\Sigma$. Define $\Sigma_* = \Sigma \cup \{*\}$. The star $*$ plays
the role of a wild card or ``don't care'' value. For a vector $\vect{\sigma}
= (\sigma_1, \ldots, \sigma_l) \in \Sigma_*^l$, we define a predicate
$f_{\vect{\sigma}}$ over $\Sigma^l$ as follows: For $\vect{x} = (x_1, \ldots,
x_l) \in \Sigma^l$, it set $f_{\vect{\sigma}}(\vect{x}) = 1$ if $\forall i :
(\sigma_i = x_i \mbox{ or } \sigma_i = *)$, it set
$f_{\vect{\sigma}}(\vect{x}) = 0$ otherwise.

\vs An HVE scheme consists of four algorithms (\textsf{Setup, GenToken,
Encrypt, Query}). Formally it is defined as:
\begin{description}
\item [\normalfont{\textsf{Setup}($1^{\lambda}$)}.] The setup algorithm
    takes as input a security parameter $1^{\lambda}$. It outputs a public
    key \textsf{PK} and a secret key \textsf{SK}.

\item [\normalfont{\textsf{GenToken}($\vect{\sigma}, \textsf{SK,PK}$)}.]
    The token generation algorithm takes as input a vector $\vect{\sigma} =
    (\sigma_1, \ldots, \sigma_l) \in \Sigma_{*}^{l}$ that corresponds to a
    predicate $f_{\vect{\sigma}}$, the secret key \textsf{SK} and the
    public key \textsf{PK}. It outputs a token
    $\textsf{TK}_{\vect{\sigma}}$ for the vector $\vect{\sigma}$.

\item [\normalfont{\textsf{Encrypt}($\vect{x}, M, \textsf{PK}$)}.] The
    encrypt algorithm takes as input a vector $\vect{x} = (x_1, \ldots,
    x_l) \in \Sigma^l$, a message $M \in \mathcal{M}$, and the public key
    \textsf{PK}. It outputs a ciphertext $\textsf{CT}$ for $\vect{x}$ and
    $M$.

\item [\normalfont{\textsf{Query}($\textsf{CT},
    \textsf{TK}_{\vect{\sigma}}, \textsf{PK}$)}.] The query algorithm takes
    as input a ciphertext $\textsf{CT}$, a token
    $\textsf{TK}_{\vect{\sigma}}$ for a vector $\vect{\sigma}$ that
    corresponds to a predicate $f_{\vect{\sigma}}$, and the public key
    \textsf{PK}. It outputs $M$ if $f_{\vect{\sigma}}(\vect{x})=1$ or
    outputs $\perp$ otherwise.
\end{description}

\noindent The scheme should satisfy the following correctness property: for
all $\vect{x} \in \Sigma^l$, $M \in \mathcal{M}$, $\vect{\sigma} \in
\Sigma_*^l$, let $(\textsf{PK},\textsf{SK}) \leftarrow
\textsf{Setup}(1^{\lambda})$, $\textsf{CT} \leftarrow
\textsf{Encrypt}(\vect{x}, M, \textsf{PK})$, and $\textsf{TK}_{\vect{\sigma}}
\leftarrow \textsf{GenToken} (\sigma, \textsf{SK,PK})$.
\begin{itemize}
\item If $f_{\vect{\sigma}}(\vect{x}) = 1$, then
    $\textsf{Query}(\textsf{CT}, \textsf{TK}_{\vect{\sigma}}, \textsf{PK})
    = M$.

\item If $f_{\vect{\sigma}}(\vect{x}) = 0$, then
    $\textsf{Query}(\textsf{CT}, \textsf{TK}_{\vect{\sigma}}, \textsf{PK})
    = \perp$ with all but negligible probability.
\end{itemize}

\section{Security Model}

\subsection{Selective Security Model}

We define the selective security model of HVE as the following game between a
challenger $\mathcal{C}$ and an adversary $\mathcal{A}$:

\begin{description}
\item[Init:] $\mathcal{A}$ submits two vectors $\vect{x}_0, \vect{x}_1 \in
    \Sigma^l$.

\item[Setup:] $\mathcal{C}$ runs the setup algorithm and keeps the secret
    key \textsf{SK} to itself, then it gives the public key \textsf{PK} to
    $\mathcal{A}$.

\item[Query 1:] $\mathcal{A}$ adaptively requests a polynomial number of
    tokens for vectors $\vect{\sigma}_1, \ldots, \vect{\sigma}_{q_1}$ that
    correspond to predicates $f_{\vect{\sigma}_1}, \ldots,
    f_{\vect{\sigma}_{q_1}}$ subject to the restriction that
    $f_{\vect{\sigma}_i}(\vect{x}_0) = f_{\vect{\sigma}_i}(\vect{x}_1)$ for
    all $i$. In responses, $\mathcal{C}$ gives the corresponding tokens
    $\textsf{TK}_{\vect{\sigma}_i}$ to $\mathcal{A}$.

\item[Challenge:] $\mathcal{A}$ submits two messages $M_0, M_1$ subject to
    the restriction that if there is an index $i$ such that
    $f_{\vect{\sigma}_i}(\vect{x}_0) = f_{\vect{\sigma}_i}(\vect{x}_1) = 1$
    then $M_0 = M_1$. $\mathcal{C}$ chooses a random coin $\gamma$ and
    gives a ciphertext $\textsf{CT}$ of $(\vect{x}_{\gamma}, M_{\gamma})$
    to $\mathcal{A}$.

\item[Query 2:] $\mathcal{A}$ continues to request tokens for vectors
    $\vect{\sigma}_{q_1 +1}, \ldots, \vect{\sigma}_{q}$ that correspond to
    predicates $f_{\vect{\sigma}_{q_1 +1}}, \ldots, f_{\vect{\sigma}_{q}}$
    subject to the two restrictions as before.

\item [Guess:] $\mathcal{A}$ outputs a guess $\gamma'$. If $\gamma =
    \gamma'$, it outputs 0. Otherwise, it outputs 1.
\end{description}

\noindent The advantage of $\mathcal{A}$ is defined as $\textsf{Adv}_{
\mathcal{A}}^{\textsf{HVE}} = \big| \Pr[\gamma = \gamma'] - 1/2 \big|$ where
the probability is taken over the coin tosses made by $\mathcal{A}$ and
$\mathcal{C}$.

\begin{definition}
We say that an HVE scheme is selectively secure if all probabilistic
polynomial-time adversaries have at most a negligible advantage in the above
game.
\end{definition}

\subsection{Full Security Model}

We define the full security model of HVE as the following game between a
challenger $\mathcal{C}$ and an adversary $\mathcal{A}$:

\begin{description}
\item[Setup:] $\mathcal{C}$ runs the setup algorithm and keeps the secret
    key \textsf{SK} to itself, then it gives the public key \textsf{PK} to
    $\mathcal{A}$.

\item[Query 1:] $\mathcal{A}$ adaptively requests a polynomial number of
    tokens for vectors $\vect{\sigma}_1, \ldots, \vect{\sigma}_{q_1}$ that
    correspond to predicates $f_{\vect{\sigma}_1}, \ldots,
    f_{\vect{\sigma}_{q_1}}$. In responses, $\mathcal{C}$ gives the
    corresponding tokens $\textsf{TK}_{\vect{\sigma}_i}$ to $\mathcal{A}$.

\item[Challenge:] $\mathcal{A}$ submits two vectors $\vect{x}_0, \vect{x}_1
    \in \Sigma^l$ and two messages $M_0, M_1$ subject to the following two
    restrictions:
    \begin{itemize}
    \item For all $i \in \{1, \ldots, q_1\}$,
        $f_{\vect{\sigma}_i}(\vect{x}_0) =
        f_{\vect{\sigma}_i}(\vect{x}_1)$.
    \item If $\exists i \in \{1, \ldots, q_1\}$ such that
        $f_{\vect{\sigma}_i}(\vect{x}_0) =
        f_{\vect{\sigma}_i}(\vect{x}_1) = 1$, then $M_0 = M_1$.
    \end{itemize}

\item[Query 2:] $\mathcal{A}$ continues to request tokens for vectors
    $\vect{\sigma}_{q_1 +1}, \ldots, \vect{\sigma}_{q}$ that correspond to
    predicates $f_{\vect{\sigma}_{q_1 +1}}, \ldots, f_{\vect{\sigma}_{q}}$
    subject to the two restrictions as before.

\item [Guess:] $\mathcal{A}$ outputs a guess $\gamma'$. If $\gamma =
    \gamma'$, it outputs 0. Otherwise, it outputs 1.
\end{description}

\noindent The advantage of $\mathcal{A}$ is defined as $\textsf{Adv}_{
\mathcal{A}}^{\textsf{HVE}} = \big| \Pr[\gamma = \gamma'] - 1/2 \big|$ where
the probability is taken over the coin tosses made by $\mathcal{A}$ and
$\mathcal{C}$.

\begin{definition}
We say that an HVE scheme is fully secure (with \textit{match concealing}) if
all probabilistic polynomial-time adversaries have at most a negligible
advantage in the above game.
\end{definition}

\begin{definition}
We say that an HVE scheme is fully secure (with \textit{match revealing}) if
all probabilistic polynomial-time adversaries have at most a negligible
advantage in the above game with restriction that the adversary can not query
predicates such that $f_{\vect{\sigma}}(\vect{x}_0) =
f_{\vect{\sigma}}(\vect{x}_1) = 1$.
\end{definition}

\section{Bilinear Groups}

\subsection{Bilinear Groups of Composite Order}

The composite order bilinear groups were first introduced in
\cite{BonehGN05}. Let $n=pqr$ where $p, q$, and $r$ are distinct prime
numbers. Let $\G$ and $\G_T$ be two multiplicative cyclic groups of composite
order $n$ and $g$ be a generator of $\G$. The bilinear map $e:\G \times \G
\rightarrow \G_{T}$ has the following properties:
\begin{enumerate}
\item Bilinearity: $\forall u, v \in \G$ and $\forall a,b \in \Z_n$,
    $e(u^a,v^b)=e(u,v)^{ab}$.
\item Non-degeneracy: $\exists g$ such that $e(g,g) \neq 1$, that is,
    $e(g,g)$ is a generator of $\G_T$.
\end{enumerate}

\noindent We say that $\G$ is a bilinear group if the group operations in
$\G$ and $\G_T$ as well as the bilinear map $e$ are all efficiently
computable. Furthermore, we assume that the description of $\G$ and $\G_T$
includes generators of $\G$ and $\G_T$ respectively.

We use the notation $\G_p, \G_q, \G_r$ to denote the subgroups of order $p,
q, r$ of $\G$ respectively. Similarly, we use the notation $\G_{T,p},
\G_{T,q}, \G_{T,r}$ to denote the subgroups of order $p, q, r$ of $\G_T$
respectively.

\subsection{Bilinear Groups of Prime Order}

Let $\G$ and $\G_{T}$ be multiplicative cyclic groups of prime $p$ order. Let
$g$ be a generator of $\G$. The bilinear map $e:\G \times \G \rightarrow
\G_{T}$ has the following properties:
\begin{enumerate}
\item Bilinearity: $\forall u,v \in \G$ and $\forall a,b \in \Z_p$,
    $e(u^a,v^b) = e(u,v)^{ab}$.
\item Non-degeneracy: $\exists g$ such that $e(g,g)$ has order $p$, that
    is, $e(g,g)$ is a generator of $\G_T$.
\end{enumerate}
We say that $(p, \G, \G_T, e)$ are bilinear groups if the group operations in
$\G$ and $\G_T$ as well as the bilinear map $e$ are all efficiently
computable.

\subsection{Asymmetric Bilinear Groups of Prime Order}

Let $\G, \hat{\G}$, and $\G_{T}$ be multiplicative cyclic groups of prime $p$
order where $\G \neq \hat{\G}$. Let $g, \hat{g}$ be generators of $\G,
\hat{\G}$, respectively. The asymmetric bilinear map $e:\G \times \hat{\G}
\rightarrow \G_{T}$ has the following properties:
\begin{enumerate}
\item Bilinearity: $\forall u \in \G, \forall v \in \hat{\G}$ and $\forall
    a,b \in \Z_p$, $e(u^a,\hat{v}^b)=e(u,\hat{v})^{ab}$.
\item Non-degeneracy: $\exists g, \hat{g}$ such that $e(g,\hat{g}) \neq 1$,
    that is, $e(g,\hat{g})$ is a generator of $\G_T$.
\end{enumerate}

\noindent We say that $\G, \hat{\G}, \G_T$ are asymmetric bilinear groups
with no efficiently computable isomorphisms if the group operations in $\G,
\hat{\G}$ and $\G_T$ as well as the bilinear map $e$ are all efficiently
computable, but there are no efficiently computable isomorphisms between $\G$
and $\hat{\G}$.

\section{Complexity Assumptions}

\subsection{Assumptions in Composite Order Bilinear Groups}
\label{assump-composite}

We introduce three assumptions under composite order bilinear groups. The
decisional composite bilinear Diffie-Hellman (cBDH) assumption was used to
construct an HVE scheme in \cite{BonehW07}. It is a natural extension of the
decisional BDH assumption in \cite{BonehF01} from prime order bilinear groups
to composite order bilinear groups. The bilinear subgroup decision (BSD)
assumption was introduced in \cite{BonehSW06} to construct a traitor tracing
scheme. The decisional composite $3$-party Diffie-Hellman (C3DH) assumption
was used to construct an HVE scheme in \cite{BonehW07}.

\vs \noindent \textbf{Decisional composite Bilinear Diffie-Hellman (cBDH)
Assumption} Let $(n, \G, \G_T, e)$ be a description of the bilinear group of
composite order $n=pqr$. Let $g_p, g_q, g_r$ be generators of subgroups of
order $p, q, r$ of $\G$ respectively. The decisional cBDH problem is stated
as follows: given a challenge tuple
    $$D = ((p,q,r, \G, \G_T, e),~
    g_p, g_q, g_r, g_p^a, g_p^b, g_p^c) \mbox{ and } T,$$
decides whether $T = e(g_p,g_p)^{abc}$ or $T = R$ with random choices of $a,
b, c \in \Z_p$, $R \in \G_{T,p}$. The advantage of $\mathcal{A}$ in solving
the decisional cBDH problem is defined as
    \begin{eqnarray*}
    \textsf{Adv}^{\textsf{cBDH}}_{\mathcal{A}} = \Big|
    \Pr \big[\mathcal{A}(D, T = e(g_p, g_p)^{abc}) = 1 \big] -
    \Pr \big[\mathcal{A}(D, T = R) = 1 \big] \Big|
    \end{eqnarray*}
where the probability is taken over the random choices of $D, T$ and the
random bits used by $\mathcal{A}$.

\begin{definition}
We say that the decisional cBDH assumption holds if no probabilistic
polynomial-time algorithm has a non-negligible advantage in solving the
decisional cBDH problem.
\end{definition}

\noindent \textbf{Bilinear Subgroup Decision (BSD) Assumption} Let $(n, \G,
\G_T, e)$ be a description of the bilinear group of composite order $n=pqr$.
Let $g_p, g_q, g_r$ be generators of subgroups of order $p, q, r$ of $\G$
respectively. The BSD problem is stated as follows: given a challenge tuple
    $$D = ((n, \G, \G_T, e),~ g_p, g_q, g_r) \mbox{ and } T,$$
decides whether $T = Q \in \G_{T,p}$ or $T = R \in \G_T$ with random choices
of $Q \in \G_{T,p}, R \in \G_T$. The advantage of $\mathcal{A}$ in solving
the BSD problem is defined as
    \begin{eqnarray*}
    \textsf{Adv}^{\textsf{BSD}}_{\mathcal{A}} = \Big|
    \Pr \big[\mathcal{A}(D, T = Q) = 1 \big] -
    \Pr \big[\mathcal{A}(D, T = R) = 1 \big] \Big|
    \end{eqnarray*}
where the probability is taken over the random choices of $D, T$ and the
random bits used by $\mathcal{A}$.

\begin{definition}
We say that the BSD assumption holds if no probabilistic polynomial-time
algorithm has a non-negligible advantage in solving the BSD problem.
\end{definition}

\noindent \textbf{Decisional Composite $3$-party Diffie-Hellman (C3DH)
Assumption} Let $(n, \G, \G_T, e)$ be a description of the bilinear group of
composite order $n=pqr$. Let $g_p, g_q, g_r$ be generators of subgroups of
order $p, q, r$ of $\G$ respectively. The decisional C3DH problem is stated
as follows: given a challenge tuple
    $$D = ((n, \G, \G_T, e),~ g_p, g_q, g_r,
    g_p^{a}, g_p^{b}, g_p^{ab} R_1, g_p^{abc} R_2) \mbox{ and } T,$$
decides whether $T = g_p^c R_3$ or $T = R$ with random choices of $R_1, R_2,
R_3 \in \G_q, R \in \G_{pq}$. The advantage of $\mathcal{A}$ in solving the
decisional C3DH problem is defined as
    \begin{eqnarray*}
    \textsf{Adv}^{\textsf{C3DH}}_{\mathcal{A}} = \Big|
    \Pr \big[\mathcal{A}(D, T = g_p^c R_3) = 1 \big] -
    \Pr \big[\mathcal{A}(D, T = R) = 1 \big] \Big|
    \end{eqnarray*}
where the probability is taken over the random choices for $D, T$ and the
random bits used by $\mathcal{A}$.

\begin{definition}
We say that the decisional C3DH assumption holds if no probabilistic
polynomial-time algorithm has a non-negligible advantage in solving the
decisional C3DH problem.
\end{definition}

\subsection{Assumptions in Prime Order Bilinear Groups}
\label{assump-prime}

We introduce two assumptions under prime order bilinear groups. The
decisional Bilinear Diffie-Hellman (BDH) assumption is well-known one and
introduced in \cite{BonehF01}. The decisional Parallel 3-party Diffie-Hellman
(P3DH) assumption is newly introduced in this paper, and its security in
generic group model is given in chapter \ref{ch:generic-model}.

\vs \noindent \textbf{Bilinear Diffie-Hellman (BDH) Assumption} Let $(p, \G,
\G_T, e)$ be a description of the bilinear group of prime order $p$. The
decisional BDH problem is stated as follows: given a challenge tuple
    $$D = \big( (p, \G, \G_T, e),~
    g, g^a, g^b, g^c \big) \mbox{ and } T,$$
decides whether $T = T_0 = e(g, g)^{abc}$ or $T = T_1 = e(g,g)^d$ with random
choices of $a, b, c, d \in \Z_p$. The advantage of $\mathcal{A}$ in solving
the decisional BDH problem is defined as
    \begin{eqnarray*}
    \textsf{Adv}^{\textsf{BDH}}_{\mathcal{A}} = \Big|
    \Pr \big[\mathcal{A}(D, T_0) = 1 \big] -
    \Pr \big[\mathcal{A}(D, T_1) = 1 \big] \Big|
    \end{eqnarray*}
where the probability is taken over the random choices of $D, T$ and the
random bits used by $\mathcal{A}$. We say that the decisional BDH assumption
holds if no probabilistic polynomial-time algorithm has a non-negligible
advantage in solving the decisional BDH problem.

\vs \noindent \textbf{Parallel 3-party Diffie-Hellman (P3DH) Assumption} Let
$(p, \G, \G_T, e)$ be a description of the bilinear group of prime order $p$.
The decisional P3DH problem is stated as follows: given a challenge tuple
    \begin{eqnarray*}
    D = \big( (p, \G, \G_T, e),~
    (g,f),(g^a, f^a), (g^b, f^b),
    (g^{ab} f^{z_1}, g^{z_1}), (g^{abc} f^{z_2}, g^{z_2}) \big)
    \mbox{ and } T,
    \end{eqnarray*}
decides whether $T = T_0 = (g^{c} f^{z_3}, g^{z_3})$ or $T = T_1 = (g^d
f^{z_3}, g^{z_3})$ with random choices of $a, b, c, d \in \Z_p$ and $z_1,
z_2, z_3 \in \Z_p$. The advantage of $\mathcal{A}$ in solving the decisional
P3DH problem is defined as
    \begin{eqnarray*}
    \textsf{Adv}^{\textsf{P3DH}}_{\mathcal{A}} = \Big|
    \Pr \big[\mathcal{A}(D, T_0) = 1 \big] -
    \Pr \big[\mathcal{A}(D, T_1) = 1 \big] \Big|
    \end{eqnarray*}
where the probability is taken over the random choices of $D, T_0, T_1$ and
the random bits used by $\mathcal{A}$. We say that the decisional P3DH
assumption holds if no probabilistic polynomial-time algorithm has a
non-negligible advantage in solving the decisional P3DH problem.

\begin{remark}
The decisional P3DH problem can be modified as follows: an adversary is given
a challenge tuple
    $D = \big( (p, \G, \G_T, e),~
    (g,f),(g^a, f^a), (g^b, f^b),
    (g^{ab} f^{z_1}, g^{z_1}), \lb (g^{c} f^{z_2}, g^{z_2}) \big)$
    and $T$,
it decides whether $T = T_0 = (g^{abc} f^{z_3}, g^{z_3})$ or $T = T_1 = (g^d
f^{z_3}, g^{z_3})$. However, this modified one is the same as the original
one by changing the position of the challenge tuple as
    $D = \big( (p, \G, \G_T, e),~
    (g,f),(g^a, f^a), (g^b, f^b),
    (g^{ab} f^{z_1}, g^{z_1}), T \big)$
    and $T' = (g^{c} f^{z_2}, g^{z_2})$,
Thus, we will use any one of challenge tuple forms for the decisional P3DH
assumption.
\end{remark}

\subsection{Assumptions in Asymmetric Bilinear Groups}
\label{assump-asym-prime}

We introduce three cryptographic assumptions that are secure under asymmetric
bilinear groups of prime order where there are no efficiently computable
isomorphisms between two groups $\G$ and $\hat{\G}$. The decisional
asymmetric bilinear Diffie-Hellman (aBDH) assumption is the same as the
decisional cBDH assumption except that it uses asymmetric bilinear groups.
The decisional asymmetric Diffie-Hellman (aDH) assumption says that the
traditional decisional DH assumption holds $\hat{\G}$ groups since there are
no efficiently computable isomorphisms between two groups. The decisional
asymmetric $3$-party Diffie-Hellman (a3DH) assumption is an asymmetric
version of the decisional C3DH assumption.

\vs \noindent \textbf{Decisional asymmetric Bilinear Diffie-Hellman (aBDH)
Assumption} Let $(p, \G, \hat{\G}, \G_T, e)$ be a description of the
asymmetric bilinear group of prime order $p$ with no efficiently computable
isomorphism from $\G$ to $\hat{\G}$. The decisional aBDH problem is stated as
follows: given a challenge tuple
    $$D = ((p, \G, \hat{\G}, \G_T, e),~
    g, g^a, g^b, g^c, \hat{g}, \hat{g}^a, \hat{g}^b) \mbox{ and } T,$$
decides whether $T = e(g, \hat{g})^{abc}$ or $T = R$ with random choices of
$a, b, c \in \Z_p$, $R \in \G_T$. The advantage of $\mathcal{A}$ in solving
the decisional aBDH problem is defined as
    \begin{eqnarray*}
    \textsf{Adv}^{\textsf{aBDH}}_{\mathcal{A}} = \Big|
    \Pr \big[\mathcal{A}(D, T = e(g,\hat{g})^{abc}) = 1 \big] -
    \Pr \big[\mathcal{A}(D, T = R) = 1 \big] \Big|
    \end{eqnarray*}
where the probability is taken over the random choices of $D, T$ and the
random bits used by $\mathcal{A}$.

\begin{definition}
We say that the decisional aBDH assumption holds if no probabilistic
polynomial-time algorithm has a non-negligible advantage in solving the
decisional aBDH problem.
\end{definition}

\noindent \textbf{Decisional asymmetric Diffie-Hellman (aDH) Assumption} Let
$(p, \G, \hat{\G}, \G_T, e)$ be a description of the asymmetric bilinear
group of prime order $p$ with no efficiently computable isomorphisms between
$\G$ and $\hat{\G}$. Let $g, \hat{g}$ be generators of $\G, \hat{\G}$
respectively. The decisional aDH problem is stated as follows: given a
challenge tuple
    $$D = ((p, \G, \hat{\G}, \G_T, e),~
    g, \hat{g}, \hat{g}^{a}, \hat{g}^{b}) \mbox{ and } T,$$
decides whether $T = \hat{g}^{ab}$ or $T = R$ with random choices of $a,b \in
\Z_p$, $R \in \hat{\G}$. The advantage of $\mathcal{A}$ in solving the
decisional aDH problem is defined as
    \begin{eqnarray*}
    \textsf{Adv}^{\textsf{aDH}}_{\mathcal{A}} = \Big|
    \Pr \big[\mathcal{A}(D, T = \hat{g}^{ab}) = 1 \big] -
    \Pr \big[\mathcal{A}(D, T = R) = 1 \big] \Big|
    \end{eqnarray*}
where the probability is taken over the random choices of $D, T$ and the
random bits used by $\mathcal{A}$.

\begin{definition}
We say that the decisional aDH assumption holds if no probabilistic
polynomial-time algorithm has a non-negligible advantage in solving the
decisional aDH problem.
\end{definition}

\noindent \textbf{Decisional asymmetric 3-party Diffie-Hellman (a3DH)
Assumption} Let $(p, \G, \hat{\G}, \G_T, e)$ be a description of the
asymmetric bilinear group of prime order $p$ with no efficiently computable
isomorphism from $\G$ to $\hat{\G}$. Let $g, \hat{g}$ be generators of $\G,
\hat{\G}$ respectively. The decisional a3DH is stated as follows: given a
challenge tuple
    $$D = ((p, \G, \hat{\G}, \G_T, e),~
    g, g^{a}, g^{b}, g^{ab}, g^{abc}, \hat{g}, \hat{g}^{a}, \hat{g}^{b})
    \mbox{ and } T,$$
decides whether $T = g^c$ or $T = R$ with random choice of $a,b,c \in \Z_p$,
$R \in \G$. The advantage of $\mathcal{A}$ in solving the decisional a3DH
problem is defined as
    \begin{eqnarray*}
    \textsf{Adv}^{\textsf{a3DH}}_{\mathcal{A}} = \Big|
    \Pr \big[\mathcal{A}(D, T = g^c) = 1 \big] -
    \Pr \big[\mathcal{A}(D, T = R) = 1 \big] \Big|
    \end{eqnarray*}
where the probability is taken over the random choices for $D, T$ and the
random bits used by $\mathcal{A}$.

\begin{definition}
We say that the decisional a3DH assumption holds if no probabilistic
polynomial-time algorithm has a non-negligible advantage in solving the
decisional a3DH problem.
\end{definition}

\begin{remark}
The decisional aDH assumption is equivalent to the external Diffie-Hellman
(XDH) assumption. In this paper, we will use aDH instead of XDH for
notational consistency.
\end{remark}

%% file: chap-previous-work.tex
In this chapter, we review the previous work of HVE. After the first
construction of HVE by Boneh and Waters, various HVE schemes were proposed in
\cite{BonehW07,ShiW08,Ducas10,LeeL11}. The design techniques of previous HVE
schemes are classified as the following three categories.

\section{Trivial Construction}

The first category is a trivial construction from public key encryption (PKE)
\cite{BonehCOP04,BonehW07,IovinoP08,KatzY09}. This method was introduced by
Boneh et al. to construct a public key encryption scheme with keyword search
(PEKS) using trapdoor permutations \cite{BonehCOP04}. After that, Boneh and
Waters showed that a searchable public key encryption for general predicates
can be constructed by this method \cite{BonehW07}. Recently, Katz and
Yerukhimovich showed that it is possible to construct predicate encryption
scheme from a CPA-secure PKE scheme if the number of predicate such that
$f(\vect{x})=0$ is less than a polynomial value of a security parameter
\cite{KatzY09}. The main idea of this method is to use a multiple instances
of key-private PKE that was introduced by Bellare et al. \cite{BellareBDP01}.
The public key of searchable public key encryption consists of the public
keys of key-private PKE, and each instance of public keys is mapped to each
predicate. However, this method has a serious problem such that the total
number of predicates is limited to the polynomial value of a security
parameter.

\subsection{PE of Boneh and Waters}

Let $\Sigma$ be a finite set of binary strings. A predicate $f$ over $\Sigma$
is a function $f:\Sigma \rightarrow \{0,1\}$. We say that $x \in \Sigma$
satisfies the predicate if $f(x) = 1$. Let $\Phi$ be the set of predicates,
that is, $\Phi = \{f_1, f_2, \ldots, f_m\}$. The trivial predicate encryption
for any set of predicates $\Phi$ using public key encryption $\mathcal{E} =
(\textsf{Setup}_{\mathcal{E}}, \textsf{Encrypt}_{\mathcal{E}},
\textsf{Decrypt}_{\mathcal{E}})$ is described as follows.

\begin{description}
\item[\normalfont{\textsf{Setup}($1^{\lambda}$)}:] The setup algorithm
    first run $\textsf{Setup}_{\mathcal{E}}$ $m$ times to obtain $(PK_1,
    SK_1), \ldots, (PK_m, SK_m)$. Then it keeps $(SK_1, \ldots, SK_m)$ as a
    private key and publishes a public key as $\textsf{PK} = (PK_1, \ldots,
    PK_m)$.

\item[\normalfont{\textsf{GenToken}($j, \textsf{SK}, \textsf{PK}$)}:] The
    token generation algorithm takes as input an index $j$ of $P$ in $\Phi$
    and the secret key $\textsf{SK}$. It outputs a token as $\textsf{TK}_j
    = (j, SK_j)$

\item[\normalfont{\textsf{Encrypt}($x, M, \textsf{PK}$)}:] The encrypt
    algorithm takes as input a string $x \in \Sigma$, a message $M$ and the
    public key $\textsf{PK}$. For $i=1,\ldots, m$, it sets $C_j \leftarrow
    \textsf{Encrypt}_{\mathcal{E}} (PK_j, M)$ if $f_j(x) = 1$, or it sets
    $C_j \leftarrow \textsf{Encrypt}_{\mathcal{E}} (PK_j, \perp)$
    otherwise. It outputs a ciphertext as $\textsf{CT} = ( C_1, \ldots, C_m
    )$.

\item[\normalfont{\textsf{Query}($\textsf{CT}, \textsf{TK}_{\vect{\sigma}},
    \textsf{PK}$)}:] The query algorithm takes as input $\textsf{CT} =
    (C_1, \ldots, C_m)$ and $\textsf{TK}_j = (j, SK_j)$. It outputs
    $\textsf{Decrypt}_{\mathcal{E}} (SK_j, C_j)$.
\end{description}

\begin{theorem} \label{thm-bw-pe}
The predicate encryption scheme of Boneh and Waters is secure if
$\mathcal{E}$ is a semantically secure public key encryption and again chosen
plaintext attacks.
\end{theorem}

\section{Extreme Generalization of AIBE}

The second category is the extreme generalization of anonymous identity-based
encryption (AIBE) \cite{BonehW07,ShiW08,Ducas10,LeeL11}. This method was
introduced by Boneh and Waters to construct an HVE scheme \cite{BonehW07}.
They used the identity-based encryption (IBE) scheme of Boneh and Boyen
\cite{BonehB04e} and composite order bilinear groups to provide the anonymity
of ciphertexts. After that, Shi and Waters constructed a delegatable HVE
(dHVE) scheme \cite{ShiW08}. In composite order bilinear groups, the random
blinding property using subgroups provides the anonymity of ciphertexts and
the orthogonal property among subgroups provides the successful decryption.
However, it is inefficient to use composite order bilinear groups since the
group order of composite order bilinear groups should be larger than 1024
bits to defeat the integer factorization attacks. One way to overcome this
problem of inefficiency is to use prime order bilinear groups. Freeman
presented a general framework that converts cryptographic schemes from
composite order bilinear groups to prime order bilinear groups
\cite{Freeman10}. Ducas also showed that HVE schemes of composite order
bilinear groups are easily converted to prime order bilinear groups
\cite{Ducas10}. However, these conversion methods have a problem such that
they work under asymmetric bilinear groups that are particular kinds of prime
order bilinear groups \cite{GalbraithPS08}.

\subsection{HVE of Boneh and Waters}

Let $\Sigma = \Z_m$ for some integer $m$ and set $\Sigma_{*} = \Z_m \cup
\{*\}$. The HVE scheme of Boneh and Waters is described as follows.

\begin{description}
\item[\normalfont{\textsf{Setup}($1^{\lambda}$)}:] The setup algorithm
    first generates the bilinear group $\G$ of composite order $n=pq$ where
    $p$ and $q$ are random primes of bit size $\Theta(\lambda)$ and $p,q >
    m$. Next, it chooses random elements $v \in \G_p$, $(u_1, h_1, w_1),
    \ldots, (u_l, h_l, w_l) \in \G_p^2$, and a random exponent $\alpha \in
    \Z_p$. It keeps these as a secret key $\textsf{SK}$. Then it chooses
    random elements $R_v \in \G_q$ and $(R_{u,1}, R_{h,1}, R_{w,1}),
    \ldots, (R_{u,l}, R_{h,l}, R_{w,l}) \in \G_q^2$, and it publishes a
    public key \textsf{PK} with the description of the bilinear group $\G$
    as follows
    \begin{align*}
    \textsf{PK} = \Big(~
    &   V = v R_v,~
        \big\{ ( U_i = u_i R_{u,i},~ H_i = h_i R_{h,i},~ W_i = w_i R_{w,i} )
        \big\}_{i=1}^l,~ g_q,~ \Omega = e(v, g)^{\alpha}
    ~\Big).
    \end{align*}

\item[\normalfont{\textsf{GenToken}($\vect{\sigma}, \textsf{SK},
    \textsf{PK}$)}:] The token generation algorithm takes as input a vector
    $\vect{\sigma} = (\sigma_1, \ldots, \sigma_l) \in \Sigma_*^l$ and the
    secret key $\textsf{SK}$. It first selects random exponents $\{r_{i,1},
    r_{i,2}\}_{i=1}^l \in \Z_p$. Let $S$ be the set of indexes that are not
    wild card positions in the vector $\vect{\sigma}$. Then it outputs a
    token as
    \begin{align*}
    \textsf{TK}_{\vect{\sigma}} = \Big(~
    &   K_0 = g^{\alpha} (\prod_{i\in S} u_i^{\sigma_i} h_i)^{r_{i,1}} w_i^{r_{i,2}},~
        \big\{ K_{i,1} = v^{r_{i,1}},~ K_{i,2} = v^{r_{i,2}} \big\}_{i\in S}
    ~\Big).
    \end{align*}

\item[\normalfont{\textsf{Encrypt}($\vect{x}, M, \textsf{PK}$)}:] The
    encrypt algorithm takes as input a vector $\vect{x} = (x_1, \ldots,
    x_l) \in \Sigma^l$, a message $M \in \mathcal{M} \subseteq \G_T$ and
    the public key $\textsf{PK}$. It first chooses a random exponent $t \in
    \Z_n$ and random elements $Z_0, (Z_{1,1}, Z_{1,2}), \ldots, (Z_{l,1},
    Z_{l,2}) \in \G_q$ by raising $g_q$ to random elements from $\Z_n$.
    Next, it outputs a ciphertext as
    \begin{align*}
    \textsf{CT} = \Big(~
    &   C = \Omega^t M,~ C_0 = V^t Z_0,~
        \big\{ C_{i,1} = (U_i^{x_i} H_i)^t Z_{i,1},~ C_{i,2} = W_i^t Z_{i,2}
        \big\}_{i=1}^l ~\Big).
    \end{align*}

\item[\normalfont{\textsf{Query}($\textsf{CT}, \textsf{TK}_{\vect{\sigma}},
    \textsf{PK}$)}:] The query algorithm takes as input a ciphertext
    $\textsf{CT}$ and a token $\textsf{TK}_{\vect{\sigma}}$ of a vector
    $\vect{\sigma}$. It first computes
    \begin{align*}
    M \leftarrow C \cdot e(C_0, K_0)^{-1} \cdot
        \prod_{i \in S} e(C_{i,1}, K_{i,1}) e(C_{i,2}, K_{i,2}).
    \end{align*}
If $M \notin \mathcal{M}$, it outputs $\perp$ indicating that the predicate
$f_{\vect{\sigma}}$ is not satisfied. Otherwise, it outputs $M$ indicating
that the predicate $f_{\vect{\sigma}}$ is satisfied.
\end{description}

\begin{theorem}[\cite{BonehW07}] \label{thm-bw-hve}
The HVE scheme of Boneh and Waters is selectively secure under the decisional
cBDH assumption, the BSD assumption, and the decisional C3DH assumption.
\end{theorem}

\subsection{dHVE of Shi and Waters}

Let $\Sigma = \Z_m$ for some integer $m$ and set $\Sigma_{*} = \Z_m \cup
\{*\}$. The dHVE scheme of Shi and Waters is described as follows.

\begin{description}
\item[\normalfont{\textsf{Setup}($1^{\lambda}$)}:] The setup algorithm
    first generates the bilinear group $\G$ of composite order $n=pqr$
    where $p, q$ and $r$ are random primes of bit size $\Theta(\lambda)$
    and $p,q,r > m$. Next, it chooses random elements $v, w_1, w_2 \in
    \G_p$, $\{ u_i, h_i \} \in \G_p^2$, and a random exponent $\alpha \in
    \Z_p$. It keeps these as a secret key $\textsf{SK}$. Then it chooses
    random elements $R_v, R_{w,1}, R_{w,2}, \{ R_{u,i}, R_{h,i} \} \in
    \G_r$, and it publishes a public key \textsf{PK} with the description
    of the bilinear group $\G$ as follows
    \begin{align*}
    \textsf{PK} = \Big(~
    &   V = v R_v,~ W_1 = w_1 R_{w,1},~ W_2 = w_2 R_{w,2},~
        \{ (U_i = u_i R_{u,i},~ H_i = h_i R_{h,i}) \}_{i=1}^l,~
        g_q,~ g_r,~ \Omega = e(v, g)^{\alpha} ~\Big).
    \end{align*}

\item[\normalfont{\textsf{GenToken}($\vect{\sigma}, \textsf{SK},
    \textsf{PK}$)}:] The token generation algorithm takes as input a vector
    $\vect{\sigma} = (\sigma_1, \ldots, \sigma_l) \in \Sigma_*^l$ and the
    secret key $\textsf{SK}$. It first selects random exponents $r_1, r_2,
    \{ r_{3,i} \} \in \Z_p$ and random elements $Y_0, Y_1, Y_2, \{ Y_{3,i}
    \} \in \G_r$ by raising $g_r$ to random exponents in $\Z_n$. Let $S$ be
    the set of indexes that are not wild card positions in the vector
    $\vect{\sigma}$. Then it outputs a token as
    \begin{align*}
    &   K_0 = g^{\alpha} w_1^{r_1} w_2^{r_2}
        \prod_{i\in S} (u_i^{\sigma_i} h_i)^{r_{3,i}} Y_0,~
        K_1 = v^{r_1} Y_1,~
        K_2 = v^{r_2} Y_2,~
        \{ K_3 = v^{r_{3,i}} Y_{3,i} \}_{i \in S}.
    \end{align*}
    Let $S_?$ be the set of indexes that are delegatable fields. It selects
    random exponents $\{ s_{i,j} \} \in \Z_p$ and random values $\{
    Y_{0,i,u}, Y_{0,i,h}, Y_{1,j}, Y_{2,j}, \{ Y_{i,j} \} \} \in \G_r$.
    Next, it computes delegation components as
    \begin{align*}
    \forall j \in S_? :
    &   L_{0,j,u} = u_i^{s_{j,j}} Y_{j,u},~
        L_{0,j,h} = {w_1}^{s_{1,j}} {w_2}^{s_{2,j}}
            \prod_{i \in S} (u_i^{\sigma_i} h_i)^{s_{j,i}}
            h_j^{s_{j,j}} Y_{j,h},~ \\
    &   L_{1,j} = v^{s_{1,j}} Y_{1,j},~
        L_{2,j} = v^{s_{2,j}} Y_{2,j},~
        \big\{ L_{3,j,i} = v^{s_{3,j,i}} Y_{j,i} \big\}_{i \in S \cup \{j\}}.
    \end{align*}
Finally, it outputs a token as
    \begin{align*}
    \textsf{TK}_{\vect{\sigma}} = \Big(~
    &   K_0,~ K_1,~ K_2,~ \{ K_{3,i} \}_{i \in S},~
        \big\{ L_{0,j,u},~ L_{0,j,h},~ L_{1,j},~ L_{2,j},~
            \{ L_{3,j,i} \}_{i \in S \cup \{j\}}
        \big\}_{j \in S_?}    ~\Big).
    \end{align*}

\item[\normalfont{\textsf{Delegate}($\vect{\sigma}',
    \textsf{TK}_{\vect{\sigma}}, \textsf{PK}$)}:] The delegation algorithm
    takes as input an attribute vector $\vect{\sigma}' = (\sigma_1, \ldots,
    \sigma_l) \in \Sigma_{?,*}^l$ and a token
    $\textsf{TK}_{\vect{\sigma}}$. Without loss of generality, we assume
    that $\sigma'$ fixes only one delegatable field of $\sigma$. It is
    clear that we can perform delegation on multiple fields if we have an
    algorithm to perform delegation on one field. Suppose $\sigma'$ fixes
    the $k$-th index of $\sigma$. If the $k$-th index of $\sigma'$ is set
    to $*$, that is, a wild-card field, then it can perform delegation by
    simply removing the delegation components that correspond to $k$-th
    index. Otherwise, that is, if the $k$-th index of $\sigma'$ is set to
    some value in $\Sigma$, then it perform delegation as follows.

    Let $S$ be the set of indexes that are not delegatable fields and
    wild-card fields in the vector $\vect{\sigma'}$. Note that $k \in S$.
    It selects random exponents $\mu \in \Z_n$ and random elements $Y_0,
    Y_1, Y_2, \{ Y_{3,i} \} \in \G_r$ and updates the token as
    \begin{align*}
    &   \tilde{K}_0 = K_0 (L_{0,k,u}^{\sigma_k} L_{0,k,h})^{\mu} Y_0,~
        \tilde{K}_1 = K_1 L_{1,k}^{\mu} Y_1,~
        \tilde{K}_2 = K_2 L_{2,k}^{\mu} Y_2,~ \\
    &   \tilde{K}_{3,k} = L_{3,k,k}^{\mu} Y_{3,k},~
        \big\{ \tilde{K}_{3,i} = K_{3,i} L_{3,k,i}^{\mu} Y_{3,i}
        \big\}_{i \in S \setminus \{k\}}.
    \end{align*}
    Let $S_?$ be the set of indexes that are delegatable fields in the
    vector $\vect{\sigma'}$. It selects random exponents $\{ \tau_j \} \in
    \Z_n$ and random elements $\{ Y_{0,j,u}, Y_{0,j,h}, Y_{1,j}, Y_{2,j},
    \lb \{ Y_{3,j,i} \}_{i\in S\cup \{j\}} \}_{j\in S_?} \in \G_r$ and
    re-randomize the delegation components of the token as
    \begin{align*}
    \forall j \in S_? :
    &   \tilde{L}_{0,j,u} = L_{0,j,u}^{\mu} Y_{0,j,u},~
        \tilde{L}_{0,j,h} = L_{0,j,h}^{\mu}
            (L_{0,k,u}^{\sigma_k} L_{0,k,h})^{\tau_j} Y_{0,j,h},~ \\
    &   \tilde{L}_{1,j} = L_{1,j}^{\mu} L_{1,j}^{\tau_j} Y_{1,j},~
        \tilde{L}_{2,j} = L_{2,j}^{\mu} L_{2,j}^{\tau_j} Y_{2,j},~ \\
    &   \tilde{L}_{3,j,j} = L_{3,j,j}^{\mu} Y_{3,j,j},~
        \tilde{L}_{3,j,k} = L_{3,j,k}^{\tau_j} Y_{3,j,k},~
        \big\{
        \tilde{L}_{3,j,i} = L_{3,j,i}^{\mu} L_{3,j,k}^{\tau_j} Y_{3,j,i}
        \big\}_{i\in S\setminus \{k\}}.
    \end{align*}
    Finally, it outputs a token as
    \begin{align*}
    \textsf{TK}_{{\sigma}'} = \Big(~
    &   \tilde{K}_0,~ \tilde{K}_1,~ \tilde{K}_2,~
        \{ \tilde{K}_{3,i} \}_{i\in S},~
        \big\{
        \tilde{L}_{0,j,h}, \tilde{L}_{0,j,u},~
        \tilde{L}_{1,j},~ \tilde{L}_{2,j},~
        \{ \tilde{L}_{3,j,i} \}_{i\in S\cup \{j\}}
        \big\}_{j\in S_?}
    ~\Big).
    \end{align*}

\item[\normalfont{\textsf{Encrypt}($\vect{x}, M, \textsf{PK}$)}:] The
    encrypt algorithm takes as input a vector $\vect{x} = (x_1, \ldots,
    x_l) \in \Sigma^l$, a message $M \in \mathcal{M} \subseteq \G_T$ and
    the public key $\textsf{PK}$. It first chooses a random exponent $t \in
    \Z_n$ and random elements $Z_0, Z_1, Z_2, \{ Z_{3,i} \} \in \G_q$.
    Next, it outputs a ciphertext as
    \begin{align*}
    \textsf{CT} = \Big(~
    &   C = \Omega^t M,~ C_0 = V^t Z_0,~ C_1 = W_1^t Z_1,~ C_2 = W_2^t Z_2,~
        \{ C_{3,i} = (U_i^{x_i} H_i)^t Z_{3,i} \}_{i=1}^l
    ~\Big).
    \end{align*}

\item[\normalfont{\textsf{Query}($\textsf{CT}, \textsf{TK}_{\vect{\sigma}},
    \textsf{PK}$)}:] The query algorithm takes as input a ciphertext
    $\textsf{CT}$ and a token $\textsf{TK}_{\vect{\sigma}}$ of a vector
    $\vect{\sigma}$. It first computes
    \begin{align*}
    M \leftarrow C \cdot e(C_0, K_0)^{-1} \cdot
        e(C_1, K_1) \cdot e(C_2, K_2) \cdot \prod_{i\in S} e(C_{3,i}, K_{3,i}).
    \end{align*}
If $M \notin \mathcal{M}$, it outputs $\perp$ indicating that the predicate
$f_{\vect{\sigma}}$ is not satisfied. Otherwise, it outputs $M$ indicating
that the predicate $f_{\vect{\sigma}}$ is satisfied.
\end{description}

\begin{theorem}[\cite{ShiW08}] \label{thm-sw-dhve}
The delegatable HVE scheme of Shi and Waters is selectively secure under the
decisional cBDH assumption, the BSD assumption, and the decisional C3DH
assumption.
\end{theorem}

\section{HVE from Inner Product Encryption}

The third category is to use inner-product encryption (IPE) \cite{KatzSW08,
OkamotoT09,Park11}. IPE is a kind of predicate encryption and it enable the
evaluation of inner-product predicates between the vector of ciphertexts and
the vector of tokens. Katz et al. constructed the first IPE scheme under
composite order bilinear groups \cite{KatzSW08}. After that Okamoto and
Takashima constructed a hierarchical IPE scheme using dual pairing vector
spaces \cite{OkamotoT09}. Recently, Park proposed an IPE scheme under prime
order bilinear groups and proved its security under the well-known
assumptions \cite{Park11}. The main idea of converting an IPE scheme to an
HVE scheme is to construct a predicate of conjunctive equality using a
predicate of inner product \cite{KatzSW08}. Though, the expressiveness of IPE
enables the evaluations of predicates like conjunctive, disjunctive,
polynomials, and CNF/DNF formula, IPE has a weakness such that the number of
pairing operations is linearly dependent on the number of attributes in the
ciphertexts.

\subsection{IPE of Katz, Sahai, and Waters}

Let $\Sigma = \Z_m$ for some integer $m$ and set $\Sigma_{*} = \Z_m \cup
\{*\}$. The IPE scheme of Katz, Sahai, and Waters is described as follows.

\begin{description}
\item[\normalfont{\textsf{Setup}($1^{\lambda}$)}:] The setup algorithm
    first generates the bilinear group $\G$ of composite order $n=pqr$
    where $p, q$ and $r$ are random primes of bit size $\Theta(\lambda)$
    and $p,q,r > m$. Let $g_p, g_q$, and $g_r$ be generators of $\G_p,
    \G_q$, and $\G_r$ respectively. It chooses random elements $\{ h_{1,i},
    h_{2,i} \}_{i=1}^l, h \in \G_p$ and a random exponent $\gamma \in
    \Z_p$. It keeps these as a secret key $\textsf{SK}$. Then it chooses
    random elements $\{ R_{1,i}, R_{2,i} \}_{i=1}^l, R_q \in \G_r$, and it
    publishes a public key \textsf{PK} with the description of the bilinear
    group $\G$ as follows
    \begin{align*}
    \textsf{PK} = \Big(~
    &   g = g_p,~ Q = g_q R_{q},~ g_r,~
        \{ (H_{1,i} = h_{1,i} R_{h,1,i},~ H_{2,i} = h_{2,i} R_{h,2,i}) \}_{i=1}^l,~
        \Omega = e(g, h)^{\gamma} ~\Big).
    \end{align*}

\item[\normalfont{\textsf{GenToken}($\vect{\sigma}, \textsf{SK},
    \textsf{PK}$)}:] The token generation algorithm takes as input a vector
    $\vect{\sigma} = (\sigma_1, \ldots, \sigma_l) \in \Sigma_*^l$ and the
    secret key $\textsf{SK}$. It first selects random exponents $\{
    r_{1,i}, r_{2,i} \}_{i=1}^l \in \Z_p$, random elements $R_5 \in \G_r,
    Q_6 \in \G_q$, and random exponents $f_1, f_2 \in \Z_q$. Then it
    outputs a token as
    \begin{align*}
    \textsf{TK}_{\vect{\sigma}} = \Big(~
    &   K_0 = h^{\gamma} R_5 Q_6
        \prod_{i=1}^l (h_{1,i}^{-r_{1,i}} h_{2,i}^{-r_{2,i}}),~
        \{ K_{1,i} = g^{r_{1,i}} g_q^{f_1 \sigma_i},~
           K_{2,i} = g^{r_{2,i}} g_q^{f_2 \sigma_i} \}_{i=1}^l ~\Big).
    \end{align*}

\item[\normalfont{\textsf{Encrypt}($\vect{x}, M, \textsf{PK}$)}:] The
    encrypt algorithm takes as input a vector $\vect{x} = (x_1, \ldots,
    x_l) \in \Sigma^l$, a message $M \in \mathcal{M} \subseteq \G_T$ and
    the public key $\textsf{PK}$. It first chooses a random exponent $t,
    \alpha, \beta \in \Z_n$ and random elements $\{ R_{3,i}, R_{4,i}
    \}_{i=1}^l \in \G_r$. Next, it outputs a ciphertext as
    \begin{align*}
    \textsf{CT} = \Big(~
    &   C = \Omega^t M,~ C_0 = g^t,~
        \{ C_{1,i} = H_{1,i}^t Q^{\alpha \sigma_i} R_{3,i},~
           C_{2,i} = H_{2,i}^t Q^{\beta \sigma_i} R_{4,i} \}_{i=1}^l ~\Big).
    \end{align*}

\item[\normalfont{\textsf{Query}($\textsf{CT}, \textsf{TK}_{\vect{\sigma}},
    \textsf{PK}$)}:] The query algorithm takes as input a ciphertext
    $\textsf{CT}$ and a token $\textsf{TK}_{\vect{\sigma}}$ of a vector
    $\vect{\sigma}$. It first computes
    \begin{align*}
    M \leftarrow C \cdot e(C_0, K_0) \cdot
        \prod_{i=1}^l \big( e(C_{1,i}, K_{1,i}) \cdot e(C_{2,i}, K_{2,i}) \big).
    \end{align*}
    If $M \notin \mathcal{M}$, it outputs $\perp$ indicating that the
    predicate $f_{\vect{\sigma}}$ is not satisfied. Otherwise, it outputs
    $M$ indicating that the predicate $f_{\vect{\sigma}}$ is satisfied.
\end{description}

\begin{theorem}[\cite{KatzSW08}] \label{thm-ksw-ipe}
The IPE scheme of Katz, Sahai, and Waters is selectively secure under the two
static assumptions.
\end{theorem}

\subsection{Conversion from IPE to HVE}

The hidden vector encryption $\mathcal{HVE} = (\textsf{Setup},
\textsf{KeyGen}, \textsf{Encrypt}, \textsf{Query})$ using inner product
encryption $\mathcal{IPE} = (\textsf{Setup}', \textsf{KeyGen}',
\textsf{Encrypt}', \textsf{Query}')$ is described as follows.

\begin{description}
\item[\normalfont{\textsf{Setup}($1^{\lambda}, l$)}:] The setup algorithm
    first run $\textsf{Setup}'(1^{\lambda}, 2l)$ and obtains
    $(\textsf{PK}', \textsf{SK}')$. Then it keeps $\textsf{SK}'$ as a
    secret key and publishes a public key as $\textsf{PK} = \textsf{PK}'$.

\item[\normalfont{\textsf{GenToken}($\vect{\sigma}, \textsf{SK},
    \textsf{PK}$)}:] The token generation algorithm takes as input a vector
    $\vect{\sigma} = (\sigma_1, \ldots, \sigma_l) \in \Sigma_{*}^l$, the
    secret key $\textsf{SK}$, and the public key $\textsf{PK}$. It first
    convert the vector $\vect{\sigma}$ to a vector $\vect{\sigma'} =
    (\sigma'_1, \ldots, \sigma'_{2l}) \in \Sigma^{2l}$ as follows:
    \begin{align*}
    &   \mbox{if } \sigma_i \neq *, \mbox{ then }
        \sigma'_{2i-1} = 1,~ \sigma'_{2i} = \sigma_i.\\
    &   \mbox{if } \sigma_i = *,    \mbox{ then }
        \sigma'_{2i-1} = 0,~ \sigma'_{2i} = 0.
    \end{align*}
    Then it outputs a token obtained by running $\textsf{GenToken}'
    (\vect{\sigma'}, \textsf{SK}', \textsf{PK}')$.

\item[\normalfont{\textsf{Encrypt}($\vect{x}, M, \textsf{PK}$)}:] The
    encrypt algorithm takes as input a vector $\vect{x} = (x_1, \ldots,
    x_l) \in \Sigma^l$, a message $M$ and the public key $\textsf{PK}$. It
    first choose random values $r_1, \ldots, r_l \in \Sigma$ and construct
    a vector $\vect{x}' = (x'_1, \ldots, x'_{2l})$ as follows:
    \begin{align*}
    &   x'_{2i-1} = -r_i \cdot x_i,~~ x'_{2i} = r_i.
    \end{align*}
    It outputs a ciphertext by running $\textsf{Encrypt}'(\vect{x}', M,
    \textsf{PK}')$.

\item[\normalfont{\textsf{Query}($\textsf{CT}, \textsf{TK}_{\vect{\sigma}},
    \textsf{PK}$)}:] The query algorithm outputs
    $\textsf{Query}'(\textsf{CT}, \textsf{TK}_{\vect{\sigma}},
    \textsf{PK})$.
\end{description}

%% file: chap-hve-short-token.tex
\section{Overview}

In this chapter, we construct efficient HVE schemes that have short tokens
and prove their selective model security under simple assumptions. Our
constructions are algebraically similar to the one of Shi and Waters, but the
cost of decrypt operation in our constructions is constant.

Previous research on HVE has mainly focused on improving the expressiveness
of predicates or providing additional properties like the delegation. To
apply HVE schemes to real applications, it is important to construct an
efficient HVE scheme. One can measure the efficiency of HVE in terms of the
ciphertext size, the token size, and the number of pairing operations in
decryption. Let $l$ be the number of attributes in the ciphertext and $s$ be
the number of attributes in the token except the wild card attribute. Then
the efficiency of previous HVE schemes is compared in Table
\ref{tab-ehve-comp}. Theoretically, the number of group elements in
ciphertext should be proportional to the number of attributes in the
ciphertexts, so the minimum size of ciphertext is $l|\G|+O(1)$. However, the
token size and the number of pairing operations in decryption can be
constant, that is, independent of $l$. Therefore constructing an HVE scheme
with the constant size of tokens and the constant number of pairing
operations is an important problem to solve.

\begin{table}[tp]
\caption{Comparison between previous HVE schemes and ours}
\label{tab-ehve-comp}
\vs \small \addtolength{\tabcolsep}{12.5pt}
\renewcommand{\arraystretch}{1.3}
    \begin{tabularx}{6.50in}{lcccc}
    \hline
    Scheme & Group Order & Ciphertext Size & Token Size & No. of Pairing \\
    \hline
    BW-HVE \cite{BonehW07}   & $pq$  & $2l |\G| +O(1)$ & $(2s+1) |\G|$ & $2s+1$ \\
    KSW-IPE \cite{KatzSW08} & $pqr$ & $4l |\G| +O(1)$ & $(4l+1) |\G|$ & $4l+1$ \\
    SW-dHVE \cite{ShiW08}  & $pqr$ & $ l |\G| +O(1)$ & $ (s+3) |\G|$ & $s+3$ \\
    IP-HVE \cite{IovinoP08}   & $p$   & $2l |\G| +O(1)$ & $  (2s) |\G|$ & $2s$ \\
    OT-IPE \cite{OkamotoT09}   & $p$   & $2l |\G| +O(1)$ & $(2l+3) |\G|$ & $2l+3$ \\
    %\hdashline
    Ours                 & $pqr$ & $ l |\G| +O(1)$ & $4 |\G|$      & $4$ \\
    Ours                 & $p$   & $ l |\G| +O(1)$ & $4 |\hat{\G}|$ & $4$ \\
    \hline
    \multicolumn{5}{l}{$p,q,r$ = prime values, $l$ = no. of attributes in
    ciphertext, $s$ = no. of attributes in token}
    \end{tabularx}
\end{table}

We propose HVE schemes that have the constant size of tokens and the constant
cost of pairing operations. Our first construction is based on composite
order bilinear groups whose order is a product of three primes. The
ciphertext consists of $l+O(1)$ group elements, the token consists of four
group elements, and the decryption requires four pairing computations. Our
second one is based on prime order asymmetric bilinear groups where
isomorphisms between two groups are not efficiently computable.

Though our construction in composite order bilinear groups is algebraically
similar to the one by Shi and Waters in \cite{ShiW08}, we achieved the
constant size of tokens and the constant cost of decryption, in contrast to
the construction of Shi and Waters. The main technique for our constructions
is to use the same random value for each attributes in the token. In
contrast, the construction of Shi and Waters used different random values for
each attributes. This technique is reminiscent of the one that enables the
design of HIBE with the constant size of ciphertexts in \cite{BonehBG05}.
However, it is not easy to prove the security of HVE when the same random
value is used in the token, since HVE should provide an additional security
property, namely \textit{attribute hiding}, that is, the ciphertext does not
reveal any information about the attributes.

\section{HVE in Composite Order Groups}

In this section, we construct an HVE scheme based on composite order bilinear
groups and prove security under the decisional cBDH, BSD, and decisional C3DH
assumptions. Our construction has a similar algebraic structure to the
construction of Shi and Waters \cite{ShiW08}, but ours has the constant size
of tokens and the constant number of pairing operations.

\subsection{Construction}

Let $\Sigma = \Z_m$ for some integer $m$ and set $\Sigma_{*} = \Z_m \cup
\{*\}$. Our scheme is described as follows.

\begin{description}
\item[\normalfont{\textsf{Setup}($1^{\lambda}$)}:] The setup algorithm
    first generates the bilinear group $\G$ of composite order $n=pqr$
    where $p, q$ and $r$ are random primes of bit size $\Theta(\lambda)$
    and $p,q,r > m$. Next, it chooses random elements $v, w_1, w_2 \in
    \G_p$, $(u_1, h_1), \ldots, (u_l, h_l) \in \G_p^2$, and exponents
    $\alpha, \beta \in \Z_p$. It keeps these as a secret key $\textsf{SK}$.
    Then it chooses random elements $R_v, R_{w,1}, R_{w,2} \in \G_q$ and
    $(R_{u,1}, R_{h,1}), \ldots, (R_{u,l}, R_{h,l}) \in \G_q^2$, and it
    publishes a public key \textsf{PK} with the description of the bilinear
    group $\G$ as follows
    \begin{align*}
    \textsf{PK} = \Big(~
    &   V = v R_v,~ W_1 = w_1 R_{w,1},~ W_2 = w_2 R_{w,2},~
        \big\{ ( U_i = u_i R_{u,i},~ H_i = h_i R_{h,i} ) \big\}_{i=1}^{l},~ \\
    &   g_q,~ g_r,~ \Omega = e(v, g)^{\alpha \beta} ~\Big).
    \end{align*}

\item[\normalfont{\textsf{GenToken}($\vect{\sigma}, \textsf{SK},
    \textsf{PK}$)}:] The token generation algorithm takes as input a vector
    $\vect{\sigma} = (\sigma_1, \ldots, \sigma_l) \in \Sigma_*^l$ and the
    secret key $\textsf{SK}$. It first selects random exponents $r_1, r_2,
    r_3 \in \Z_p$ and random elements $Y_0, Y_1, Y_2, Y_3 \in \G_r$ by
    raising $g_r$ to random exponents in $\Z_n$. Let $S$ be the set of
    indexes that are not wild card positions in the vector $\vect{\sigma}$.
    Then it outputs a token as
    \begin{align*}
    \textsf{TK}_{\vect{\sigma}} = \Big(~
    &   K_0 = g^{\alpha \beta} w_1^{r_1} w_2^{r_2}
        (\prod_{i\in S} u_i^{\sigma_i} h_i)^{r_3} Y_0,~
        K_1 = v^{r_1} Y_1,~ K_2 = v^{r_2} Y_2,~
        K_3 = v^{r_3} Y_3
    ~\Big).
    \end{align*}

\item[\normalfont{\textsf{Encrypt}($\vect{x}, M, \textsf{PK}$)}:] The
    encrypt algorithm takes as input a vector $\vect{x} = (x_1, \ldots,
    x_l) \in \Sigma^l$, a message $M \in \mathcal{M} \subseteq \G_T$ and
    the public key $\textsf{PK}$. It first chooses a random exponent $t \in
    \Z_n$ and random elements $Z_0, Z_1, Z_2, Z_{3,1}, \ldots, Z_{3,l} \in
    \G_q$ by raising $g_q$ to random elements from $\Z_n$. Next, it outputs
    a ciphertext as
    \begin{align*}
    \textsf{CT} = \Big(~
    &   C = \Omega^t M,~
        C_0 = V^t Z_0,~ C_1 = W_1^t Z_1,~ C_2 = W_2^t Z_2,~
        \big\{ C_{3,i} = (U_i^{x_i} H_i)^t Z_{3,i} \big\}_{i=1}^l
    ~\Big).
    \end{align*}

\item[\normalfont{\textsf{Query}($\textsf{CT}, \textsf{TK}_{\vect{\sigma}},
    \textsf{PK}$)}:] The query algorithm takes as input a ciphertext
    $\textsf{CT}$ and a token $\textsf{TK}_{\vect{\sigma}}$ of a vector
    $\vect{\sigma}$. It first computes
    \begin{align*}
    M \leftarrow C \cdot e(C_0, K_0)^{-1} \cdot
        e(C_1, K_1) \cdot e(C_2, K_2) \cdot e(\prod_{i \in S} C_{3,i}, K_3).
    \end{align*}
If $M \notin \mathcal{M}$, it outputs $\perp$ indicating that the predicate
$f_{\vect{\sigma}}$ is not satisfied. Otherwise, it outputs $M$ indicating
that the predicate $f_{\vect{\sigma}}$ is satisfied.
\end{description}

\begin{remark}
In our construction, we limited the finite set $\Sigma$ of attributes to be
$\Z_m$. If we use a collision-resistant hash function, then we can easily
expand this space to all of $\{0,1\}^*$ when $m$ is large enough to contain
the range of the hash function.
\end{remark}

\subsection{Correctness}

If $f_{\vect{\sigma}}(\vect{x}) = 1$, then the following simple calculation
shows that
$\textsf{Query}(\textsf{CT},\textsf{TK}_{\vect{\sigma}},\textsf{PK}) = M$ as
    \begin{align*}
    \lefteqn{ e(C_0, K_0)^{-1} \cdot e(C_1, K_1) \cdot e(C_2, K_2) \cdot
    e(\prod_{i \in S} C_{3,i}, K_3) } \\
    &=  e(v^t, g^{\alpha \beta} w_1^{r_1} w_2^{r_2}
        (\prod_{i\in S} u_i^{\sigma_i} h_i)^{r_3})^{-1}
        \cdot e(w_1^t, v^{r_1}) \cdot e(w_2^t, v^{r_2})
        \cdot e(\prod_{i \in S} (u_i^{x_i} h_i)^t, v^{r_3}) \\
    &=  e(v^t, g^{\alpha \beta})^{-1} \cdot
        e((\prod_{i \in S} u_i^{(-\sigma_i+x_i)})^{r_3}, v^t)
     =  e(v^t, g^{\alpha \beta})^{-1}.
    \end{align*}
Otherwise, that is $f_{\vect{\sigma}}(\vect{x}) = 0$, then we can use Lemma
5.2 in \cite{BonehW07} to show that the probability of $\textsf{Query}
(\textsf{CT}, \textsf{TK}_{\vect{\sigma}}, \lb \textsf{PK}) \neq \perp$ is
negligible by limiting $|\mathcal{M}|$ to less than $|\G_T|^{1/4}$.

\subsection{Security}

\begin{theorem} \label{thm-ehve-comp}
The above HVE construction is selectively secure under the decisional cBDH
assumption, the BSD assumption, and the decisional C3DH assumption.
\end{theorem}

\begin{proof}
Suppose there exists an adversary that distinguishes the original selective
security game. Then the adversary commits two vectors $\vect{x}_0 = (x_{0,1},
\ldots, x_{0,l})$ and $\vect{x}_1 =(x_{1,1}, \ldots, x_{1,l}) \in \Sigma^l$
at the beginning of the game. Let $X$ be the set of indexes $i$ such that
$x_{0,i} = x_{1,i}$ and $\overline{X}$ be the set of indexes $i$ such that
$x_{0,i} \neq x_{1,i}$. The proof uses a sequence of four games to argue that
the adversary cannot win the original security game. Each individual game is
described as follows.

\vs \noindent $\textsf{Game}_0$. This game denotes the original selective
security game that is defined in Section \ref{back-hve}.

\svs \noindent $\textsf{Game}_1$. We first modify $\textsf{Game}_0$ slightly
into a new game $\textsf{Game}_1$. $\textsf{Game}_1$ is almost identical to
$\textsf{Game}_0$ except in the way the challenge ciphertext elements are
generated. In $\textsf{Game}_1$, if $M_0 \neq M_1$, then the simulator
generates the challenge ciphertext element $C$ by  multiplying a random
element in $\G_T$, and it generates the rest of the ciphertext elements as
usual. If $M_0 = M_1$, then the challenge ciphertext is generated correctly.

\svs \noindent $\textsf{Game}_2$. Next, we modify $\textsf{Game}_1$ into a
new game $\textsf{Game}_2$. $\textsf{Game}_2$ is almost identical to
$\textsf{Game}_1$ except in the way the tokens are generated. Let $S$ be the
set of indexes that are not wild card positions of the token query vector
$\vect{\sigma}$. Then any token query by the adversary must satisfy one of
the following two cases:
\begin{itemize}
\item Type 1 $f_{\vect{\sigma}}(\vect{x}_0) = f_{\vect{\sigma}}(\vect{x}_1)
    = 1$. In this case, $S \cap \overline{X} = \emptyset$ and $\sigma_j =
    x_{0,j} = x_{1,j}$ for all index $j \in S \cap X$.
\item Type 2 $f_{\vect{\sigma}}(\vect{x}_0) = f_{\vect{\sigma}}(\vect{x}_1)
    = 0$. In this case, there exists an index $j \in S$ such that $\sigma_j
    \neq x_{\gamma,j}$ for all $\gamma \in \{0,1\}$.
\end{itemize}
In $\textsf{Game}_2$, if the adversary requests the Type 1 token query, then
the simulator chooses two exponents $r_1$ and $r_2$ not independently at
random, but in a correlated way as $r_1 = \pi r_2$ for a fixed value $\pi$.
The simulator can use this correlation to simulate this game. However, the
adversary cannot distinguish this correlation because of random blinding
elements $\G_r$ in the token.

\svs \noindent $\textsf{Game}_3$. We modify $\textsf{Game}_2$ into a game
$\textsf{Game}_3$. $\textsf{Game}_2$ and $\textsf{Game}_3$ are identical
except in the challenge ciphertext. In $\textsf{Game}_3$, the simulator
creates the ciphertext according to the following distribution as
    \begin{align*}
    C_1 = W_1^t g_p^{\rho} Z_1,~~ C_2 = W_2^t g_p^{-\rho \pi}Z_2,
    \end{align*}
where $\rho$ is a random value in $\Z_p$ and $\pi$ is the fixed value in
$\Z_p$ but $\pi$ is hidden from the adversary.

\svs \noindent $\textsf{Game}_4$. We now define a new game $\textsf{Game}_4$.
$\textsf{Game}_4$ differs from $\textsf{Game}_3$ in that for all $i \in
\overline{X}$, the ciphertext component $C_i$ is replaced by a random element
from $\G_{pq}$. Note that in $\textsf{Game}_4$, the ciphertext gives no
information about the vector $\vect{x}_{\gamma}$ or the message $M_{\gamma}$
encrypted. Therefore, the adversary can win $\textsf{Game}_4$ with
probability at most $1/2$.

\vs Through the following four lemmas, we will prove that it is hard to
distinguish $\textsf{Game}_{i-1}$ from $\textsf{Game}_{i}$ under the given
assumptions. Thus, the proof is easily obtained by the following four lemmas.
This completes our proof.
\end{proof}

\begin{lemma} \label{lem-ehve-comp-1}
If the decisional cBDH assumption and the BSD assumption hold, then no
polynomial-time adversary can distinguish between $\textsf{Game}_0$ and
$\textsf{Game}_1$ with a non-negligible advantage.
\end{lemma}

\begin{proof}
For this lemma, we additionally define a sequence of games
$\textsf{Game}_{0,0}, \textsf{Game}_{0,1}$, and $\textsf{Game}_{0,2}$ where
$\textsf{Game}_{0,0} = \textsf{Game}_{0}$. $\textsf{Game}_{0,1}$ and
$\textsf{Game}_{0,2}$ are almost identical to $\textsf{Game}_{0,0}$ except in
the way the challenge ciphertext is generated. In $\textsf{Game}_{0,1}$, if
$M_0 \neq M_1$, then the simulator generates the challenge ciphertext element
$C$ by multiplying a random element in $\G_{T,p}$, and it generates the rest
of the ciphertext elements as usual. If $M_0 = M_1$, then the challenge
ciphertext is generated correctly. In $\textsf{Game}_{0,2}$, if $M_0 \neq
M_1$, then the simulator generates the challenge ciphertext element $C$ as a
random elements from $\G_T$ instead of $\G_{T,p}$, and it generates the rest
of the ciphertext elements as usual. If $M_0 = M_1$, then the challenge
ciphertext is generated correctly. It is not hard to see that
$\textsf{Game}_{0,2}$ is identical to $\textsf{Game}_1$.

\vs Suppose there exists an adversary $\mathcal{A}$ that distinguishes
between $\textsf{Game}_{0,0}$ and $\textsf{Game}_{0,1}$ with a non-negligible
advantage. A simulator $\mathcal{B}$ that solves the decisional cBDH
assumption using $\mathcal{A}$ is given: a challenge tuple $D = ((p,q,r, \G,
\G_T, e), g_p, g_q, \lb g_r, g_p^a, g_p^b, g_p^c)$ and $T$ where $T = e(g_p,
g_p)^{abc}$ or $T = R \in \G_{T,p}$. Then $\mathcal{B}$ that interacts with
$\mathcal{A}$ is described as follows.

\begin{description}
\item [Init:] $\mathcal{A}$ gives two vectors $\vect{x}_0 = (x_{0,1},
    \ldots, x_{0,l}), \vect{x}_1 =(x_{1,1}, \ldots, x_{1,l}) \in \Sigma^l$.
    $\mathcal{B}$ then flips a random coin $\gamma$ internally.

\item [Setup:] $\mathcal{B}$ first chooses random elements $R_v, R_{w,1},
    R_{w,2} \in \G_q$, $(R_{u,1}, R_{h,1}), \ldots, (R_{u,l}, R_{h,l}) \in
    \G_q^{2}$, and random exponents $v', w'_1, w'_2 \in \Z_n$, $(u'_1,
    h'_1), \ldots, (u'_l, h'_l) \in \Z_n^2$. Next, it publishes the group
    description $(n, \G, \G_T, e)$ and a public key as
    \begin{align*}
    &   V = g_p^{v'}R_v,~ W_1 = g_p^{w'_1}R_{w,1},~ W_2 = g_p^{w'_2}R_{w,2},~
        \{(U_i = (g_p^a)^{u'_i}R_{u,i},~
           H_i = g_p^{h'_i}(g_p^a)^{-u'_i x_{\gamma,i}}R_{h,i} ) \},~ \\
    &   g_q,~ g_r,~ \Omega = e(g_p^a, g_p^b)^{v'}.
    \end{align*}

\item [Query 1:] $\mathcal{A}$ adaptively requests a token for a vector
    $\vect{\sigma} = (\sigma_1, \ldots, \sigma_l) \in \Sigma_*^l$ to
    $\mathcal{B}$. Let $S$ be the set of indexes that are not wild card
    positions.

\begin{description}
\item [Type 1] If $\mathcal{A}$ requests a Type 1 query, then
    $\mathcal{B}$ simply aborts and takes a random guess. The reason for
    this is by our definition such as if a Type 1 query is made then the
    challenge messages $M_0, M_1$ will be equal. However, in this case
    the games $\textsf{Game}_0$ and $\textsf{Game}_1$ are identical, so
    there can be no difference in the adversary's advantage.

\item [Type 2] If $\mathcal{A}$ requests a Type 2 query, then there
    exists an index $j \in S$ such that $\sigma_j \neq x_{\gamma,j}$. Let
    $\Delta = \sum_{i \in S} u'_i (\sigma_i - x_{\gamma,i}) \in \Z_p$.
    Note that $\Delta \neq 0$ except with negligible probability. If
    $\Delta \neq 0$, then $\mathcal{B}$ chooses random exponents $r'_1,
    r'_2, r'_3 \in \Z_p$ and random elements $Y_0, Y_1, Y_2, Y_3 \in
    \G_r$. Next, it creates a token as
    \begin{align*}
    K_0 =& g_p^{w'_1 r'_1} g_p^{w'_2 r'_2}
           (g_p^b)^{-{\sum_{i \in S} h'_i}/{\Delta}}
           \prod_{i \in S} ((g_p^a)^{u'_i (\sigma_i - x_{\gamma,i})}
           g_p^{h'_i})^{r'_3} Y_0, \\
    K_1 =& g_p^{v' r'_1} Y_1,~~ K_2 = g_p^{v' r'_2} Y_2,~~
    K_3 =  g_p^{v' r'_3} (g_p^b)^{-{v'}/{\Delta}} Y_3.
    \end{align*}
    Note that it can compute $\Delta^{-1}$ since it knows $p$. To show
    that the above token is the same as the real scheme, we define the
    randomness of the token as
    \begin{align*}
    r_1 = r'_1 \mod p,~~ r_2 = r'_2 \mod p,~~ r_3 = r'_3 - b/\Delta \mod p.
    \end{align*}
    It is obvious that $r_1, r_2, r_3$ are all uniformly distributed if
    $r'_1, r'_2, r'_3$ are independently chosen at random. The following
    calculation shows that the above token is correctly distributed as
    the token in the real scheme as
    \begin{align*}
    K_0 =& g_p^{ab} w_1^{r'_1} w_2^{r'_2} \prod_{i \in S}
           \big( (g_p^a)^{u'_i(\sigma_i - x_{\gamma,i})} g_p^{h'_i}
           \big)^{r'_3 - b/\Delta} Y_0 \\
        =& g_p^{ab} w_1^{r'_1} w_2^{r'_2} g_p^{-ab} (g_p^b)^{-{\sum_{i \in S} h'_i}/{\Delta}}
        \prod_{i \in S} \big( (g_p^a)^{u'_i(\sigma_i - x_{\gamma,i})} g_p^{h'_i}
        \big)^{r'_3} Y_0.
    \end{align*}
\end{description}

\item [Challenge:] $\mathcal{A}$ gives two messages $M_0, M_1$ to
    $\mathcal{B}$. If $M_0 = M_1$, then $\mathcal{B}$ aborts and takes a
    random guess. Otherwise, it chooses random elements $Z_0, Z_1, Z_2,
    Z_{3,1}, \ldots, Z_{3,l} \in \G_q$ and outputs a challenge ciphertext
    as
    \begin{align*}
    &   C = T^{v'} M_{\gamma},~ C_0 = (g^c)^{v'} Z_0,~ C_1 = (g^c)^{w'_1} Z_1,~
        C_2 = (g^c)^{w'_2} Z_2,~
        \forall i : C_{3,i} = (g^c)^{h'_i} Z_{3,i}.
    \end{align*}
    If $T$ is a valid cBDH tuple, then $\mathcal{B}$ is playing
    $\textsf{Game}_{0,0}$. Otherwise, it is playing $\textsf{Game}_{0,1}$.

\item [Query 2:] Same as Query Phase 1.

\item [Guess:] $\mathcal{A}$ outputs a guess $\gamma'$. If $\gamma =
    \gamma'$, it outputs 0. Otherwise, it outputs 1.
\end{description}

Suppose there exists an adversary $\mathcal{A}$ that distinguishes between
$\textsf{Game}_{0,1}$ and $\textsf{Game}_{0,2}$ with a non-negligible
advantage. A simulator $\mathcal{B}$ that solves the BSD assumption using
$\mathcal{A}$ is given: a tuple $\vect{D} = ((n, \G, \G_T, e), g_p, g_q,
g_r)$ and $T$ where $T = Q \in \G_{T,p}$ or $T = R \in \G_T$. Then
$\mathcal{B}$ that interacts with $\mathcal{A}$ is described as follows.

\begin{description}
\item [Init:] $\mathcal{A}$ gives two vectors $\vect{x}_0, \vect{x}_1 \in
    \Sigma^l$. $\mathcal{B}$ then flips a random coin $\gamma$ internally.

\item [Setup:] $\mathcal{B}$ sets up the public key as the real setup
    algorithm using $g_p, g_q, g_r$ from the assumption.

\item [Query 1:] $\mathcal{B}$ answers token queries by running the real
    token generation algorithm except that it chooses random exponents from
    $\Z_n$ instead of $\Z_p$. However, this does not affect the simulation
    since it will raise the elements from $\G_p$ to the exponents.

\item [Challenge:] $\mathcal{A}$ gives two messages $M_0, M_1$ to
    $\mathcal{B}$. If $M_0 = M_1$, then $\mathcal{B}$ encrypts the message
    to the vector $\vect{x}_{\gamma}$. Otherwise, it creates the challenge
    ciphertext of message $M_{\gamma}$ to $\vect{x}_{\gamma}$ as normal
    with except that $C$ is multiplied by $T$. If $T \in \G_{T,p}$, then
    $\mathcal{B}$ is playing $\textsf{Game}_{0,1}$. Otherwise, it is
    playing $\textsf{Game}_{0,2}$.

\item [Query 2:] Same as Query Phase 1.

\item [Guess:] $\mathcal{A}$ outputs a guess $\gamma'$. If $\gamma =
    \gamma'$, it outputs 0. Otherwise, it outputs 1.
\end{description}

\noindent This completes our proof.
\end{proof}

\begin{lemma} \label{lem-ehve-comp-2}
If the decisional C3DH assumption holds, then no polynomial-time adversary
can distinguish between $\textsf{Game}_1$ and $\textsf{Game}_2$ with a
non-negligible advantage.
\end{lemma}

\begin{proof}
Let $q_1$ denote the maximum number of Type 1 queries made by the adversary.
We define a sequence of games $\textsf{Game}_{1,0}, \textsf{Game}_{1,1},
\ldots, \textsf{Game}_{1,q_1}$ where $\textsf{Game}_{1,0} = \textsf{Game}_1$.
In $\textsf{Game}_{1,i}$, for all $k$-th Type-1 queries such that $k > i$,
the simulator creates the token as usual using three independent random
exponents $r_1, r_2, r_3 \in \Z_n$. However, for all $k$-th Type-1 queries
such that $k \leq i$, the simulator creates token components using the
correlated random exponents such as $r_1 = \pi r_2$ for a fixed value $\pi$.
It is obvious that $\textsf{Game}_{1,q_1}$ is equal with $\textsf{Game}_{2}$.

\vs Before proving this lemma, we introduce the decisional Composite 2-party
Diffie-Hellman (C2DH) assumption as follows: Let $(n, \G, \G_T, e)$ be a
description of the bilinear group of composite order $n=pqr$. Let $g_p, g_q,
g_r$ be generators of subgroups of order $p, q, r$ of $\G$ respectively. The
decisional C2DH problem is stated as follows: given a challenge tuple $D =
((n, \G, \G_T, e),~ g_p, g_q, g_r, g_p^a R_1, g_p^b R_2)$ and $T$, decides
whether $T = g_p^{ab} R_3$ or $T = R$ with random choices of $R_1, R_2, R_3
\in \G_q, R \in \G_{pq}$. It is easy to show that if there exists an
adversary that breaks the decisional C2DH assumption, then it can break the
decisional C3DH assumption.

\vs Suppose there exists an adversary $\mathcal{A}$ that distinguishes
between $\textsf{Game}_{1,d-1}$ and $\textsf{Game}_{1,d}$ with a
non-negligible advantage. A simulator $\mathcal{B}$ that solves the
decisional C2DH assumption using $\mathcal{A}$ is given: a challenge tuple $D
= ((n, \G, \G_T, e), g_p, g_q, g_r, g_p^a Y_1, g_p^b Y_2)$ and $T$ where $T =
g_p^{ab} Y_3$ or $T = R$ with random choices of $Y_1, Y_2, Y_3 \in \G_r$, $R
\in \G_{pr}$. Then $\mathcal{B}$ that interacts with $\mathcal{A}$ is
described as follows.

\begin{description}
\item [Init:] $\mathcal{A}$ gives two vectors $\vect{x}_0, \vect{x}_1\in
    \Sigma^l$. $\mathcal{B}$ then flips a random coin $\gamma$ internally.

\item [Setup:] $\mathcal{B}$ first chooses random exponents $v', w'_1,
    w'_2, \alpha, \beta \in \Z_n$, $(u'_1, h'_1), \ldots, (u'_l, h'_l) \in
    \Z_n^2$, then it sets $v = g_p^{v'}, w_1 = g_p^{w'_1}, w_2 =
    g_p^{w'_2}, u_i = g_p^{u'_i}, h_i = g_p^{h'_i}$. Next, it chooses
    random elements $R_v, R_{w,1}, R_{w,2} \in \G_q$, $(R_{u,1}, R_{h,1}),
    \ldots, (R_{u,l}, R_{h,l}) \in \G_q^{2}$, and it publishes the group
    description and a public key as
    \begin{align*}
    &   V = v R_v,~ W_1 = w_1 R_{w,1},~ W_2 = w_2 R_{w,2},~
        \{(U_i = u_i R_{u,i},~ H_i = h_i R_{h,i} ) \},~
        g_q,~ g_r,~ \Omega = e(v, g_p)^{\alpha \beta}.
    \end{align*}

\item [Query 1:] $\mathcal{A}$ adaptively requests a token for a vector
    $\vect{\sigma} = (\sigma_1, \ldots, \sigma_l) \in \Sigma_*^l$ to
    $\mathcal{B}$. Let $S$ be the set of indexes that are not wild card
    positions.

\begin{description}
\item [Type 1] Let $k$ be the index of Type 1 queries. If $\mathcal{A}$
    requests a Type 1 query, then $\mathcal{B}$ chooses random exponents
    $r_1, r_2, r_3 \in \Z_n$ and random elements $Y'_0, Y'_1, Y'_2, Y'_3
    \in \G_r$. Next, it creates a token depending on the $k$ value as
    \begin{align*}
    k<d : &~~
        K_0 = g_p^{\alpha \beta} (g_p^a Y_1)^{w'_1 r_2} w_2^{r_2}
              (\prod_{i\in S} u_i^{\sigma_i} h_i)^{r_3} Y'_0,~
        K_1 = (g_p^a Y_1)^{v' r_2} Y'_1,~
        K_2 = v^{r_2} Y'_2,~ K_3 = v^{r_3} Y'_3, \\
    k=d : & ~~
        K_0 = g_p^{\alpha \beta} T^{w'_1} (g_p^b Y_2)^{w'_2}
              (\prod_{i\in S} u_i^{\sigma_i} h_i)^{r_3} Y'_0,~
        K_1 = T^{v'} Y'_1,~
        K_2 = (g_p^b Y_2)^{v'} Y'_2,~ K_3 = v^{r_3} Y'_3, \\
    k>d : & ~~
        K_0 = g_p^{\alpha \beta} w_1^{r_1} w_2^{r_2}
              (\prod_{i\in S} u_i^{\sigma_i} h_i)^{r_3} Y'_0,~
        K_1 = v^{r_1} Y'_1,~
        K_2 = v^{r_2} Y'_2,~ K_3 = v^{r_3} Y'_3.
    \end{align*}
    If $T$ is not a valid C2DH tuple, then $\mathcal{B}$ is playing
    $\textsf{Game}_{1,d-1}$. Otherwise, it is playing
    $\textsf{Game}_{1,d}$ as
    \begin{align*}
    K_0 =&  g_p^{\alpha \beta} (g_p^{ab} Y_3)^{w'_1} (g_p^b Y_2)^{w'_2}
            (\prod_{i\in S} u_i^{\sigma_i} h_i)^{r_3} Y'_0
        =   g_p^{\alpha \beta} w_1^{ab} w_2^{b}
            (\prod_{i\in S} u_i^{\sigma_i} h_i)^{r_3} \widetilde{Y}_0 \\
        =&  g_p^{\alpha \beta} w_1^{\pi r_2} w_2^{r_2}
            (\prod_{i\in S} u_i^{\sigma_i} h_i)^{r_3} \widetilde{Y}_0,~ \\
    K_1 =&  (g_p^{ab} Y_3)^{v'} Y'_1 = v^{ab} \widetilde{Y}_1
        =   v^{\pi r_2} \widetilde{Y}_1,~
    K_2 =   (g_p^b Y_2)^{v'} Y'_2 = v^{b} \widetilde{Y}_2
        =   v^{r_2} \widetilde{Y}_2,~
    \end{align*}
    where $\pi = a$ and $r_2 = b$.

\item [Type 2] If $\mathcal{A}$ requests a Type 2 query, then
    $\mathcal{B}$ creates the token as the real token generation
    algorithm since it knows all values that are needed.
\end{description}

\item [Challenge:] $\mathcal{A}$ gives two messages $M_0, M_1$ to
    $\mathcal{B}$. $\mathcal{B}$ creates the ciphertext for $M_{\gamma}$
    and $\vect{x}_{\gamma}$ as the real encrypt algorithm by choosing a
    random exponent $t \in \Z_n$ and random elements in $\G_q$.

\item [Query 2:] Same as Query Phase 1.

\item [Guess:] $\mathcal{A}$ outputs a guess $\gamma'$. If $\gamma =
    \gamma'$, it outputs 0. Otherwise, it outputs 1.
\end{description}

\noindent This completes our proof.
\end{proof}

\begin{lemma} \label{lem-ehve-comp-3}
If the decisional C3DH assumption holds, then no polynomial-time adversary
can distinguish between $\textsf{Game}_2$ and $\textsf{Game}_3$ with a
non-negligible advantage.
\end{lemma}

\begin{proof}
Suppose there exists an adversary $\mathcal{A}$ that distinguishes between
$\textsf{Game}_2$ and $\textsf{Game}_3$ with a non-negligible advantage. A
simulator $\mathcal{B}$ that solves the decisional C3DH assumption using
$\mathcal{A}$ is given: a challenge tuple $D = ((n, \G, \G_T, e),~ g_p, g_q,
g_r, \lb g_p^a, g_p^b,~ g_p^{ab}R_1, g_p^{abc}R_2)$ and $T$ where $T = g_p^c
R_3$ or $T = g_p^d R_3$ for a random exponent $d \in \Z_p$. Then
$\mathcal{B}$ that interacts with $\mathcal{A}$ is described as follows.

\begin{description}
\item [Init:] $\mathcal{A}$ gives two vectors $\vect{x}_0 = (x_{0,1},
    \ldots, x_{0,l}), \vect{x}_1 =(x_{1,1}, \ldots, x_{1,l}) \in \Sigma^l$.
    $\mathcal{B}$ then flips a random coin $\gamma$ internally.

\item [Setup:] $\mathcal{B}$ first chooses random exponents $w'_1, w'_2,
    \alpha, \beta \in \Z_n$, $(u'_1, h'_1), \ldots, (u'_l, h'_l) \in
    \Z_n^2$, and random elements $R_v, R_{w,1}, R_{w,2} \in \G_q$,
    $(R_{u,1}, R_{h,1}), \ldots, (R_{u,l}, R_{h,l}) \in \G_q^{2}$. Next, it
    publishes a public key as
    \begin{align*}
    &   V = (g_p^{ab}R_1) R_v,~
        W_1 = (g_p^{ab}R_1 \cdot g_p)^{w'_1}R_{w,1},~ W_2 = g_p^{w'_2}R_{w,2}, \\
    &   \{ (U_i = (g_p^b)^{u'_i} R_{u,i},~
            H_i = (g_p^b)^{-u'_i x_{\gamma,i}} (g_p^{ab}R_1)^{h'_i} R_{h,i} )
        \}_{1 \leq i \leq l},~
        g_q,~ g_r,~ \Omega = e(g_p^{ab}R_1, g_p)^{\alpha \beta}.
    \end{align*}

\item [Query 1:] $\mathcal{A}$ adaptively requests a token for a vector
    $\vect{\sigma} = (\sigma_1, \ldots, \sigma_l) \in \Sigma_*^l$ to
    $\mathcal{B}$. Let $S$ be the set of indexes that are not wild card
    positions.

\begin{description}
\item [Type 1] If $\mathcal{A}$ requests a Type 1 query, then
    $\mathcal{B}$ chooses random exponents $r'_1, r'_3 \in \Z_n$ and
    random elements $Y_0, Y_1, Y_2, Y_3 \in \G_r$. Next, it creates a
    token as
    \begin{align*}
    K_0 =&  g_p^{\alpha \beta} \big( g_p^a \big)^{w'_1 w'_2 r'_1}
            g_p^{\sum_{i \in S} h'_i r'_3} Y_0,
    K_1 =   \big( g_p^a \big)^{w'_2 r'_1} Y_1,~~
    K_2 =   \big( g_p^a \big)^{-w'_1 r'_1} Y_2,~~
    K_3 =   g_p^{r'_3} Y_3.
    \end{align*}
    To show that the above token is the same as the token in
    $\textsf{Game}_3$, we define the randomness of the token as
    \begin{align*}
    r_1 = {w'_2 r'_1}/{b} \mod p,~~
    r_2 = -{w'_1 r'_1}/{b} \mod p,~~
    r_3 = {r'_3}/{ab} \mod p.
    \end{align*}
    It is obvious that two random $r_1$ and $r_2$ are correlated as $r_1
    = \pi r_2$ where $\pi = - w'_2/w'_1$. The distribution of the above
    token is correct as follows
    \begin{align*}
    K_0 =&
        g_p^{\alpha \beta}
        \big( g_p^{(ab + 1)w'_1} \big)^{ {w'_2 r'_1}/{b} }
        \big( g_p^{w'_2} \big)^{ -{w'_1 r'_1}/{b} }
        \big( g_p^{\sum_{i \in S} (bu'_i(\sigma_i - x_{\gamma,i}) + abh'_i)}
            \big)^{ {r'_3}/{ab} } Y_0 \\
    =&  g_p^{\alpha \beta} g_p^{a w'_1 w'_2 r'_1}
        g_p^{\sum_{i \in S} h'_i r'_3} Y_0.
    \end{align*}

\item [Type 2] If $\mathcal{A}$ requests a Type 2 query, then there
    exists an index $j \in S$ such that $\sigma_j \neq x_{\gamma,j}$. Let
    $\Delta = \sum_{i \in S} u'_i (\sigma_i - x_{\gamma,i}) \in \Z_p$.
    Note that $\Delta \neq 0$ except with negligible probability.
    $\mathcal{B}$ first chooses random exponents $r'_1, r'_2, r'_3 \in
    \Z_n$ and random elements $Y_0, Y_1, Y_2, Y_3 \in \G_r$, then it
    creates a token as
    \begin{align*}
    K_0 =&  g_p^{\alpha \beta} \big( g_p^a \big)^{w'_1 w'_2 r'_1}
            g_p^{\Delta w'_2 r'_3}
            \big( g_p^a \big)^{\sum_{i \in S} h'_i w'_2 r'_3}
            g_p^{\sum_{i \in S} h'_i w'_2 r'_2} Y_0, \\
    K_1 =&  \big( g_p^a \big)^{w'_2 r'_1} Y_1,~
    K_2 =   \big( g_p^a \big)^{- w'_1 r'_1}
            \big( g_p^b \big)^{- \Delta r'_2} Y_2,~
    K_3 =   \big( g_p^a \big)^{w'_2 r'_3} g_p^{w'_2 r'_2} Y_3.
    \end{align*}
    To show that the above token is the same as the token in
    $\textsf{Game}_3$, we define the randomness of the token as
    \begin{align*}
    &   r_1 = {w'_2 r'_1}/{b} \mod p,~~
        r_2 = -{w'_1 r'_1}/{b} - {b \Delta r'_2}/{ab} \mod p,~~ \\
    &   r_3 = {w'_2 r'_3}/{b} + {w'_2 r'_2}/{ab} \mod p.
    \end{align*}
    It is not hard to see that $r_1, r_2, r_3$ are independent random
    values since $\Delta \neq 0$ except with negligible probability. The
    distribution of the above token is correct as follows
    \begin{align*}
    K_0 =&
        g_p^{\alpha \beta}
        \big( g_p^{(ab + 1)w'_1} \big)^{ {w'_2 r'_1}/{b} }
        \big( g_p^{w'_2} \big)^{ -{w'_1 r'_1}/{b} -
            {\sum_{i \in S} b u'_i(\sigma_i - x_{\gamma,i}) r'_2}/{ab}}
        \big( g_p^{\sum_{i \in S} (b u'_i(\sigma_i - x_{\gamma,i}) + ab h'_i)} \big)^{
            {w'_2 r'_3}/{b} + {w'_2 r'_2}/{ab} } Y_0 \\
    =&  g_p^{\alpha \beta} g_p^{a w'_1 w'_2 r'_1}
        g_p^{\Delta w'_2 r'_3}
        g_p^{\sum_{i \in S} (a h'_i w'_2 r'_3 + h'_i w'_2 r'_2)} Y_0.
    \end{align*}
\end{description}

\item [Challenge:] $\mathcal{A}$ gives two messages $M_0, M_1$ to
    $\mathcal{B}$. If $M_0 = M_1$, then $\mathcal{B}$ computes $C =
    e(g_p^{abc} R_2, g_p)^{\alpha \beta} M_{\gamma}$. Otherwise, it chooses
    a random elements in $\G_T$ for $C$. Next, it chooses random elements
    $Z_0, Z_1, Z_2, Z_{3,1}, \ldots, \lb Z_{3,l} \in \G_q$ and outputs a
    challenge ciphertext as
    \begin{align*}
    &   C_0 = (g_p^{abc} R_2) Z_0,~ C_1 = (g_p^{abc} R_2 \cdot T)^{w'_1} Z_1,~
        C_2 = T^{w'_2} Z_2,~
        \forall i : C_{3,i} = (g_p^{abc} R_2)^{h'_i} Z_{3,i}.
    \end{align*}
    If $T$ is a valid C3DH tuple, then $\mathcal{B}$ is playing
    $\textsf{Game}_2$. Otherwise, it is playing $\textsf{Game}_3$ as follows
    \begin{align*}
    C_1  =& (g_p^{abc} R_2 \cdot g_p^d R_3)^{w'_1} Z_1
         =  (g_p^{abc} \cdot g_p^{c-c} \cdot g_p^d)^{w'_1} Z_1
         =  (g_p^{abc} g_p^c)^{w'_1} \cdot g_p^{(-c + d) w'_1} Z_1
         =  W_1^c g_p^{\rho} Z'_1, \\
    C_2  =& (g_p^d R_3)^{w'_2} Z_2
         =  g_p^{(\rho/w'_1 + c) w'_2} Z_2
         =  g_p^{c w'_2} \cdot g_p^{\rho \cdot w'_2/w'_1} Z_2
         =  W_2^c g_p^{- \rho \cdot \pi} Z'_2
    \end{align*}
    where $T = g_p^d R_3$, $\rho = (-c + d)w'_1$ and $\pi = -w'_2/w'_1$.

\item [Query 2:] Same as Query Phase 1.

\item [Guess:] $\mathcal{A}$ outputs a guess $\gamma'$. If $\gamma =
    \gamma'$, it outputs 0. Otherwise, it outputs 1.
\end{description}

\noindent This completes our proof.
\end{proof}

\begin{lemma} \label{lem-ehve-comp-4}
If the decisional C3DH assumption holds, then no polynomial-time adversary
can distinguish between $\textsf{Game}_3$ and $\textsf{Game}_4$ with a
non-negligible advantage.
\end{lemma}

\begin{proof}
Let $\overline{X}$ denote the set of indexes $i$ where two committed vectors
$\vect{x}_0, \vect{x}_1$ are not equal. We define a sequence of games
$\textsf{Game}_{3,0}, \textsf{Game}_{3,1}, \ldots,
\textsf{Game}_{3,|\overline{X}|}$ where $\textsf{Game}_{3,0} =
\textsf{Game}_3$. Let $\overline{X}_i \subseteq \overline{X}$ denote the set
of first $i$ indexes in $\overline{X}$. In $\textsf{Game}_{3,i}$, the
simulator creates ciphertext elements $C, C_0$, and $C_j$ normally for all $j
\notin \overline{X}_i$. For all $j \in \overline{X}_i$, the simulator
replaces $C_j$ with random elements in $\G_{pq}$. For $C_1, C_2$, the
simulator creates the following ciphertext elements like in game
$\textsf{Game}_4$ as
    \begin{align*}
    C_1 = W_1^t g_p^{\rho} Z_1,~~ C_2 = W_2^t g_p^{-\rho \pi} Z_2
    \end{align*}
where $\rho$ is a random element from $\Z_p$. Note that it is not hard to see
that $\textsf{Game}_{3,|\overline{X}|} = \textsf{Game}_4$.

\vs Suppose there exists an adversary $\mathcal{A}$ that distinguishes
between $\textsf{Game}_{3,d-1}$ and $\textsf{Game}_{3,d}$ with a
non-negligible advantage. A simulator $\mathcal{B}$ that solves the C3DH
assumption using $\mathcal{A}$ is given: a challenge tuple $D = ((n, \G,
\G_T, e),~ g_p, g_q, g_r, g_p^a, g_p^b, \lb g_p^{ab}R_1, g_p^{abc}R_2)$ and
$T$ where $T = g_p^c R_3$ or $T = R$. Then $\mathcal{B}$ that interacts with
$\mathcal{A}$ is described as follows.

\begin{description}
\item [Init:] $\mathcal{A}$ gives two vectors $\vect{x}_0 = (x_{0,1},
    \ldots, x_{0,l}), \vect{x}_1 =(x_{1,1}, \ldots, x_{1,l}) \in \Sigma^l$.
    $\mathcal{B}$ then flips a random coin $\gamma$ internally.

\item [Setup:] $\mathcal{B}$ first chooses random exponents $w'_1, w'_2,
    \alpha, \beta \in \Z_n$, $(u'_1, h'_1), \ldots, (u'_l, h'_l) \in
    \Z_n^2$, and random elements $R_v, R_{w,1}, R_{w,2} \in \G_q$,
    $(R_{u,1}, R_{h,1}), \ldots, (R_{u,l}, R_{h,l}) \in \G_q^{2}$. Next, it
    publishes a public key as
    \begin{align*}
    &   V = (g_p^{ab}R_1) R_v,~
        W_1 = (g_p^{ab}R_1 \cdot g_p)^{w'_1}R_{w,1},~ W_2 = g_p^{w'_2}R_{w,2},
        (U_d = (g_p^b)^{u'_d} R_{u,d},~
         H_d = (g_p^b)^{-u'_d x_{\gamma,d}} (g_p)^{h'_d} R_{h,d} ), \\
    &   \{ (U_i = (g_p^b)^{u'_i} R_{u,i},~
            H_i = (g_p^b)^{-u'_i x_{\gamma,i}} (g_p^{ab}R_1)^{h'_i} R_{h,i})
        \}_{1 \leq i \neq d \leq l},~
        g_q,~ g_r,~ \Omega = e(g_p^{ab}R_1, g_p)^{\alpha \beta}.
    \end{align*}

\item [Query 1:] $\mathcal{A}$ adaptively requests a token for a vector
    $\vect{\sigma} = (\sigma_1, \ldots, \sigma_l) \in \Sigma_*^l$ to
    $\mathcal{B}$. Let $S$ be the set of indexes that are not wild card
    positions.

\begin{description}
\item [Type 1] For Type 1 queries, it is guaranteed that $d \notin S$
    since $S \cap \overline{X} = \emptyset$ and $d \in \overline{X}$. If
    $\mathcal{A}$ requests a Type 1 query, then $\mathcal{B}$ chooses
    random exponents $r'_1, r'_3 \in \Z_n$ and random elements $Y_0, Y_1,
    Y_2, Y_3 \in \G_r$. Next, it creates a token as
    \begin{align*}
    K_0 &= g_p^{\alpha \beta} \big( g_p^a \big)^{w'_1 w'_2 r'_1}
           g_p^{\sum_{i \in S} h'_i r'_3} Y_0,~
    K_1  = \big( g_p^a \big)^{w'_2 r'_1} Y_1,~
    K_2  = \big( g_p^a \big)^{-w'_1 r'_1} Y_2,~
    K_3  = g_p^{r'_3} Y_3.
    \end{align*}
    Note that it is the same as the simulation of the Type 1 token in
    $\textsf{Game}_3$ if the randomness of the token are defined as
    \begin{align*}
    r_1 &= {w'_2 r'_1}/{b} \mod p,~
    r_2  = -{w'_1 r'_1}/{b} \mod p,~
    r_3  = {r'_3}/{ab} \mod p.
    \end{align*}

\item [Type 2] For Type 2 queries, there exists an index $j \in S$ such
    that $\sigma_j \neq x_{\gamma,j}$ and there exists two cases such
    that $d \notin S$ or $d \in S$. Let $\Delta = \sum_{i \in S} u'_i
    (\sigma_i - x_{\gamma,i}) \in \Z_p$. Note that $\Delta \neq 0$ except
    with negligible probability.

    \vs In case of $d \notin S$, $\mathcal{B}$ chooses random exponents
    $r'_1, r'_2, r'_3 \in \Z_n$ and random elements $Y_0, Y_1, Y_2, Y_3
    \in \G_r$, then it creates a token as
    \begin{align*}
    K_0 &= g_p^{\alpha \beta} \big( g_p^a \big)^{w'_1 w'_2 r'_1}
           g_p^{\Delta w'_2 r'_3} \big( g_p^a \big)^{\sum_{i \in S} h'_i w'_2 r'_3}
           g_p^{\sum_{i \in S} h'_i w'_2 r'_2} Y_0, \\
    K_1 &= \big( g_p^a \big)^{w'_2 r'_1} Y_1,~
    K_2  = \big( g_p^a \big)^{- w'_1 r'_1} \big( g_p^b \big)^{-\Delta r'_2} Y_2,~
    K_3  = \big( g_p^a \big)^{w'_2 r'_3} g_p^{w'_2 r'_2} Y_3.
    \end{align*}
    Note that it is the same as the simulation of the Type 2 token in
    $\textsf{Game}_3$ if the randomness of the token are defined as
    \begin{align*}
    &   r_1 = {w'_2 r'_1}/{b} \mod p,~~
        r_2 = -{w'_1 r'_1}/{b} - {b \Delta r'_2 }/{ab} \mod p, \\
    &   r_3 = {w'_2 r'_3}/{b} + {w'_2 r'_2}/{ab} \mod p.
    \end{align*}

    \vs In case of $d \in S$, $\mathcal{B}$ chooses random exponents
    $r'_1, r'_2, r'_3 \in \Z_n$ and random elements $Y_0, Y_1, Y_2, Y_3
    \in \G_r$, then it creates a token as
    \begin{align*}
    K_0 &= g_p^{\alpha \beta} \big( g_p^a \big)^{w'_1 w'_2 r'_1}
           g_p^{\Delta w'_2 r'_3}
           \big( g_p^a \big)^{\sum_{i \in S \setminus \{d\}} h'_i w'_2 r'_3}
           g_p^{\sum_{i \in S \setminus \{d\}} h'_i w'_2 r'_2} Y_0, \\
    K_1 &= \big( g_p^a \big)^{w'_2 r'_1} Y_1,~
    K_2  = \big( g_p^a \big)^{- w'_1 r'_1}
           \big( g_p^b \big)^{- \Delta r'_2} Y_2,~
    K_3  = \big( g_p^a \big)^{w'_2 r'_3} g_p^{w'_2 r'_2} Y_3.
    \end{align*}
    To show that the above token is the same as the token in
    $\textsf{Game}_3$, we define the randomness of the token as
    \begin{align*}
    &   r_1 = {w'_2 r'_1}/{b} \mod p,~~
        r_2 = -(w'_1 r'_1 + h'_d r'_3)/{b} - {(b \Delta + h'_d) r'_2 }/{ab}
        \mod p, \\
    &   r_3 = {w'_2 r'_3}/{b} + {w'_2 r'_2}/{ab} \mod p.
    \end{align*}
    It is not hard to see that $r_1, r_2, r_3$ are independent random
    values since $\Delta \neq 0$ except with negligible probability.
    Therefore, the distribution of the above token is correct as follows
    \begin{align*}
    K_0
     =& g_p^{\alpha \beta} w_1^{r_1} w_2^{r_2} (u_d^{\sigma_d} h_d)^{r_3}
        \big( \prod_{i \in S \setminus \{d\}} u_i^{\sigma_i} h_i \big)^{r_3} Y_0 \\
     =& g_p^{\alpha \beta} \big( g_p^{(ab + 1)w'_1} \big)^{ {w'_2 r'_1}/{b} }
        \big( g_p^{w'_2} \big)^{ -{(w'_1 r'_1 + h'_d r'_3)}/{b} -
            {(b \Delta + h'_d)r'_2}/{ab}}
        \big( g_p^{(b u'_d(\sigma_d - x_{\gamma,d}) + h'_d)} \big)^{
            {w'_2 r'_3}/{b} + {w'_2 r'_2}/{ab} } \\
      & \big( g_p^{\sum_{i \in S \setminus \{d\}} (b u'_i(\sigma_i
            - x_{\gamma,i}) + ab h'_i)} \big)^{
            {w'_2 r'_3}/{b} + {w'_2 r'_2}/{ab} } Y_0 \\
     =& g_p^{\alpha \beta} g_p^{a w'_1 w'_2 r'_1}
        g_p^{\Delta w'_2 r'_3} g_p^{\sum_{i \in S \setminus \{d\}}
            (a h'_i w'_2 r'_3 + h'_i w'_2 r'_2)} Y_0.
    \end{align*}
\end{description}

\item [Challenge:] $\mathcal{A}$ gives two messages $M_0, M_1$ to
    $\mathcal{B}$. If $M_0 = M_1$, then $\mathcal{B}$ computes $C =
    e(g_p^{abc} R_2, g_p)^{\alpha \beta} M_{\gamma}$. Otherwise, it chooses
    a random elements in $\G_T$ for $C$. Next, it chooses random elements
    $P, P_{3,1}, \ldots, P_{3,d-1} \in \G_p$ and $Z_0, Z_1, Z_2, Z_{3,1},
    \ldots, Z_{3,l} \in \G_q$, then it outputs a challenge ciphertext as
    \begin{align*}
    &   C_0 = (g_p^{abc} R_2) Z_0,~ C_1 = (g_p^{abc} R_2 \cdot P)^{w'_1} Z_1,~
        C_2 = P^{w'_2} Z_2, \\
    &   \forall i < d : C_{3,i} = P_{3,i} Z_{3,i},~
        C_{3,d} = T^{h'_d} Z_{3,d},~
        \forall i > d : C_{3,i} = (g_p^{abc} R_2)^{h'_i} Z_{3,i}.
    \end{align*}
    If $T$ is a valid C3DH tuple, then $\mathcal{B}$ is playing
    $\textsf{Game}_{3,d-1}$. Otherwise, it is playing
    $\textsf{Game}_{3,d}$.

\item [Query 2:] Same as Query Phase 1.

\item [Guess:] $\mathcal{A}$ outputs a guess $\gamma'$. If $\gamma =
    \gamma'$, it outputs 0. Otherwise, it outputs 1.
\end{description}

\noindent This completes our proof.
\end{proof}

\section{HVE in Asymmetric Bilinear Groups}

In this section, we construct an HVE scheme in asymmetric bilinear groups of
prime order where there are no efficiently computable isomorphisms between
two groups $\G$ and $\hat{\G}$. This construction is algebraically similar to
our construction in composite order bilinear groups. In the composite order
setting, the subgroups $\G_q$ and $\G_r$ were used to provide the anonymity
of ciphertexts and to hide the correlation between two random values
respectively. However, in the prime order asymmetric setting, the
non-existence of efficiently computable isomorphisms provides the anonymity
of ciphertexts and hides the correlation of two random values in tokens.

\subsection{Construction}

Let $\Sigma = \Z_m$ for some integer $m$ and set $\Sigma_{*} = \Z_m \cup
\{*\}$. Our scheme is described as follows.

\begin{description}
\item[\normalfont{\textsf{Setup}($1^{\lambda}$)}:] The setup algorithm
    first generates the asymmetric bilinear group $\G, \hat{\G}$ of prime
    order $p$ where $p$ is a random prime of bit size $\Theta(\lambda)$ and
    $p > m$. Let $g, \hat{g}$ be the generators of $\G, \hat{\G}$
    respectively. Next, it chooses random exponents $v', w'_1, w'_2 \in
    \Z_p$, $(u'_1, h'_1), \ldots, (u'_l, h'_l) \in \Z_p$, and $\alpha,
    \beta \in \Z_p$. It keeps these as a secret key $\textsf{SK}$ and
    outputs a public key \textsf{PK} with the description of the asymmetric
    bilinear group $\G, \hat{\G}$ as follows
    \begin{align*}
    \textsf{PK} = \Big(~
    &   v=g^{v'}, w_1=g^{w'_1}, w_2=g^{w'_2},~
        \{(u_i=g^{u'_i}, h_i=g^{h'_i})\}_{i=1}^{l},~
        \Omega = e(v, \hat{g})^{\alpha \beta} ~\Big).
    \end{align*}

\item[\normalfont{\textsf{GenToken}($\vect{\sigma}, \textsf{SK},
    \textsf{PK}$)}:] The token generation algorithm takes as input a vector
    $\vect{\sigma} = (\sigma_1, \ldots, \sigma_l) \in \Sigma_*^l$ and the
    secret key $\textsf{SK}$. It first selects random exponents $r_1, r_2,
    r_3 \in \Z_p$ and computes $\hat{v} = \hat{g}^{v'}, \hat{w}_1 =
    \hat{g}^{w'_1}, \hat{w}_2 = \hat{g}^{w'_2}, \hat{u}_i = \hat{g}^{u'_i},
    \hat{h}_i = \hat{g}^{h'_i}$. Next, it outputs a token as
    \begin{align*}
    \textsf{TK}_{\vect{\sigma}} = \Big(~
    &   K_0 = \hat{g}^{\alpha \beta} \hat{w}_1^{r_1} \hat{w}_2^{r_2}
            (\prod_{i \in S} \hat{u}_i^{\sigma_i} \hat{h}_i)^{r_3},~
        K_1 = \hat{v}^{r_1},~ K_2 = \hat{v}^{r_2},~
        K_3 = \hat{v}^{r_3} ~\Big).
    \end{align*}

\item[\normalfont{\textsf{Encrypt}($\vect{x}, M, \textsf{PK}$)}:] The
    encrypt algorithm takes as input a vector $\vect{x} = (x_1, \ldots,
    x_l) \in \Sigma^l$, a message $M \in \mathcal{M} \subseteq \G_T$ and
    the public key $\textsf{PK}$. It chooses a random exponent $t \in \Z_p$
    and outputs a ciphertext as
    \begin{align*}
    \textsf{CT} = \Big(~
    &   C = \Omega^t M,~ C_0 = v^t,~ C_1 = w_1^t,~ C_2 = w_2^t,~
        \{ C_{3,i} = (u_i^{x_i} h_i)^t \}_{i=1}^l ~\Big).
    \end{align*}

\item[\normalfont{\textsf{Query}($\textsf{CT}, \textsf{TK}_{\vect{\sigma}},
    \textsf{PK}$)}:] The query algorithm takes as input a ciphertext
    $\textsf{CT}$ and a token $\textsf{TK}_{\vect{\sigma}}$ with a vector
    $\vect{\sigma}$. It first computes
    \begin{align*}
    M \leftarrow C \cdot e(C_0, K_0)^{-1} \cdot e(C_1, K_1) \cdot e(C_2, K_2) \cdot
        e(\prod_{i \in S} C_{3,i}, K_3).
    \end{align*}
    If $M \notin \mathcal{M}$, it outputs $\perp$ indicating that the
    predicate $f_{\vect{\sigma}}$ is not satisfied. Otherwise, it outputs
    $M$ indicating that the predicate $f_{\vect{\sigma}}$ is satisfied.
\end{description}

\begin{remark}
We can expand the finite space $\Sigma$ from $\Z_m$ to all of $\{0,1\}^*$ by
using a collision-resistant hash function for the vector of attributes.
\end{remark}

\subsection{Security}

\begin{theorem} \label{thm-ehve-prime}
The above HVE construction is selectively secure under the decisional aBDH
assumption, the decisional aDH assumption, and the decisional a3DH
assumption.
\end{theorem}

\begin{proof}
The main structure of this proof is almost the same as the proof of Theorem
\ref{thm-ehve-comp}. That is, it consists of a sequence of $\textsf{Game}_0$,
$\textsf{Game}_1$, $\textsf{Game}_2$, $\textsf{Game}_3$, $\textsf{Game}_4$
games, and we prove that there is no probabilistic polynomial-time adversary
that distinguishes between $\textsf{Game}_{i-1}$ and $\textsf{Game}_{i}$.
These games are nearly the same as those in the proof of Theorem
\ref{thm-ehve-comp}. The difference is that the ciphertext elements and the
token elements are represented in prime order groups, whereas those elements
were represented in composite order groups in the proof of Theorem
\ref{thm-ehve-comp}. For instance, $C_1, C_2$ elements of the challenge
ciphertext are replaced by $C_1 = w_1^t g^{\rho}, C_2 = w_2^t g^{-\rho \pi}$
in $\textsf{Game}_3$, and the $C_i$ elements of the challenge ciphertext in
$\textsf{Game}_4$ are replaced with random values in $\G$.

First, the indistinguishability between $\textsf{Game}_0$ and
$\textsf{Game}_1$ can be proven using the decisional aBDH assumption. The
proof is almost the same as Lemma \ref{lem-ehve-comp-1}, since the main
components of the decisional aBDH assumption under prime order asymmetric
bilinear groups are the same as the decisional cBDH assumption. Note that the
BSD assumption for Theorem \ref{thm-ehve-comp} is not needed. Second, the
indistinguishability between $\textsf{Game}_1$ and $\textsf{Game}_2$ can be
proven using the decisional aDH assumption for $\hat{\G}$ under prime order
asymmetric bilinear groups. The proof is the same as Lemma
\ref{lem-ehve-comp-2}, since the decisional C2DH assumption in Lemma
\ref{lem-ehve-comp-2} is converted to the decisional aDH assumption in prime
order asymmetric bilinear groups. Finally, the indistinguishability between
$\textsf{Game}_2$ and $\textsf{Game}_3$, (the indistinguishability between
$\textsf{Game}_3$ and $\textsf{Game}_4$, respectively) can be proven under
the decisional a3DH assumption. The proof is the same as Lemma
\ref{lem-ehve-comp-3} (Lemma \ref{lem-ehve-comp-4} respectively) except using
the decisional a3DH instead of the decisional C3DH assumption, since the
decisional C3DH assumption can be converted to the decisional a3DH in prime
order asymmetric bilinear groups. This completes our proof.
\end{proof}

\subsection{Discussion}

\subsection{Freeman Method}

Recently, a heuristic methodology that converts cryptosystems from composite
order bilinear groups to prime order asymmetric bilinear groups was proposed
by Freeman in \cite{Freeman10}. The main idea of Freeman's method is
constructing a product group $\G^n$ that has orthogonal subgroups by applying
the direct product to a prime order bilinear group $\G$ where $n$ is the
number of subgroups. Our construction in composite order bilinear groups is
also converted to a new construction in prime order asymmetric bilinear
groups by applying Freeman's method. However, the new construction requires
three group elements of the prime order group to represent one element in the
composite order group since Freeman's method converts one element of
composite order groups with three subgroups to three elements of prime order
groups. That is, the number of groups elements in ciphertexts and tokens, and
the number of pairing operations in decryption increase by three times.

%% file: chap-convert-hve-prime.tex
\section{Overview}

In this chapter, we construct HVE schemes that are secure under any kind of
pairing types and prove their selective model security. To achieve our goals,
we first presents a framework that converts HVE schemes that are the extreme
generalization of AIBE from composite order bilinear groups to prime order
bilinear groups.

The previous conversion methods that convert cryptographic schemes from
composite order bilinear groups to prime order bilinear groups are Freeman's
method and Ducas' method \cite{Freeman10,Ducas10}. The conversion method of
Ducas is that random blinding elements in ciphertexts can be eliminated in
asymmetric bilinear groups of prime order since the decisional Diffie-Hellman
(DDH) assumption holds in asymmetric bilinear groups. Using this method,
Ducas converted some anonymous hierarchical IBE (AHIBE) and HVE schemes from
bilinear groups of composite order to asymmetric bilinear groups of prime
order. The conversion method of Freeman is that product groups using a direct
product of groups and vector orthogonality using an inner product operation
provide the subgroup decision hardness and the subgroup orthogonality
properties in prime order bilinear groups, respectively. The merit of
Freeman's method is that it convert almost all cryptographic schemes from
bilinear groups of composite order to asymmetric bilinear groups of prime
order. The demerits of Freeman's method are that the resulting schemes work
in asymmetric bilinear groups and use complex assumptions that depend on
complex basis vectors.

The conversion method of this paper is similar to the conversion method of
Freeman in terms of using product groups and vector orthogonality, but it has
the following three differences. The first difference is that Freeman's
method is related to the subgroup decision (SGD) assumption in prime order
bilinear groups, whereas our method is not related to the SGD assumption. The
second difference is that Freeman's method only works in asymmetric bilinear
groups of prime order, whereas our method works in any bilinear groups of
prime order. The third difference is that the cryptographic schemes from
Freeman's method use complex assumptions that depend on complex basis
vectors, whereas the HVE schemes from our method use simple assumptions that
are independent of basis vectors.

\begin{table}[tp]
\caption{Comparison between previous HVE schemes and ours}
\label{tab-chve-comp}
\vs \small \addtolength{\tabcolsep}{11.0pt}
\renewcommand{\arraystretch}{1.3}
    \begin{tabularx}{6.50in}{lcccc}
    \hline
    Scheme  & Group Order  & Pairing Type & Ciphertext Size & No. of Pairing \\
    \hline
    BW-HVE \cite{BonehW07}   & $p_1 p_2$     & Type 1 & $(2l+1)|\G|+|\G_T|$ & $2s+1$ \\
    KSW-IPE \cite{KatzSW08} & $p_1 p_2 p_3$ & Type 1 & $(4l+1)|\G|+|\G_T|$ & $4l+1$ \\
    SW-dHVE \cite{ShiW08}  & $p_1 p_2 p_3$ & Type 1 & $(l+3)|\G|+|\G_T|$  & $s+3$  \\
    OT-IPE \cite{OkamotoT09}   & $p$       & Type 1,2,3 & $(2l+3)|\G|+|\G_T|$ & $2l+3$ \\
    Duc-dHVE \cite{Ducas10} & $p$          & Type 3 & $(l+3)|\G_1|+|\G_T|$ & $s+3$ \\
    Par-IPE \cite{Park11} & $p$       & Type 1,2,3 & $(8l+2)|\G|+|\G_T|$ & $8s+2$ \\
    LL-HVE \cite{LeeL11}   & $p_1 p_2 p_3$ & Type 1 & $(l+3)|\G|+|\G_T|$  & $4$    \\
    LL-HVE \cite{LeeL11}   & $p$           & Type 3 & $(l+3)|\G_1|+|\G_T|$ & $4$   \\
    %\hdashline
    Ours (BW-HVE)        & $p$  & Type 1,2,3 & $(4l+2)|\G|+|\G_T|$ & $4s+2$  \\
    Ours (SW-dHVE)       & $p$  & Type 1,2,3 & $(3l+9)|\G|+|\G_T|$ & $3s+9$  \\
    Ours (LL-HVE)        & $p$  & Type 1,2,3 & $(3l+9)|\G|+|\G_T|$ & $12$    \\
    \hline
    \multicolumn{5}{l}{$p$ = prime value, $l$ = no. of attributes in
    ciphertext, s = no. of attributes in token}
    \end{tabularx}
\end{table}

We first convert the HVE scheme of Boneh and Waters, the delegatable HVE
scheme of Shi and Waters, and the efficient HVE scheme with constant cost of
pairing of Lee and Lee from composite order bilinear groups to prime order
bilinear groups. Next we prove that these converted HVE schemes are
selectively secure under the decisional Bilinear Diffie-Hellman (BDH) and the
decisional Parallel 3-party Diffie-Hellman (P3DH) assumptions. Through these
conversion, we constructed the first delegatable HVE scheme and efficient HVE
scheme with constant cost of pairing in any bilinear groups of prime order.
The previous HVE schemes and ours are compared in Table \ref{tab-chve-comp}.
In Table \ref{tab-chve-comp}, HVE schemes from IPE schemes are also included
since IPE imply HVE. Finally, we prove that the new decisional P3DH
assumption is secure in generic group model that was introduced by Shoup in
\cite{Shoup97}.

\section{Framework}

The basic idea to convert HVE schemes from composite order bilinear groups to
prime order bilinear groups is to use bilinear product groups that are
extended from bilinear groups using the direct product operation. Bilinear
product groups were widely used in dual system encryption of Waters
\cite{Waters09,LewkoW10}, private linear broadcast encryption of Garg et al.
\cite{GargKSW10}, and the conversion method of Freeman \cite{Freeman10}. The
product groups extended from multiplicative cyclic groups represent an
exponent as a vector. Thus vector operations in product groups should be
defined. Since bilinear groups have bilinear maps, the bilinear maps on
bilinear product groups should be defined. Definition \ref{def-vec-ops} and
Definition \ref{def-bil-prod} define the vector operations in product groups
and bilinear product groups, respectively.

\begin{definition}[Vector Operations] \label{def-vec-ops}
Let $\G$ be multiplicative cyclic groups of prime $p$ order. Let $g$ be a
generator of $\G$. We define vector operations over $\G$ as follows:
\begin{enumerate}
\item For a vector $\vect{b} = (b_1, \ldots, b_n) \in \Z_p^n$, define
    $g^{\vect{b}} := (g^{b_1}, \ldots, g^{b_n}) \in \G^n$.
\item For a vector $\vect{b} = (b_1, \ldots, b_n) \in \Z_p^n$ and a scalar
    $c \in \Z_p$, define $(g^{\vect{b}})^c := (g^{b_1 c}, \ldots, g^{b_n
    c}) \in \G^n$.
\item For two vectors $\vect{a} = (a_1, \ldots, a_n), \vect{b} = (b_1,
    \ldots, b_n) \in \Z_p^n$, define $g^{\vect{a}} g^{\vect{b}} :=
(g^{a_1 + b_1}, \lb \ldots, g^{a_n + b_n}) \in \G^n$.
\end{enumerate}
\end{definition}

\begin{definition}[Bilinear Product Groups] \label{def-bil-prod}
Let $(p, \G, \G_{T}, \hat{e})$ be bilinear groups of prime order. Let $g$ be
a generator of $\G$. For integers $n$ and $m$, the bilinear product groups
$((p, \G, \G_T, e), g^{\vect{b}_1}, \ldots, g^{\vect{b}_m})$ of basis vectors
$\vect{b}_1, \ldots, \vect{b}_m$ is defined as follows
\begin{enumerate}
\item The basis vectors $\vect{b}_1, \ldots, \vect{b}_m$ are random vectors
    such that $\vect{b}_i = (b_{i,1}, \ldots, b_{i,n}) \in \Z_p^n$.
\item The bilinear map $e:\G^n \times \G^n \rightarrow \G_T$ is defined as
    $e(g^{\vect{a}},g^{\vect{b}}) := \prod_{i=1}^n \hat{e}(g^{a_i},
    g^{b_i}) = \hat{e}(g,g)^{\vect{a} \cdot \vect{b}}$ where $\cdot$ is the
    inner product operation.
\end{enumerate}
\end{definition}

To guarantee the correctness of cryptographic schemes in bilinear product
groups, the orthogonal property of composite order bilinear groups should be
implemented in bilinear product groups. The previous researches
\cite{Waters09,GargKSW10,Freeman10,LewkoW10} showed that the orthogonal
property can be implemented in bilinear product groups. The idea is that the
orthogonality between vectors can be defined using the inner-product
operation such that $\vect{x} \cdot \vect{y} = 0$ since the bilinear map
provides the inner-product operation. Definition \ref{def-orth} define the
orthogonality in bilinear product groups.

\begin{definition}[Orthogonality] \label{def-orth}
Let $((p, \G, \G_{T}, e), g^{\vect{b}_1}, \ldots, g^{\vect{b}_m})$ be
bilinear product groups with $n,m$ parameters. Let $G_i, G_j$ be subgroups
spanned by $g^{\vect{b}_i}, g^{\vect{b}_j}$, respectively. That is, $G_i =
\langle g^{\vect{b}_i} \rangle$ and $G_j = \langle g^{\vect{b}_j} \rangle$.
Then the two subgroups $G_i$ and $G_j$ are orthogonal to each other if
$e(\vect{A}, \vect{B}) = 1$ for all $\vect{A} \in G_i$ and $\vect{B} \in
G_j$.
\end{definition}

The main idea of our method that convert HVE schemes from composite order
bilinear groups to prime order bilinear groups is that the previous HVE
schemes \cite{BonehW07,ShiW08,LeeL11} in composite order bilinear groups use
the decisional Composite 3-party Diffie-Hellman (C3DH) assumption that is not
a kind of the subgroup decision (SGD) assumption.

The SGD assumption is to distinguish whether $h \in \G$ or $h \in \G_1$ where
$\G$ is a group and $\G_1$ is a subgroup of $\G$ \cite{BonehGN05}. In product
groups $\G^n$, a subgroup $G$ is defined as a vector space spanned by some
basis vectors $\vect{b}_1, \ldots, \vect{b}_m$ such that $G = \langle
g^{\vect{b}_1}, \ldots, g^{\vect{b}_m} \rangle$. If a subgroup is constructed
from one basis vector, then the SGD assumption is related to the DDH
assumption. If a subgroup is constructed from $k$ number of basis vectors,
then the SGD assumption is related to the decisional $k$-Linear ($k$-DLIN)
assumption \cite{Freeman10}. In symmetric bilinear groups of prime order, a
subgroup should be constructed from two basis vectors since the DDH
assumption is not valid \cite{Waters09,GargKSW10}. If a subgroup is
constructed from two basis vectors, then cryptographic schemes become
complicated and there is no generic conversion method from composite order
groups to prime order groups. In asymmetric bilinear groups of prime order, a
subgroup can be constructed from one basis vector since the DDH assumption is
valid \cite{Freeman10, LewkoW10}. If a subgroup is constructed from one basis
vector, then there is a generic conversion method of Freeman, but it only
works in asymmetric bilinear groups.

The decisional C3DH assumption is defined in Definition \ref{def-c3dh}. The
notable properties of the decisional C3DH assumption are that the $T$ value
is always an element of $\G_{p_1 p_2}$ in contrast to the SGD assumption, and
the subgroup $\G_{p_2}$ plays the role of random blinding. From these
properties of the C3DH assumption, it is possible to use just one basis
vector to construct a subgroup. Additionally, it is possible to use simple
basis vectors for cryptographic schemes since ciphertexts and tokens can use
different subgroups that are not orthogonal.

\begin{definition}[Composite 3-party Diffie-Hellman (C3DH) Assumption]
\label{def-c3dh} Let $(n, \G, \G_{T}, e)$ be a description of bilinear groups
of composite order $n=p_1 \cdots p_m$ where $p_i$ is a random prime. Let
$g_{p_i}$ be a generator of the subgroup $\G_{p_i}$. The decisional C3DH
assumption is stated as follows: given a challenge tuple
    $D = \big( (n, \G, \G_T, e),
    g_{p_1}, \ldots, g_{p_m},
    g_{p_1}^a, g_{p_1}^b, g_{p_1}^{ab} R_1, g_{p_1}^{abc} R_2 \big)$
    and $T$,
decides whether $T = T_0 = g_{p_1}^{c} R_3$ or $T = T_1 = g_{p_1}^d R_3$ with
random choices of $a,b,c,d \in \Z_{p_1}$ and $R_1,R_2,R_3 \in \G_{p_2}$.
\end{definition}

For instance, we selects basis vectors $\vect{b}_{1,1} = (1,0),
\vect{b}_{1,2} = (1,a), \vect{b}_{2} = (a,-1)$ for the conversion from
bilinear groups of composite $n = p_1 p_2$ order. For the conversion from
bilinear groups of composite $n = p_1 p_2 p_3$ order, we selects basis
vectors $\vect{b}_{1,1} = (1,0,a_1), \vect{b}_{1,2} = (1,a_2,0), \vect{b}_{2}
= (a_2,-1,a_1a_2-a_3), \vect{b}_3 = (a_1,a_3,-1)$. Though different basis
vectors for the structure of composite order were selected, the assumption
for the security proof is the simple one that is independent of basis
vectors.

\section{Conversion 1: BW-HVE} \label{sec-bw-hve}

In this section, we convert the HVE scheme of Boneh and Waters
\cite{BonehW07} to prime order bilinear groups and prove its selective model
security under the decisional BDH and P3DH assumptions.

\subsection{Construction}

\begin{description}
\item[\normalfont{\textsf{Setup}($1^{\lambda}, l$)}:] The setup algorithm
    first generates the bilinear group $\G$ of prime order $p$ of bit size
    $\Theta(\lambda)$. It chooses a random value $a \in \Z_p$ and sets
    basis vectors for bilinear product groups as $\vect{b}_{1,1} = (1, 0),~
    \vect{b}_{1,2} = (1, a),~ \vect{b}_{2} = (a, -1)$. Next, it chooses
    random exponents $v', \{ u'_i, h'_i, w'_i \}_{i=1}^l, \alpha \in \Z_p$,
    and it computes the following values using the basis vectors
    \begin{align*}
    &   \vect{B}_{1,1} = g^{\vect{b}_{1,1}},~
        \vect{B}_{1,2} = g^{\vect{b}_{1,2}},~
        \vect{B}_2 = g^{\vect{b}_2},~ \\
    &   g^{\vect{v}_1} = \vect{B}_{1,1}^{v'},~
        \big\{
        g^{\vect{u}_{1,i}} = \vect{B}_{1,1}^{u'_i},~
        g^{\vect{h}_{1,i}} = \vect{B}_{1,1}^{h'_i},~
        g^{\vect{w}_{1,i}} = \vect{B}_{1,1}^{w'_i}
        \big\}_{i=1}^l,~ \\
    &   g^{\vect{v}_2} = \vect{B}_{1,2}^{v'},~
        \big\{
        g^{\vect{u}_{2,i}} = \vect{B}_{1,2}^{u'_i},~
        g^{\vect{h}_{2,i}} = \vect{B}_{1,2}^{h'_i},~
        g^{\vect{w}_{2,i}} = \vect{B}_{1,2}^{w'_i}
        \big\}_{i=1}^l.
    \end{align*}
    It keeps $g^{\vect{v}_2}, \{ g^{\vect{u}_{2,i}}, g^{\vect{h}_{2,i}},
    g^{\vect{w}_{2,i}} \}_{i=1}^l, (g^{\vect{b}_{1,2}})^{\alpha}$ as a
    secret key $\textsf{SK}$. Then it publishes a public key \textsf{PK}
    using random blinding values $z_v, \{ z_{u,i}, z_{h,i}, z_{w,i}
    \}_{i=1}^l \in \Z_p$ as follows
    \begin{align*}
    \textsf{PK} = \Big(~
    &   \vect{B}_{1,1},~ \vect{B}_{1,2},~ \vect{B}_2,~
        \vect{V} = g^{\vect{v}_1} \vect{B}_2^{z_v},~
        \big\{
        \vect{U}_i = g^{\vect{u}_{1,i}} \vect{B}_2^{z_{u,i}},~
        \vect{H}_i = g^{\vect{h}_{1,i}} \vect{B}_2^{z_{h,i}},~
        \vect{W}_i = g^{\vect{w}_{1,i}} \vect{B}_2^{z_{w,i}}
        \big\}_{i=1}^{l},~ \\
    &   \Omega = e(g^{\vect{v}_1},g^{\vect{b}_{1,2}})^{\alpha} ~\Big).
    \end{align*}

\item[\normalfont{\textsf{GenToken}($\vect{\sigma}, \textsf{SK},
    \textsf{PK}$)}:] The token generation algorithm takes as input an
    attribute vector $\vect{\sigma} = (\sigma_1, \ldots, \sigma_l) \in
    \Sigma_*^l$ and the secret key $\textsf{SK}$. Let $S$ be the set of
    indexes that are not wild-card fields in the vector $\vec{\sigma}$. It
    selects random exponents $\{ r_{1,i}, r_{2,i} \}_{i \in S} \in \Z_p$
    and outputs a token as
    \begin{align*}
    \textsf{TK}_{\vec{\sigma}} = \Big(~
    &   \vect{K}_1 = (g^{\vect{b}_{1,2}})^{\alpha}
        \prod_{i \in S} ((g^{\vect{u}_{2,i}})^{\sigma_i}
            g^{\vect{h}_{2,i}})^{r_{1,i}} (g^{\vect{w}_{2,i}})^{r_{2,i}},~
        \big\{
        \vect{K}_{2,i} = (g^{\vect{v}_2})^{-r_{1,i}},~
        \vect{K}_{3,i} = (g^{\vect{v}_2})^{-r_{2,i}}
        \big\}_{i \in S}
    ~\Big).
    \end{align*}

\item[\normalfont{\textsf{Encrypt}($\vect{x}, M, \textsf{PK}$)}:] The
    encryption algorithm takes as input an attribute vector $\vect{x} =
    (x_1, \ldots, x_l) \in \Sigma^l$, a message $M \in \mathcal{M}
    \subseteq \G_T$, and the public key $\textsf{PK}$. It first chooses a
    random exponent $t \in \Z_p$ and random blinding values $z_1, \{
    z_{2,i}, z_{3,i} \}_{i=1}^l \in \Z_p$. Then it outputs a ciphertext as
    \begin{align*}
    \textsf{CT} = \Big(~
    &   C_0 = \Omega^t M,~
        \vect{C}_1 = \vect{V}^t \vect{B}_2^{z_1},~
        \big\{
        \vect{C}_{2,i} = (\vect{U}_i^{x_i} \vect{H}_i)^t \vect{B}_2^{z_{2,i}},~
        \vect{C}_{3,i} = \vect{W}_i^t \vect{B}_2^{z_{3,i}}
        \big\}_{i=1}^{l}
    ~\Big).
    \end{align*}

\item[\normalfont{\textsf{Query}($\textsf{CT}, \textsf{TK}_{\vect{\sigma}},
    \textsf{PK}$)}:] The query algorithm takes as input a ciphertext
    $\textsf{CT}$ and a token $\textsf{TK}_{\vect{\sigma}}$ of a vector
    $\vect{\sigma}$. It first computes
    \begin{align*}
    M \leftarrow C_0 \cdot \Big( e(\vect{C}_1, \vect{K}_1) \cdot
        \prod_{i \in S} \big( e(\vect{C}_{2,i}, \vect{K}_{2,i}) \cdot
            e(\vect{C}_{3,i}, \vect{K}_{3,i}) \big) \Big)^{-1}.
    \end{align*}
    If $M \notin \mathcal{M}$, it outputs $\perp$ indicating that the
    predicate $f_{\vec{\sigma}}$ is not satisfied. Otherwise, it outputs
    $M$ indicating that the predicate $f_{\vec{\sigma}}$ is satisfied.
\end{description}

\subsection{Correctness}

If $f_{\vect{\sigma}}(\vect{x}) = 1$, then the following calculations shows
that $\textsf{Query}(\textsf{CT}, \textsf{TK}_{\vect{\sigma}}, \textsf{PK}) =
M$ using the orthogonality of basis vectors such that $e(g^{\vect{b}_2},
g^{\vect{b}_{1,2}})=1$.
    \begin{align*}
    \lefteqn{ e(\vect{C}_1, \vect{K}_1) \cdot
        \prod_{i \in S} \big( e(\vect{C}_{2,i}, \vect{K}_{2,i}) \cdot
            e(\vect{C}_{3,i}, \vect{K}_{3,i}) \big) } \\
    &=  e((g^{\vect{v}_1})^{t}, (g^{\vect{b}_{1,2}})^{\alpha}
            \prod_{i \in S} ((g^{\vect{u}_{2,i}})^{\sigma_i}
            g^{\vect{h}_{2,i}})^{r_{1,i}} (g^{\vect{w}_{2,i}})^{r_{2,i}}) \cdot
        \prod_{i \in S} \big(
            e(((g^{\vect{u}_{1,i}})^{x_i} g^{\vect{h}_{1,i}})^t,
            (g^{\vect{v}_2})^{-r_{1,i}}) \cdot
            e((g^{\vect{w}_i})^t, (g^{\vect{v}_2})^{-r_{2,i}}) \big) \\
    &=  e((g^{\vect{v}_1})^t, (g^{\vect{b}_{1,2}})^{\alpha}) \cdot
        \prod_{i \in S} e(g^{v'}, (g^{u'_i})^{(\sigma_i - x_i)})^{t \cdot r_{1,i}}
     =  e(g^{\vect{v}_1}, g^{\vect{b}_{1,2}})^{\alpha t}.
    \end{align*}
Otherwise, that is $f_{\vect{\sigma}}(\vect{x}) = 0$, then the probability of
$\textsf{Query}(\textsf{CT},\textsf{TK}_{\vect{\sigma}},\textsf{PK}) \neq
\perp$ is negligible by limiting $|\mathcal{M}|$ to less than $|\G_T|^{1/4}$.

\subsection{Security}

\begin{theorem} \label{thm-bw-hve}
The above HVE construction is selectively secure under the decisional BDH and
P3DH assumptions.
\end{theorem}

\noindent The proof of this theorem is easily obtained from the following
four Lemma \ref{lem-bw-hve}, \ref{lem-bw-hve-a1}, \ref{lem-bw-hve-a2}, and
\ref{lem-bw-hve-a3}. Before presenting the four lemmas, we first introduce
the following three assumptions. The HVE scheme of Boneh and Waters
constructed in bilinear groups of composite $n = p_1 p_2$ order, and its
security was proven under the decisional BDH, Bilinear Subgroup Decision
(BSD), and the decisional C3DH assumptions \cite{BonehW07}. These assumptions
in composite order bilinear groups are converted to the following Assumptions
\ref*{sec-bw-hve}-1, \ref*{sec-bw-hve}-2, and \ref*{sec-bw-hve}-3 using our
conversion method.

\vs \noindent \textbf{Assumption \ref*{sec-bw-hve}-1} Let $((p, \G, \G_T, e),
g^{\vect{b}_{1,1}}, g^{\vect{b}_{1,2}}, g^{\vect{b}_{2}})$ be the bilinear
product group of basis vectors $\vect{b}_{1,1} = (1,0), \vect{b}_{1,2} =
(1,a)$, and $\vect{b}_{2} = (a,-1)$. The Assumption \ref*{sec-bw-hve}-1 is
stated as follows: given a challenge tuple
    \begin{align*}
    D = \big( (p, \G, \G_T, e),~
        g^{\vect{b}_{1,1}}, g^{\vect{b}_{1,2}}, g^{\vect{b}_2},
        (g^{\vect{b}_{1,1}})^{c_1}, (g^{\vect{b}_{1,1}})^{c_2},
        (g^{\vect{b}_{1,2}})^{c_1}, (g^{\vect{b}_{1,2}})^{c_2},
        (g^{\vect{b}_{1,1}})^{c_3} \big) \mbox{ and } T,
    \end{align*}
decides whether $T = T_0 = e(g, g)^{c_1 c_2 c_3}$ or $T = T_1 = e(g,g)^d$
with random choices of $c_1, c_2, c_3, d \in \Z_p$.

\vs \noindent \textbf{Assumption \ref*{sec-bw-hve}-2} Let $((p, \G, \G_T, e),
g^{\vect{b}_{1,1}}, g^{\vect{b}_{1,2}}, g^{\vect{b}_{2}})$ be the bilinear
product group of basis vectors $\vect{b}_{1,1} = (1,0), \vect{b}_{1,2} =
(1,a)$, and $\vect{b}_{2} = (a,-1)$. The Assumption \ref*{sec-bw-hve}-2 is
stated as follows: given a challenge tuple
    $$D = \big( (p, \G, \G_T, e),~
    g^{\vect{b}_{1,1}}, g^{\vect{b}_{1,2}}, g^{\vect{b}_2}
    \big) \mbox{ and } T,$$
decides whether $T = T_0 = e((g^{\vect{b}_{1,1}})^{c_1}
(g^{\vect{b}_2})^{c_3}, (g^{\vect{b}_{1,2}})^{c_2})$ or $T = T_1 =
e((g^{\vect{b}_{1,1}})^{c_1}, (g^{\vect{b}_{1,2}})^{c_2})$ with random
choices of $c_1, c_2, c_3 \in \Z_p$.

\vs \noindent \textbf{Assumption \ref*{sec-bw-hve}-3} Let $((p, \G, \G_T, e),
g^{\vect{b}_{1,1}}, g^{\vect{b}_{1,2}}, g^{\vect{b}_{2}})$ be the bilinear
product group of basis vectors $\vect{b}_{1,1} = (1,0), \vect{b}_{1,2} =
(1,a)$, and $\vect{b}_{2} = (a,-1)$. The Assumption \ref*{sec-bw-hve}-3 is
stated as follows: given a challenge tuple
    \begin{align*}
    D = \big( (p, \G, \G_T, e),~
        g^{\vect{b}_{1,1}}, g^{\vect{b}_{1,2}}, g^{\vect{b}_2},
        (g^{\vect{b}_{1,2}})^{c_1}, (g^{\vect{b}_{1,2}})^{c_2},
        (g^{\vect{b}_{1,1}})^{c_1 c_2} (g^{\vect{b}_2})^{z_1},
        (g^{\vect{b}_{1,1}})^{c_1 c_2 c_3} (g^{\vect{b}_2})^{z_2}
        \big) \mbox{ and } T,
    \end{align*}
decides whether $T = T_0 = (g^{\vect{b}_{1,1}})^{c_3} (g^{\vect{b}_2})^{z_3}$
or $T = T_1 = (g^{\vect{b}_{1,1}})^d (g^{\vect{b}_2})^{z_3}$ with random
choices of $c_1, c_2, c_3, d \in \Z_p$ and $z_1, z_2, z_3 \in \Z_p$.

\begin{lemma} \label{lem-bw-hve}
The above HVE construction is selectively secure under the Assumptions
\ref*{sec-bw-hve}-1, \ref*{sec-bw-hve}-2, and \ref*{sec-bw-hve}-3.
\end{lemma}

\begin{proof}
The proof of this lemma is directly obtained from \cite{BonehW07} since the
Assumptions \ref*{sec-bw-hve}-1, \ref*{sec-bw-hve}-2, and \ref*{sec-bw-hve}-2
in prime order bilinear groups are correspond to the Bilinear Diffie-Hellman
(BDH), Bilinear Subgroup Decision (BSD), and Composite 3-party Diffie-Hellman
(C3DH) assumptions in composite order bilinear groups. That is, the proof of
\cite{BonehW07} can be exactly simulated using the vector operations in the
Definition \ref{def-vec-ops} and the Assumptions \ref*{sec-bw-hve}-1,
\ref*{sec-bw-hve}-2, and \ref*{sec-bw-hve}-3.
\end{proof}

\begin{lemma} \label{lem-bw-hve-a1}
If the decisional BDH assumption holds, then the Assumption
\ref*{sec-bw-hve}-1 also holds.
\end{lemma}

\begin{proof}
Suppose there exists an adversary $\mathcal{A}$ that breaks the Assumption
\ref*{sec-bw-hve}-1 with a non-negligible advantage. An algorithm
$\mathcal{B}$ that solves the decisional BDH assumption using $\mathcal{A}$
is given: a challenge tuple $D = ((p, \G, \G_T, e), g, g^{c_1}, g^{c_2},
g^{c_3})$ and $T$ where $T = T_0 = e(g,g)^{c_1 c_2 c_3}$ or $T = T_1 =
e(g,g)^d$. $\mathcal{B}$ first chooses random values $a \in \Z_p$ and
computes
    \begin{align*}
    &   g^{\vect{b}_{1,1}} = (g, 1),~
        g^{\vect{b}_{1,2}} = (g, g^a),~
        g^{\vect{b}_2} = (g^a, g^{-1}),~ \\
    &   (g^{\vect{b}_{1,1}})^{c_1} = (g^{c_1}, 1),~
        (g^{\vect{b}_{1,1}})^{c_2} = (g^{c_2}, 1),~
        (g^{\vect{b}_{1,1}})^{c_3} = (g^{c_3}, 1),~ \\
    &   (g^{\vect{b}_{1,2}})^{c_1} = (g^{c_1}, (g^{c_1})^a),~
        (g^{\vect{b}_{1,2}})^{c_2} = (g^{c_2}, (g^{c_2})^a).
    \end{align*}
Next, it gives the tuple
    $D' = ((p,\G, \G_T, e),
    g^{\vect{b}_{1,1}}, g^{\vect{b}_{1,2}}, g^{\vect{b}_2},
    (g^{\vect{b}_{1,1}})^{c_1}, (g^{\vect{b}_{1,1}})^{c_2},
    (g^{\vect{b}_{1,2}})^{c_1}, \lb (g^{\vect{b}_{1,2}})^{c_2},
    (g^{\vect{b}_{1,1}})^{c_3})$ and $T$
to $\mathcal{A}$. Then $\mathcal{A}$ outputs a guess $\gamma'$. $\mathcal{B}$
also outputs $\gamma'$. If the advantage of $\mathcal{A}$ is $\epsilon$, then
the advantage of $\mathcal{B}$ is greater than $\epsilon$ since the
distribution of the challenge tuple to $\mathcal{A}$ is equal to the
Assumption \ref*{sec-bw-hve}-1.
\end{proof}

\begin{lemma} \label{lem-bw-hve-a2}
The Assumption \ref*{sec-bw-hve}-2 holds for all adversaries.
\end{lemma}

\begin{proof}
The equation $e((g^{\vect{b}_{1,1}})^{c_1} (g^{\vect{b}_2})^{c_3},
(g^{\vect{b}_{1,2}})^{c_2}) = e((g^{\vect{b}_{1,1}})^{c_1},
(g^{\vect{b}_{1,2}})^{c_2})$ holds by the orthogonality of basis vectors such
that $e(g^{\vect{b}_{2}}, g^{\vect{b}_{1,2}}) = 1$. Therefore, any adversary
can not break the Assumption \ref*{sec-bw-hve}-2.
\end{proof}

\begin{lemma} \label{lem-bw-hve-a3}
If the decisional P3DH assumption holds, then the Assumption
\ref*{sec-bw-hve}-3 also holds.
\end{lemma}

\begin{proof}
Suppose there exists an adversary $\mathcal{A}$ that breaks the Assumption
\ref*{sec-bw-hve}-3 with a non-negligible advantage. An algorithm
$\mathcal{B}$ that solves the decisional P3DH assumption using $\mathcal{A}$
is given: a challenge tuple
    $D = ((p, \G, \G_T, e), (g,f),
    (g^{c_1}, f^{c_1}), \lb (g^{c_2}, f^{c_2}),
    (g^{c_1 c_2} f^{z_1}, g^{z_1}), (g^{c_1 c_2 c_3} f^{z_2}, g^{z_2}))$
and $T$ where $T = T_0 = (g^{c_3} f^{z_3}, g^{z_3})$ or $T = T_1 = (g^d
f^{z_3}, g^{z_3})$. $\mathcal{B}$ first computes
    \begin{align*}
    &   g^{\vect{b}_{1,1}} = (g, 1),~
        g^{\vect{b}_{1,2}} = (g, f),~
        g^{\vect{b}_2}     = (f, g^{-1}),~
        (g^{\vect{b}_{1,2}})^{c_1} = (g^{c_1}, f^{c_1}),~
        (g^{\vect{b}_{1,2}})^{c_2} = (g^{c_2}, f^{c_2}),~ \\
    &   (g^{\vect{b}_{1,1}})^{c_1 c_2} (g^{\vect{b}_2})^{z_1}
            = (g^{c_1 c_2} f^{z_1}, (g^{z_1})^{-1}),~
        (g^{\vect{b}_{1,1}})^{c_1 c_2 c_3} (g^{\vect{b}_2})^{z_2}
            = (g^{c_1 c_2 c_3} f^{z_2}, (g^{z_2})^{-1}).
    \end{align*}
Intuitively, it sets $a = \text{dlog}(f)$. Next, it gives the tuple
    $D' = ((p,\G, \G_T, e),
    g^{\vect{b}_{1,1}}, g^{\vect{b}_{1,2}}, \lb g^{\vect{b}_2},
    (g^{\vect{b}_{1,1}})^{c_1}, (g^{\vect{b}_{1,1}})^{c_2}, \lb
    (g^{\vect{b}_{1,2}})^{c_1}, (g^{\vect{b}_{1,2}})^{c_2},
    (g^{\vect{b}_{1,1}})^{c_1 c_2 c_3})$ and $T$
to $\mathcal{A}$. Then $\mathcal{A}$ outputs a guess $\gamma'$. $\mathcal{B}$
also outputs $\gamma'$. If the advantage of $\mathcal{A}$ is $\epsilon$, then
the advantage of $\mathcal{B}$ is greater than $\epsilon$ since the
distribution of the challenge tuple to $\mathcal{A}$ is equal to the
Assumption \ref*{sec-bw-hve}-3.
\end{proof}

\section{Conversion 2: LL-HVE} \label{sec-ll-hve}

In this section, we convert the HVE scheme of Lee and Lee \cite{LeeL11} to
prime order bilinear groups and prove its selective model security under the
decisional BDH and P3DH assumptions.

\subsection{Construction}

\begin{description}
\item[\normalfont{\textsf{Setup}($1^{\lambda}, l$)}:] The setup algorithm
    first generates the bilinear group $\G$ of prime order $p$ of bit size
    $\Theta(\lambda)$. It chooses random values $a_1, a_2, a_3 \in \Z_p$
    and sets basis vectors for bilinear product groups as $\vect{b}_{1,1} =
    (1, 0, a_1),~ \vect{b}_{1,2} = (1, a_2, 0),~ \vect{b}_2 = (a_2, -1, a_1
    a_2 - a_3),~ \vect{b}_3 = (a_1, a_3, -1)$. Next, it chooses random
    exponents $v', w'_1, w'_2, \{ u'_i, h_i \}_{i=1}^l, \alpha \in \Z_p$,
    and it computes the following values using the basis vectors
    \begin{align*}
    &   \vect{B}_{1,1} = g^{\vect{b}_{1,1}},~
        \vect{B}_{1,2} = g^{\vect{b}_{1,2}},~
        \vect{B}_2 = g^{\vect{b}_2},~
        \vect{B}_3 = g^{\vect{b}_3},~ \\
    &   g^{\vect{v}_1} = \vect{B}_{1,1}^{v'},~
        g^{\vect{w}_{1,1}} = \vect{B}_{1,1}^{w'_1},~
        g^{\vect{w}_{1,2}} = \vect{B}_{1,1}^{w'_2},~
        \big\{
        g^{\vect{u}_{1,i}} = \vect{B}_{1,1}^{u'_i},~
        g^{\vect{h}_{1,i}} = \vect{B}_{1,1}^{h'_i}
        \big\}_{i=1}^l,~ \\
    &   g^{\vect{v}_2} = \vect{B}_{1,2}^{v'},~
        g^{\vect{w}_{2,1}} = \vect{B}_{1,2}^{w'_1},~
        g^{\vect{w}_{2,2}} = \vect{B}_{1,2}^{w'_2},~
        \big\{
        g^{\vect{u}_{2,i}} = \vect{B}_{1,2}^{u'_i},~
        g^{\vect{h}_{2,i}} = \vect{B}_{1,2}^{h'_i}
        \big\}_{i=1}^l.
    \end{align*}
    It keeps $g^{\vect{v}_2}, g^{\vect{w}_{2,1}}, g^{\vect{w}_{2,2}}, \{
    g^{\vect{u}_{2,i}}, g^{\vect{h}_{2,i}} \}_{i=1}^l,
    (g^{\vect{b}_{1,2}})^{\alpha}$ as a secret key $\textsf{SK}$. Then it
    publishes a public key \textsf{PK} using random blinding values $z_v,
    z_{w,1}, z_{w,2}, \{ z_{u,i}, z_{h,i} \}_{i=1}^l \in \Z_p$ as follows
    \begin{align*}
    \textsf{PK} = \Big(~
    &   \vect{B}_{1,1},~ \vect{B}_{1,2},~ \vect{B}_2,~ \vect{B}_3,~
        \vect{V} = g^{\vect{v}_1} \vect{B}_2^{z_v},~
        \vect{W}_1 = g^{\vect{w}_{1,1}} \vect{B}_2^{z_{w,1}},~
        \vect{W}_2 = g^{\vect{w}_{1,2}} \vect{B}_2^{z_{w,2}},~ \\
    &   \big\{
        \vect{U}_i = g^{\vect{u}_{1,i}} \vect{B}_2^{z_{u,i}},~
        \vect{H}_i = g^{\vect{h}_{1,i}} \vect{B}_2^{z_{h,i}}
        \big\}_{i=1}^{l},~
        \Omega = e(g^{\vect{v}_1}, g^{\vect{b}_{1,2}})^{\alpha} ~\Big).
    \end{align*}

\item[\normalfont{\textsf{GenToken}($\vect{\sigma}, \textsf{SK},
    \textsf{PK}$)}:] The token generation algorithm takes as input an
    attribute vector $\vect{\sigma} = (\sigma_1, \ldots, \sigma_l) \in
    \Sigma_*^l$ and the secret key $\textsf{SK}$. Let $S$ be the set of
    indexes that are not wild-card fields in the vector $\vec{\sigma}$. It
    selects random exponents $r_1, r_2, r_3 \in \Z_p$ and random blinding
    values $y_1, y_2, y_3, y_4 \in \Z_p$. Next it outputs a token as
    \begin{align*}
    \textsf{TK}_{\vec{\sigma}} = \Big(~
    &   \vect{K}_1 = (g^{\vect{b}_{1,2}})^{\alpha}
            (g^{\vect{w}_{2,1}})^{r_1} (g^{\vect{w}_{2,2}})^{r_2}
            \big( \prod_{i \in S} ((g^{\vect{u}_{2,i}})^{\sigma_i}
                g^{\vect{h}_{2,i}}) \big)^{r_3}
            \vect{B}_3^{y_1},~ \\
    &   \vect{K}_2 = (g^{\vect{v}_2})^{-r_1} \vect{B}_3^{y_2},~
        \vect{K}_3 = (g^{\vect{v}_2})^{-r_2} \vect{B}_3^{y_3},~
        \vect{K}_4 = (g^{\vect{v}_2})^{-r_3} \vect{B}_3^{y_4}
    ~\Big).
    \end{align*}

\item[\normalfont{\textsf{Encrypt}($\vect{x}, M, \textsf{PK}$)}:] The
    encryption algorithm takes as input an attribute vector $\vect{x} =
    (x_1, \ldots, x_l) \in \Sigma^l$, a message $M \in \mathcal{M}
    \subseteq \G_T$, and the public key $\textsf{PK}$. It first chooses a
    random exponent $t \in \Z_p$ and random blinding values $z_1, z_2, z_3,
    \{ z_{4,i} \}_{i=1}^l \in \Z_p$. Then it outputs a ciphertext as
    \begin{align*}
    \textsf{CT} = \Big(~
    &   C_0 = \Omega^t M,~
        \vect{C}_1 = \vect{V}^t \vect{B}_2^{z_1},~
        \vect{C}_2 = \vect{W}_1^t \vect{B}_2^{z_2},~
        \vect{C}_3 = \vect{W}_2^t \vect{B}_2^{z_3},~
        \big\{
        \vect{C}_{4,i} = (\vect{U}_i^{x_i} \vect{H}_i)^t \vect{B}_2^{z_{4,i}}
        \big\}_{i=1}^{l}
    ~\Big).
    \end{align*}

\item[\normalfont{\textsf{Query}($\textsf{CT}, \textsf{TK}_{\vec{\sigma}},
    \textsf{PK}$)}:] The query algorithm takes as input a ciphertext
    $\textsf{CT}$ and a token $\textsf{TK}_{\vec{\sigma}}$ of a vector
    $\vec{\sigma}$. It first computes
    \begin{align*}
    M \leftarrow C_0 \cdot \Big(
        e(\vect{C}_1, \vect{K}_1) \cdot e(\vect{C}_2, \vect{K}_2) \cdot
        e(\vect{C}_3, \vect{K}_3) \cdot
        e(\prod_{i \in S} \vect{C}_{4,i}, \vect{K}_4) \Big)^{-1}.
    \end{align*}
    If $M \notin \mathcal{M}$, it outputs $\perp$ indicating that the
    predicate $f_{\vec{\sigma}}$ is not satisfied. Otherwise, it outputs
    $M$ indicating that the predicate $f_{\vec{\sigma}}$ is satisfied.
\end{description}

\subsection{Correctness}

If $f_{\vec{\sigma}}(\vec{x}) = 1$, then the following calculation shows that
$\textsf{Query}(\textsf{CT},\textsf{TK}_{\vec{\sigma}},\textsf{PK}) = M$ by
the orthogonality of basis vectors such that $e(g^{\vect{b}_{1,1}},
g^{\vect{b}_3}) = 1, e(g^{\vect{b}_{1,2}}, g^{\vect{b}_2}) = 1,
e(g^{\vect{b}_2}, g^{\vect{b}_3}) = 1$.
    \begin{align*}
    \lefteqn{ e(\vect{C}_1, \vect{K}_1) \cdot e(\vect{C}_2, \vect{K}_2) \cdot
        e(\vect{C}_3, \vect{K}_3) \cdot
        e(\prod_{i \in S} \vect{C}_{4,i}, \vect{K}_4) } \\
    &=  e((g^{\vect{v}_1})^t, (g^{\vect{b}_{1,2}})^{\alpha}
            (g^{\vect{w}_{2,1}})^{r_1} (g^{\vect{w}_{2,2}})^{r_2}
            \big( \prod_{i \in S} ((g^{\vect{u}_{2,i}})^{\sigma_i}
                g^{\vect{h}_{2,i}}) \big)^{r_3}) \cdot \\
    &   e((g^{\vect{w}_{1,1}})^t, (g^{\vect{v}_2})^{-r_1}) \cdot
        e((g^{\vect{w}_{1,2}})^t, (g^{\vect{v}_2})^{-r_2}) \cdot
        e(\prod_{i \in S} ((g^{\vect{u}_{1,i}})^{x_i} g^{\vect{h}_{1,i}})^t,
            (g^{\vect{v}_2})^{-r_3}) \\
    &=  e((g^{\vect{v}_1})^t, (g^{\vect{b}_{1,2}})^{\alpha}) \cdot
        e(g^{v'}, \prod_{i \in S} (g^{u'_i})^{(\sigma_i - x_i)})^{t r_3}
     =  e(g^{\vect{v}_1}, g^{\vect{b}_{1,2}})^{\alpha t}.
    \end{align*}
Otherwise, that is $f_{\vec{\sigma}}(\vec{x}) = 0$, the probability of
$\textsf{Query}(\textsf{CT},\textsf{TK}_{\vec{\sigma}},\textsf{PK}) \neq
\perp$ is negligible by limiting $|\mathcal{M}|$ to less than $|\G_T|^{1/4}$.

\subsection{Security}

\begin{theorem} \label{thm-ll-hve}
The above HVE construction is selectively secure under the decisional BDH and
P3DH assumptions.
\end{theorem}

\noindent The proof of this theorem is easily obtained from the following
five Lemma \ref{lem-ll-hve}, \ref{lem-ll-hve-a1}, \ref{lem-ll-hve-a2},
\ref{lem-ll-hve-a3} and \ref{lem-ll-hve-a4}. Before presenting the five
lemmas, we first introduce the following four assumptions. The HVE scheme of
Lee and Lee constructed in bilinear groups of composite $n = p_1 p_2 p_3$
order, and its security was proven under the decisional BDH, Bilinear
Subgroup Decision (BSD), and the decisional C3DH assumptions \cite{ShiW08}.
In composite order bilinear groups, the decisional C3DH assumption imply the
decisional C2DH assumption that was introduced in \cite{LeeL11}. However, in
prime order bilinear groups, this implication is not valid since the basis
vectors for ciphertexts and tokens are different. Thus the decisional C3DH
assumption for ciphertexts and the decisional C2DH assumption for tokens
should be treated as differently. These assumptions in composite order
bilinear groups are converted to the following Assumptions
\ref*{sec-ll-hve}-1, \ref*{sec-ll-hve}-2, \ref*{sec-ll-hve}-3, and
\ref*{sec-ll-hve}-4 using our conversion method.

\vs \noindent \textbf{Assumption \ref*{sec-ll-hve}-1} Let $((p, \G, \G_T, e),
g^{\vect{b}_{1,1}}, g^{\vect{b}_{1,2}}, g^{\vect{b}_2}, g^{\vect{b}_3})$ be
the bilinear product group of basis vectors $\vect{b}_{1,1} = (1,0,a_1),
\vect{b}_{1,2} = (1,a_2,0), \vect{b}_2 = (a_2,-1,a_1 a_2 - a_3)$, and
$\vect{b}_3 = (a_1, a_3, -1)$. The Assumption \ref*{sec-ll-hve}-1 is stated
as follows: given a challenge tuple
    \begin{align*}
    D = \big( (p, \G, \G_T, e),~
        g^{\vect{b}_{1,1}}, g^{\vect{b}_{1,2}}, g^{\vect{b}_2},
        g^{\vect{b}_3},
        (g^{\vect{b}_{1,1}})^{c_1}, (g^{\vect{b}_{1,1}})^{c_2},
        (g^{\vect{b}_{1,2}})^{c_1}, (g^{\vect{b}_{1,2}})^{c_2},
        (g^{\vect{b}_{1,1}})^{c_3} \big) \mbox{ and } T,
    \end{align*}
decides whether $T = T_0 = e(g, g)^{c_1 c_2 c_3}$ or $T = T_1 = e(g,g)^d$
with random choices of $c_1, c_2, c_3, d \in \Z_p$.

\vs \noindent \textbf{Assumption \ref*{sec-ll-hve}-2} Let $((p, \G, \G_T, e),
g^{\vect{b}_{1,1}}, g^{\vect{b}_{1,2}}, g^{\vect{b}_2}, g^{\vect{b}_3})$ be
the bilinear product group of basis vectors $\vect{b}_{1,1} = (1,0,a_1),
\vect{b}_{1,2} = (1,a_2,0), \vect{b}_2 = (a_2,-1,a_1 a_2 - a_3)$, and
$\vect{b}_3 = (a_1, a_3, -1)$. The Assumption \ref*{sec-ll-hve}-2 is stated
as follows: given a challenge tuple
    $$D = \big( (p, \G, \G_T, e),~
    g^{\vect{b}_{1,1}}, g^{\vect{b}_{1,2}}, g^{\vect{b}_2}, g^{\vect{b}_3}
    \big) \mbox{ and } T,$$
decides whether $T = T_0 = e((g^{\vect{b}_{1,1}})^{c_1}
(g^{\vect{b}_2})^{c_3}, (g^{\vect{b}_{1,2}})^{c_2} (g^{\vect{b}_3})^{c_4})$
or $T = T_1 = e((g^{\vect{b}_{1,1}})^{c_1}, \lb (g^{\vect{b}_{1,2}})^{c_2})$
with random choices of $c_1, c_2, c_3, c_4 \in \Z_p$.

\vs \noindent \textbf{Assumption \ref*{sec-ll-hve}-3} Let $((p, \G, \G_T, e),
g^{\vect{b}_{1,1}}, g^{\vect{b}_{1,2}}, g^{\vect{b}_2}, g^{\vect{b}_3})$ be
the bilinear product group of basis vectors $\vect{b}_{1,1} = (1,0,a_1),
\vect{b}_{1,2} = (1,a_2,0), \vect{b}_2 = (a_2,-1,a_1 a_2 - a_3)$, and
$\vect{b}_3 = (a_1, a_3, -1)$. The Assumption \ref*{sec-ll-hve}-3 is stated
as follows: given a challenge tuple
    \begin{align*}
    D = \big( (p, \G, \G_T, e),~
        g^{\vect{b}_{1,1}}, g^{\vect{b}_{1,2}}, g^{\vect{b}_2},
        g^{\vect{b}_3},
        (g^{\vect{b}_{1,2}})^{c_1}, (g^{\vect{b}_{1,2}})^{c_2},
        (g^{\vect{b}_{1,1}})^{c_1 c_2} (g^{\vect{b}_2})^{z_1},
        (g^{\vect{b}_{1,1}})^{c_1 c_2 c_3} (g^{\vect{b}_2})^{z_2}
        \big) \mbox{ and } T,
    \end{align*}
decides whether $T = T_0 = (g^{\vect{b}_{1,1}})^{c_3} (g^{\vect{b}_2})^{z_3}$
or $T = T_1 = (g^{\vect{b}_{1,1}})^d (g^{\vect{b}_2})^{z_3}$ with random
choices of $c_1, c_2, c_3, d \in \Z_p$ and $z_1, z_2, z_3 \in \Z_p$.

\vs \noindent \textbf{Assumption \ref*{sec-ll-hve}-4} Let $((p, \G, \G_T, e),
g^{\vect{b}_{1,1}}, g^{\vect{b}_{1,2}}, g^{\vect{b}_2}, g^{\vect{b}_3})$ be
the bilinear product group of basis vectors $\vect{b}_{1,1} = (1,0,a_1),
\vect{b}_{1,2} = (1,a_2,0), \vect{b}_2 = (a_2,-1,a_1 a_2 - a_3)$, and
$\vect{b}_3 = (a_1, a_3, -1)$. The Assumption \ref*{sec-ll-hve}-4 is stated
as follows: given a challenge tuple
    \begin{align*}
    D = \big( (p, \G, \G_T, e),~
        g^{\vect{b}_{1,1}}, g^{\vect{b}_{1,2}}, g^{\vect{b}_2},
        g^{\vect{b}_3},
        (g^{\vect{b}_{1,2}})^{c_1} (g^{\vect{b}_3})^{z_1},
        (g^{\vect{b}_{1,2}})^{c_2} (g^{\vect{b}_3})^{z_2}
        \big) \mbox{ and } T,
    \end{align*}
decides whether $T = T_0 = (g^{\vect{b}_{1,2}})^{c_1 c_2}
(g^{\vect{b}_3})^{z_3}$ or $T = T_1 = (g^{\vect{b}_{1,2}})^d
(g^{\vect{b}_3})^{z_3}$ with random choices of $c_1, c_2, d \in \Z_p$ and
$z_1, z_2, z_3 \in \Z_p$.

\begin{lemma} \label{lem-ll-hve}
The above HVE construction is selectively secure under the Assumptions
\ref*{sec-ll-hve}-1, \ref*{sec-ll-hve}-2, \ref*{sec-ll-hve}-3, and
\ref*{sec-ll-hve}-4.
\end{lemma}

\begin{proof}
The proof of this lemma is directly obtained from \cite{LeeL11} since the
Assumptions \ref*{sec-ll-hve}-1, \ref*{sec-ll-hve}-2, \ref*{sec-ll-hve}-3,
and \ref*{sec-ll-hve}-4 in prime order bilinear groups are corresponds to the
Bilinear Diffie-Hellman (BDH), Bilinear Subgroup Decision (BSD), Composite
3-party Diffie-Hellman (C3DH), and Composite 2-party Diffie-Hellman (C2DH)
assumptions in composite order bilinear groups.
\end{proof}

\begin{lemma} \label{lem-ll-hve-a1}
If the decisional BDH assumption holds, then the Assumption
\ref*{sec-ll-hve}-1 also holds.
\end{lemma}

\begin{proof}
Suppose there exists an adversary $\mathcal{A}$ that breaks the Assumption
\ref*{sec-ll-hve}-1 with a non-negligible advantage. An algorithm
$\mathcal{B}$ that solves the decisional BDH assumption using $\mathcal{A}$
is given: a challenge tuple $D = ((p, \G, \G_T, e), g, g^{c_1}, g^{c_2},
g^{c_3})$ and $T$ where $T = T_0 = e(g,g)^{c_1 c_2 c_3}$ or $T = T_1 =
e(g,g)^d$. $\mathcal{B}$ first chooses random values $a_1, a_2, a_3 \in \Z_p$
and sets
    \begin{align*}
    &   g^{\vect{b}_{1,1}} = (g, 1, g^{a_1}),~
        g^{\vect{b}_{1,2}} = (g, g^{a_2}, 1),~
        g^{\vect{b}_2} = (g^{a_2}, g^{-1}, g^{a_1 a_2 - a_3}),~
        g^{\vect{b}_3} = (g^{a_1}, g^{a_3}, g^{-1}),~ \\
    &   (g^{\vect{b}_{1,1}})^{c_1} = (g^{c_1}, 1, (g^{c_1})^{a_1}),~
        (g^{\vect{b}_{1,1}})^{c_2} = (g^{c_2}, 1, (g^{c_2})^{a_1}),~
        (g^{\vect{b}_{1,1}})^{c_3} = (g^{c_3}, 1),~ \\
    &   (g^{\vect{b}_{1,2}})^{c_1} = (g^{c_1}, (g^{c_1})^{a_2}, 1),~
        (g^{\vect{b}_{1,2}})^{c_2} = (g^{c_2}, (g^{c_2})^{a_2}, 1).
    \end{align*}
Next, it gives the tuple
    $D' = ((p,\G, \G_T, e),
    g^{\vect{b}_{1,1}}, g^{\vect{b}_{1,2}}, g^{\vect{b}_2},
    (g^{\vect{b}_{1,1}})^{c_1}, (g^{\vect{b}_{1,1}})^{c_2},
    (g^{\vect{b}_{1,2}})^{c_1}, \lb (g^{\vect{b}_{1,2}})^{c_2},
    (g^{\vect{b}_{1,1}})^{c_3})$ and $T$
to $\mathcal{A}$. Then $\mathcal{A}$ outputs a guess $\gamma'$. $\mathcal{B}$
also outputs $\gamma'$. If the advantage of $\mathcal{A}$ is $\epsilon$, then
the advantage of $\mathcal{B}$ is greater than $\epsilon$ since the
distribution of the challenge tuple to $\mathcal{A}$ is equal to the
Assumption \ref*{sec-ll-hve}-1.
\end{proof}

\begin{lemma} \label{lem-ll-hve-a2}
The Assumption \ref*{sec-ll-hve}-2 holds for all adversaries.
\end{lemma}

\begin{proof}
The equation $e((g^{\vect{b}_{1,1}})^{c_1} (g^{\vect{b}_2})^{c_3},
(g^{\vect{b}_{1,2}})^{c_2} (g^{\vect{b}_3})^{c_4}) =
e((g^{\vect{b}_{1,1}})^{c_1}, (g^{\vect{b}_{1,2}})^{c_2})$ holds by the
orthogonality of basis vectors such that $e(g^{\vect{b}_{1,1}},
g^{\vect{b}_3}) = 1, e(g^{\vect{b}_2}, g^{\vect{b}_{1,2}}) = 1$, and
$e(g^{\vect{b}_2}, g^{\vect{b}_3}) = 1$. Therefore, any adversary can not
break the Assumption \ref*{sec-ll-hve}-2.
\end{proof}

\begin{lemma} \label{lem-ll-hve-a3}
If the decisional P3DH assumption holds, then the Assumption
\ref*{sec-ll-hve}-3 also holds.
\end{lemma}

\begin{proof}
Suppose there exists an adversary $\mathcal{A}$ that breaks the Assumption
\ref*{sec-ll-hve}-3 with a non-negligible advantage. An algorithm
$\mathcal{B}$ that solves the decisional P3DH assumption using $\mathcal{A}$
is given: a challenge tuple
    $D = ((p, \G, \G_T, e), (g,f),
    (g^{c_1}, f^{c_1}), \lb (g^{c_2}, f^{c_2}),
    (g^{c_1 c_2} f^{z_1}, g^{z_1}), (g^{c_1 c_2 c_3} f^{z_2}, g^{z_2}))$
and $T = T_{\gamma} = (T_{\gamma,1}, T_{\gamma,2})$ where $T = T_0 = (g^{c_3}
f^{z_3}, g^{z_3})$ or $T = T_1 = (g^d f^{z_3}, g^{z_3})$. $\mathcal{B}$ first
chooses random values $a_1, a_3 \in \Z_p$ and sets
    \begin{align*}
    &   g^{\vect{b}_{1,1}} = (g, 1, g^{a_1}),~
        g^{\vect{b}_{1,2}} = (g, f, 1),~
        g^{\vect{b}_2} = (f, g^{-1}, f^{a_1} g^{-a_3}),~
        g^{\vect{b}_3} = (g^{a_1}, g^{a_3}, g^{-1}),~ \\
    &   (g^{\vect{b}_{1,2}})^{c_1} = (g^{c_1}, f^{c_1}, 1),~
        (g^{\vect{b}_{1,2}})^{c_2} = (g^{c_2}, f^{c_2}, 1),~ \\
    &   (g^{\vect{b}_{1,1}})^{c_1 c_2} (g^{\vect{b}_2})^{z_1}
        = (g^{c_1 c_2} f^{z_1}, (g^{z_1})^{-1}, (g^{c_1 c_2} f^{z_1})^{a_1}
        (g^{z_1})^{-a_3}),~ \\
    &   (g^{\vect{b}_{1,1}})^{c_1 c_2 c_3} (g^{\vect{b}_2})^{z_2}
        = (g^{c_1 c_2 c_3} f^{z_2}, (g^{z_2})^{-1},
        (g^{c_1 c_2 c_3} f^{z_2})^{a_1} (g^{z_2})^{-a_3}),~ \\
    &   T' = (T_{\gamma,1}, T_{\gamma,2}, (T_{\gamma,1})^{a_1}
        (T_{\gamma,2})^{-a_3}).
    \end{align*}
Intuitively, it sets $a_2 = \text{dlog}(f)$. Next, it gives the tuple
    $D' = ((p,\G, \G_T, e),
    g^{\vect{b}_{1,1}}, g^{\vect{b}_{1,2}}, \lb g^{\vect{b}_2}, g^{\vect{b}_3},
    (g^{\vect{b}_{1,2}})^{c_1}, (g^{\vect{b}_{1,2}})^{c_2}, \lb
    (g^{\vect{b}_{1,1}})^{c_1 c_2} (g^{\vect{b}_2})^{z_1},
    (g^{\vect{b}_{1,1}})^{c_1 c_2 c_3} (g^{\vect{b}_2})^{z_2} )$ and $T'$
to $\mathcal{A}$. Then $\mathcal{A}$ outputs a guess $\gamma'$. $\mathcal{B}$
also outputs $\gamma'$. If the advantage of $\mathcal{A}$ is $\epsilon$, then
the advantage of $\mathcal{B}$ is greater than $\epsilon$ since the
distribution of the challenge tuple to $\mathcal{A}$ is equal to the
Assumption \ref*{sec-ll-hve}-3.
\end{proof}

\begin{lemma} \label{lem-ll-hve-a4}
If the decisional P3DH assumption holds, then the Assumption
\ref*{sec-ll-hve}-4 also holds.
\end{lemma}

\begin{proof}
Suppose there exists an adversary $\mathcal{A}$ that breaks the Assumption
\ref*{sec-ll-hve}-4 with a non-negligible advantage. An algorithm
$\mathcal{B}$ that solves the decisional P3DH assumption using $\mathcal{A}$
is given: a challenge tuple
    $D = ((p, \G, \G_T, e), (g,f),
    (g^{c_1}, f^{c_1}), \lb (g^{c_2}, f^{c_2}),
    (g^{c_1 c_2} f^{z_1}, g^{z_1}), (g^{c_3} f^{z_2}, g^{z_2}))$
and $T = T_{\gamma} = (T_{\gamma,1}, T_{\gamma,2})$ where $T = T_0 = (g^{c_1
c_2 c_3} f^{z_3}, g^{z_3})$ or $T = T_1 = (g^d f^{z_3}, g^{z_3})$.
$\mathcal{B}$ first chooses random values $a_2, a_3 \in \Z_p$ and sets
    \begin{align*}
    &   g^{\vect{b}_{1,1}} = (g, 1, f),~
        g^{\vect{b}_{1,2}} = (g, g^{a_2}, 1),~
        g^{\vect{b}_2} = (g^{a_2}, g^{-1}, g^{a_3}),~
        g^{\vect{b}_3} = (f, f^{a_2} g^{-a_3}, g^{-1}),~ \\
    &   (g^{\vect{b}_{1,2}})^{c'_1} (g^{\vect{b}_3})^{z_1}
        = (g^{c_1 c_2} f^{z_1}, (g^{c_1 c_2} f^{z_1})^{a_2} (g^{z_1})^{-a_3},
        (g^{z_1})^{-1}),~ \\
    &   (g^{\vect{b}_{1,2}})^{c'_2} (g^{\vect{b}_3})^{z_2}
        = (g^{c_3} f^{z_2}, (g^{c_3} f^{z_2})^{a_2} (g^{z_2})^{-a_3}, (g^{z_2})^{-1}),~ \\
    &   T' = (T_{\gamma,1}, (T_{\gamma,1})^{a_2} (T_{\gamma,2})^{-a_3},
        (T_{\gamma,2})^{-1}).
    \end{align*}
Intuitively, it sets $a'_1 = \text{dlog}(f), a'_2 = a_2, a'_3 = a_1 a_2 -
a_3$ and $c'_1 = c_1 c_2, c'_2 = c_3$ where $a'_1, a'_2, a'_3$ are elements
of basis vectors for the Assumption 5-4. Next, it gives the tuple
    $D' = ((p,\G, \G_T, e),
    g^{\vect{b}_{1,1}}, g^{\vect{b}_{1,2}}, g^{\vect{b}_2}, g^{\vect{b}_3},
    (g^{\vect{b}_{1,1}})^{c'_1} \lb (g^{\vect{b}_2})^{z_1}, \lb
    (g^{\vect{b}_{1,1}})^{c'_2} (g^{\vect{b}_2})^{z_2} )$ and $T'$
to $\mathcal{A}$. Then $\mathcal{A}$ outputs a guess $\gamma'$. $\mathcal{B}$
also outputs $\gamma'$. If the advantage of $\mathcal{A}$ is $\epsilon$, then
the advantage of $\mathcal{B}$ is greater than $\epsilon$ since the
distribution of the challenge tuple to $\mathcal{A}$ is equal to the
Assumption \ref*{sec-ll-hve}-4.
\end{proof}

\section{Conversion 3: SW-dHVE} \label{sec-sw-dhve}

In this section, we convert the delegatable HVE scheme of Shi and Waters
\cite{ShiW08} to prime order bilinear groups and prove its selective model
security under the decisional BDH and P3DH assumptions.

\subsection{Construction}

Let $\Sigma$ be a finite set of attributes and let $?, *$ be two special
symbol not in $\Sigma$. Define $\Sigma_{?,*} = \Sigma \cup \{?,*\}$. The
symbol $?$ denotes a delegatable field, i.e., a field where one is allowed to
fill in an arbitrary value and perform delegation. The symbol $*$ denotes a
wild-card field or ``don't care'' field.

\begin{description}
\item[\normalfont{\textsf{Setup}($1^{\lambda}, l$)}:] The setup algorithm
    first generates the bilinear group $\G$ of prime order $p$ of bit size
    $\Theta(\lambda)$. It chooses random values $a_1, a_2, a_3 \in \Z_p$
    and sets basis vectors for bilinear product groups as $\vect{b}_{1,1} =
    (1, 0, a_1),~ \vect{b}_{1,2} = (1, a_2, 0),~ \vect{b}_2 = (a_2, -1, a_1
    a_2 - a_3),~ \vect{b}_3 = (a_1, a_3, -1)$. Next, it chooses random
    exponents $v', w'_1, w'_2, \{ u'_i, h'_i \}_{i=1}^l, \alpha \in \Z_p$,
    and it computes the following values using the basis vectors
    \begin{align*}
    &   \vect{B}_{1,1} = g^{\vect{b}_{1,1}},~
        \vect{B}_{1,2} = g^{\vect{b}_{1,2}},~
        \vect{B}_2 = g^{\vect{b}_2},~
        \vect{B}_3 = g^{\vect{b}_3},~ \\
    &   g^{\vect{v}_1} = \vect{B}_{1,1}^{v'},~
        g^{\vect{w}_{1,1}} = \vect{B}_{1,1}^{w'_1},~
        g^{\vect{w}_{1,2}} = \vect{B}_{1,1}^{w'_2},~
        \big\{
        g^{\vect{u}_{1,i}} = \vect{B}_{1,1}^{u'_i},~
        g^{\vect{h}_{1,i}} = \vect{B}_{1,1}^{h'_i}
        \big\}_{i=1}^l,~ \\
    &   g^{\vect{v}_2} = \vect{B}_{1,2}^{v'},~
        g^{\vect{w}_{2,1}} = \vect{B}_{1,2}^{w'_1},~
        g^{\vect{w}_{2,2}} = \vect{B}_{1,2}^{w'_2},~
        \big\{
        g^{\vect{u}_{2,i}} = \vect{B}_{1,2}^{u'_i},~
        g^{\vect{h}_{2,i}} = \vect{B}_{1,2}^{h'_i}
        \big\}_{i=1}^l.
    \end{align*}
    It keeps $g^{\vect{v}_2}, \{ g^{\vect{u}_{2,i}}, g^{\vect{h}_{2,i}},
    g^{\vect{w}_{2,i}} \}_{i=1}^l, (g^{\vect{b}_{1,2}})^{\alpha}$ as a
    secret key $\textsf{SK}$. Then it publishes a public key \textsf{PK}
    using random blinding values $z_v, z_{w,1}, z_{w,2}, \{ z_{u,i},
    z_{h,i} \}_{i=1}^l \in \Z_p$ as follows
    \begin{align*}
    \textsf{PK} = \Big(~
    &   \vect{B}_{1,1},~ \vect{B}_{1,2},~ \vect{B}_2,~ \vect{B}_3,~
        \vect{V} = g^{\vect{v}_1} \vect{B}_2^{z_v},~
        \vect{W}_1 = g^{\vect{w}_{1,1}} \vect{B}_2^{z_{w,1}},~
        \vect{W}_2 = g^{\vect{w}_{1,2}} \vect{B}_2^{z_{w,2}},~ \\
    &   \big\{
        \vect{U}_i = g^{\vect{u}_{1,i}} \vect{B}_2^{z_{u,i}},~
        \vect{H}_i = g^{\vect{h}_{1,i}} \vect{B}_2^{z_{h,i}}
        \big\}_{i=1}^{l},~
        \Omega = e(g^{\vect{v}_1}, g^{\vect{b}_{1,2}})^{\alpha} ~\Big).
    \end{align*}

\item[\normalfont{\textsf{GenToken}($\vect{\sigma}, \textsf{SK},
    \textsf{PK}$)}:] The token generation algorithm takes as input an
    attribute vector $\vect{\sigma} = (\sigma_1, \ldots, \sigma_l) \in
    \Sigma_{?,*}^l$ and the secret key $\textsf{SK}$. Let $S$ be the set of
    indexes that are not delegatable fields and wild-card fields in the
    vector $\vec{\sigma}$. It first selects random exponents $r_1, r_2, \{
    r_{3,i} \}_{i \in S} \in \Z_p$ and random blinding values $y_1, y_2,
    y_3, \{ y_{4,i} \}_{i \in S} \in \Z_p$. Then it computes decryption
    components as
    \begin{align*}
    &   \vect{K}_1 = (g^{\vect{b}_{1,2}})^{\alpha}
        (g^{\vect{w}_{2,1}})^{r_1} (g^{\vect{w}_{2,2}})^{r_2}
        \prod_{i \in S} ((g^{\vect{u}_{2,i}})^{\sigma_i} g^{\vect{h}_{2,i}})^{r_{3,i}}
        \vect{B}_3^{y_1},~ \\
    &   \vect{K}_2 = (g^{\vect{v}_2})^{-r_1} \vect{B}_3^{y_2},~
        \vect{K}_3 = (g^{\vect{v}_2})^{-r_2} \vect{B}_3^{y_3},~
        \big\{
        \vect{K}_{4,i} = (g^{\vect{v}_2})^{-r_{3,i}} \vect{B}_3^{y_{4,i}}
        \big\}_{i \in S}.
    \end{align*}
    Let $S_?$ be the set of indexes that are delegatable fields. It selects
    random exponents $\{ s_{1,j}, s_{2,j}, \{ s_{3,j,i} \} \} \in \Z_p$ and
    random blinding values $\{ y_{1,j,u}, y_{1,j,h}, y_{2,j}, y_{3,j}, \{
    y_{4,j,i} \} \} \in \Z_p$. Next, it computes delegation components as
    \begin{align*}
    \forall j \in S_? :
    &   \vect{L}_{1,j,u} = (g^{\vect{u}_{2,i}})^{s_{3,j,j}}
        \vect{B}_3^{y_{1,j,u}},~ \\
    &   \vect{L}_{1,j,h} =
            (g^{\vect{w}_{2,1}})^{s_{1,j}} (g^{\vect{w}_{2,2}})^{s_{2,j}}
            \prod_{i \in S} ((g^{\vect{u}_{2,i}})^{\sigma_i}
                g^{\vect{h}_{2,i}})^{s_{3,j,i}}
            (g^{\vect{h}_{2,j}})^{s_{3,j,j}} \vect{B}_3^{y_{1,j,h}},~ \\
    &   \vect{L}_{2,j} = (g^{\vect{v}_2})^{-s_{1,j}} \vect{B}_3^{y_{2,j}},~
        \vect{L}_{3,j} = (g^{\vect{v}_2})^{-s_{2,j}} \vect{B}_3^{y_{3,j}},~
        \big\{ \vect{L}_{4,j,i} = (g^{\vect{v}_2})^{-s_{3,j,i}}
            \vect{B}_3^{y_{4,j,i}}
        \big\}_{i \in S \cup \{j\}}.
    \end{align*}
    Finally, it outputs a token as
    \begin{align*}
    \textsf{TK}_{\vect{\sigma}} = \Big(~
    &   \vect{K}_1,~ \vect{K}_2,~ \vect{K}_3,~ \{ \vect{K}_{4,i} \}_{i \in
        S},~
        \big\{ \vect{L}_{1,j,u},~ \vect{L}_{1,j,h},~ \vect{L}_{2,j},~
            \vect{L}_{3,j},~ \{ \vect{L}_{4,j,i} \}_{i \in S \cup \{j\}}
        \big\}_{j \in S_?}
    ~\Big).
    \end{align*}

\item[\normalfont{\textsf{Delegate}($\vect{\sigma}',
    \textsf{TK}_{\vect{\sigma}}, \textsf{PK}$)}:] The delegation algorithm
    takes as input an attribute vector $\vect{\sigma}' = (\sigma_1, \ldots,
    \sigma_l) \in \Sigma_{?,*}^l$ and a token
    $\textsf{TK}_{\vect{\sigma}}$. Without loss of generality, we assume
    that $\sigma'$ fixes only one delegatable field of $\sigma$. It is
    clear that we can perform delegation on multiple fields if we have an
    algorithm to perform delegation on one field. Suppose $\sigma'$ fixes
    the $k$-th index of $\sigma$. If the $k$-th index of $\sigma'$ is set
    to $*$, that is, a wild-card field, then it can perform delegation by
    simply removing the delegation components that correspond to $k$-th
    index. Otherwise, that is, if the $k$-th index of $\sigma'$ is set to
    some value in $\Sigma$, then it perform delegation as follows.

    Let $S$ be the set of indexes that are not delegatable fields and
    wild-card fields in the vector $\vect{\sigma'}$. Note that $k \in S$.
    It selects random exponents $\mu, y_1, y_2, y_3, \lb \{y_{4,i}\}_{i \in
    S} \in \Z_p$ and updates the token as
    \begin{align*}
    &   \tilde{\vect{K}}_1 = \vect{K}_1
        (\vect{L}_{1,k,u}^{\sigma_k} \vect{L}_{1,k,h})^{\mu} \vect{B}_3^{y_1},~
        \tilde{\vect{K}}_2 = \vect{K}_2 \vect{L}_{2,k}^{\mu} \vect{B}_3^{y_2},~
        \tilde{\vect{K}}_3 = \vect{K}_3 \vect{L}_{3,k}^{\mu} \vect{B}_3^{y_3},~ \\
    &   \tilde{\vect{K}}_{4,k} = \vect{L}_{4,k,k}^{\mu} \vect{B}_3^{y_{4,k}},~
        \big\{
        \tilde{\vect{K}}_{4,i} = \vect{K}_{4,i} \vect{L}_{4,k,i}^{\mu}
        \vect{B}_3^{y_{4,i}}
        \big\}_{i \in S \setminus \{k\}}.
    \end{align*}
    Let $S_?$ be the set of indexes that are delegatable fields in the
    vector $\vect{\sigma'}$. It selects random exponents $\{ \tau_j,
    y_{1,j,u}, y_{1,j,h}, y_{2,j}, y_{3,j}, \{ y_{4,j,i} \}_{i \in S \cup
    \{j\}} \}_{j \in S_?} \in \Z_p$ and re-randomize the delegation
    components of the token as
    \begin{align*}
    \forall j \in S_? :
    &   \tilde{\vect{L}}_{1,j,u} = \vect{L}_{1,j,u}^{\mu} \vect{B}_3^{y_{1,j,u}},~
        \tilde{\vect{L}}_{1,j,h} = \vect{L}_{1,j,h}^{\mu}
            (\vect{L}_{1,k,u}^{\sigma_k} \vect{L}_{1,k,h})^{\tau_j}
            \vect{B}_3^{y_{1,j,h}},~ \\
    &   \tilde{\vect{L}}_{2,j} = \vect{L}_{2,j}^{\mu} \vect{L}_{2,j}^{\tau_j}
            \vect{B}_3^{y_{2,j}},~
        \tilde{\vect{L}}_{3,j} = \vect{L}_{3,j}^{\mu} \vect{L}_{3,j}^{\tau_j}
            \vect{B}_3^{y_{3,j}},~ \\
    &   \tilde{\vect{L}}_{4,j,j} = \vect{L}_{4,j,j}^{\mu} \vect{B}_3^{y_{4,j,j}},~
        \tilde{\vect{L}}_{4,j,k} = \vect{L}_{4,j,k}^{\tau_j}
        \vect{B}_3^{y_{4,j,k}},~
        \big\{
        \tilde{\vect{L}}_{4,j,i} = \vect{L}_{4,j,i}^{\mu} \vect{L}_{4,j,k}^{\tau_j}
            \vect{B}_3^{y_{4,j,i}}
        \big\}_{i \in S \setminus \{k\}}.
    \end{align*}
    Finally, it outputs a token as
    \begin{align*}
    \textsf{TK}_{\vect{\sigma}'} = \Big(~
    &   \tilde{\vect{K}}_1,~ \tilde{\vect{K}}_2,~ \tilde{\vect{K}}_3,~
        \{ \tilde{\vect{K}}_{4,i} \}_{i \in S},~
        \big\{
        \tilde{\vect{L}}_{1,j,h}, \tilde{\vect{L}}_{1,j,u},~
        \tilde{\vect{L}}_{2,j},~ \tilde{\vect{L}}_{3,j},~
        \{ \tilde{\vect{L}}_{4,j,i} \}_{i \in S \cup \{j\}}
        \big\}_{j \in S_?}
    ~\Big).
    \end{align*}

\item[\normalfont{\textsf{Encrypt}($\vect{x}, M, \textsf{PK}$)}:] The
    encryption algorithm takes as input an attribute vector $\vect{x} =
    (x_1, \ldots, x_l) \in \Sigma^l$, a message $M \in \mathcal{M}
    \subseteq \G_T$, and the public key $\textsf{PK}$. It first chooses a
    random exponent $t \in \Z_p$ and random blinding values $z_1, z_2, z_3,
    \{ z_{4,i} \}_{i=1}^l \in \Z_p$. Then it outputs a ciphertext as
    \begin{align*}
    \textsf{CT} = \Big(~
    &   C_0 = \Omega^t M,~
        \vect{C}_1 = \vect{V}^t \vect{B}_2^{z_1},~
        \vect{C}_2 = \vect{W}_1^t \vect{B}_2^{z_2},~
        \vect{C}_3 = \vect{W}_2^t \vect{B}_2^{z_3},~
        \big\{
        \vect{C}_{4,i} = (\vect{U}_i^{x_i} \vect{H}_i)^t \vect{B}_2^{z_{4,i}}
        \big\}_{i=1}^l
    ~\Big).
    \end{align*}

\item[\normalfont{\textsf{Query}($\textsf{CT}, \textsf{TK}_{\vec{\sigma}},
    \textsf{PK}$)}:] The query algorithm takes as input a ciphertext
    $\textsf{CT}$ and a token $\textsf{TK}_{\vec{\sigma}}$ of a vector
    $\vec{\sigma}$. It first computes
    \begin{align*}
    M \leftarrow C_0 \cdot \Big( e(\vect{C}_1, \vect{K}_1) \cdot
        e(\vect{C}_2, \vect{K}_2) \cdot e(\vect{C}_3, \vect{K}_3) \cdot
        \prod_{i \in S} e(\vect{C}_{4,i}, \vect{K}_{4,i}) \Big)^{-1}.
    \end{align*}
    If $M \notin \mathcal{M}$, it outputs $\perp$ indicating that the
    predicate $f_{\vec{\sigma}}$ is not satisfied. Otherwise, it outputs
    $M$ indicating that the predicate $f_{\vec{\sigma}}$ is satisfied.
\end{description}

\subsection{Correctness}

If $f_{\vec{\sigma}}(\vec{x}) = 1$, then the following calculation shows that
$\textsf{Query}(\textsf{CT},\textsf{TK}_{\vec{\sigma}},\textsf{PK}) = M$ by
the orthogonality of basis vectors such that $e(g^{\vect{b}_{1,1}},
g^{\vect{b}_3}) = 1, e(g^{\vect{b}_{1,2}}, g^{\vect{b}_2}) = 1$, and
$e(g^{\vect{b}_2}, g^{\vect{b}_3}) = 1$.
    \begin{align*}
    \lefteqn{ e(\vect{C}_1, \vect{K}_1) \cdot e(\vect{C}_2, \vect{K}_2) \cdot
        e(\vect{C}_3, \vect{K}_3) \cdot
        \prod_{i \in S} e(\vect{C}_{4,i}, \vect{K}_{4,i}) } \\
    &=  e((g^{\vect{v}_1})^t, (g^{\vect{b}_{1,2}})^{\alpha}
            (g^{\vect{w}_{2,1}})^{r_1} (g^{\vect{w}_{2,2}})^{r_2}
            \prod_{i \in S} ((g^{\vect{u}_{2,i}})^{\sigma_i}
                g^{\vect{h}_{2,i}})^{r_{3,i}}) \cdot \\
    &\quad
        e((g^{\vect{w}_{1,1}})^t, (g^{\vect{v}_2})^{-r_1}) \cdot
        e((g^{\vect{w}_{1,2}})^t, (g^{\vect{v}_2})^{-r_2}) \cdot
        \prod_{i \in S} e(((g^{\vect{u}_{1,i}})^{x_i} g^{\vect{h}_{1,i}})^t,
            (g^{\vect{v}_2})^{-r_{3,i}}) \\
    &=  e((g^{\vect{v}_1})^t, (g^{\vect{b}_{1,2}})^{\alpha}) \cdot
        \prod_{i \in S} e(g^{v'}, (g^{u'_i})^{(\sigma_i - x_i)})^{t r_{3,i}}
     =  e(g^{\vect{v}_1}, g^{\vect{b}_{1,2}})^{\alpha t}.
    \end{align*}
Otherwise, that is $f_{\vec{\sigma}}(\vec{x}) = 0$, the probability of
$\textsf{Query}(\textsf{CT},\textsf{TK}_{\vec{\sigma}},\textsf{PK}) \neq
\perp$ is negligible by limiting $|\mathcal{M}|$ to less than $|\G_T|^{1/4}$.

\subsection{Security}

\begin{theorem} \label{thm-sw-dhve}
The above dHVE construction is selectively secure under the decisional BDH
and P3DH assumptions.
\end{theorem}

\noindent The proof of this theorem is easily obtained from the following
five Lemma \ref{lem-sw-dhve}, \ref{lem-sw-dhve-a1}, \ref{lem-sw-dhve-a2},
\ref{lem-sw-dhve-a3} and \ref{lem-sw-dhve-a4}. Before presenting the five
lemmas, we first introduce the following four assumptions. The HVE scheme of
Shi and Waters constructed in bilinear groups of composite $n = p_1 p_2 p_3$
order, and its security was proven under the decisional BDH, Bilinear
Subgroup Decision (BSD), and the decisional C3DH assumptions \cite{ShiW08}.
In composite order bilinear groups, the decisional C3DH assumption imply the
decisional $l$-C3DH assumption that was introduced in \cite{ShiW08}. However,
in prime order bilinear groups, this implication is not valid since the basis
vectors for ciphertexts and tokens are different. Thus the decisional C3DH
assumption for ciphertexts and the decisional C3DH assumption for tokens
should be treated as differently. These assumptions in composite order
bilinear groups are converted to the following Assumptions
\ref*{sec-sw-dhve}-1, \ref*{sec-sw-dhve}-2, \ref*{sec-sw-dhve}-3, and
\ref*{sec-sw-dhve}-4 using our conversion method.

\vs \noindent \textbf{Assumption \ref*{sec-sw-dhve}-1} Let $((p, \G, \G_T,
e), g^{\vect{b}_{1,1}}, g^{\vect{b}_{1,2}}, g^{\vect{b}_2}, g^{\vect{b}_3})$
be the bilinear product group of basis vectors $\vect{b}_{1,1} = (1,0,a_1),
\vect{b}_{1,2} = (1,a_2,0), \vect{b}_2 = (a_2,-1,a_1 a_2 - a_3)$, and
$\vect{b}_3 = (a_1, a_3, -1)$. The Assumption \ref*{sec-sw-dhve}-1 is stated
as follows: given a challenge tuple
    \begin{align*}
    D = \big( (p, \G, \G_T, e),~
        g^{\vect{b}_{1,1}}, g^{\vect{b}_{1,2}}, g^{\vect{b}_2},
        g^{\vect{b}_3},
        (g^{\vect{b}_{1,1}})^{c_1}, (g^{\vect{b}_{1,1}})^{c_2},
        (g^{\vect{b}_{1,2}})^{c_1}, (g^{\vect{b}_{1,2}})^{c_2},
        (g^{\vect{b}_{1,1}})^{c_3} \big) \mbox{ and } T,
    \end{align*}
decides whether $T = T_0 = e(g, g)^{c_1 c_2 c_3}$ or $T = T_1 = e(g,g)^d$
with random choices of $c_1, c_2, c_3, d \in \Z_p$.

\vs \noindent \textbf{Assumption \ref*{sec-sw-dhve}-2} Let $((p, \G, \G_T,
e), g^{\vect{b}_{1,1}}, g^{\vect{b}_{1,2}}, g^{\vect{b}_2}, g^{\vect{b}_3})$
be the bilinear product group of basis vectors $\vect{b}_{1,1} = (1,0,a_1),
\vect{b}_{1,2} = (1,a_2,0), \vect{b}_2 = (a_2,-1,a_1 a_2 - a_3)$, and
$\vect{b}_3 = (a_1, a_3, -1)$. The Assumption \ref*{sec-sw-dhve}-2 is stated
as follows: given a challenge tuple
    $$D = \big( (p, \G, \G_T, e),~
    g^{\vect{b}_{1,1}}, g^{\vect{b}_{1,2}}, g^{\vect{b}_2}, g^{\vect{b}_3}
    \big) \mbox{ and } T,$$
decides whether $T = T_0 = e((g^{\vect{b}_{1,1}})^{c_1}
(g^{\vect{b}_2})^{c_3}, (g^{\vect{b}_{1,2}})^{c_2} (g^{\vect{b}_3})^{c_4})$
or $T = T_1 = e((g^{\vect{b}_{1,1}})^{c_1}, \lb (g^{\vect{b}_{1,2}})^{c_2})$
with random choices of $c_1, c_2, c_3, c_4 \in \Z_p$.

\vs \noindent \textbf{Assumption \ref*{sec-sw-dhve}-3} Let $((p, \G, \G_T,
e), g^{\vect{b}_{1,1}}, g^{\vect{b}_{1,2}}, g^{\vect{b}_2}, g^{\vect{b}_3})$
be the bilinear product group of basis vectors $\vect{b}_{1,1} = (1,0,a_1),
\vect{b}_{1,2} = (1,a_2,0), \vect{b}_2 = (a_2,-1,a_1 a_2 - a_3)$, and
$\vect{b}_3 = (a_1, a_3, -1)$. The Assumption \ref*{sec-sw-dhve}-3 is stated
as follows: given a challenge tuple
    \begin{align*}
    D = \big( (p, \G, \G_T, e),~
        g^{\vect{b}_{1,1}}, g^{\vect{b}_{1,2}}, g^{\vect{b}_2},
        g^{\vect{b}_3},
        (g^{\vect{b}_{1,2}})^{c_1}, (g^{\vect{b}_{1,2}})^{c_2},
        (g^{\vect{b}_{1,1}})^{c_1 c_2} (g^{\vect{b}_2})^{z_1},
        (g^{\vect{b}_{1,1}})^{c_1 c_2 c_3} (g^{\vect{b}_2})^{z_2}
    \big) \mbox{ and } T,
    \end{align*}
decides whether $T = T_0 = (g^{\vect{b}_{1,1}})^{c_3} (g^{\vect{b}_2})^{z_3}$
or $T = T_1 = (g^{\vect{b}_{1,1}})^d (g^{\vect{b}_2})^{z_3}$ with random
choices of $c_1, c_2, c_3, d \in \Z_p$, and $z_1, z_2, z_3 \in \Z_p$.

\vs \noindent \textbf{Assumption \ref*{sec-sw-dhve}-4} Let $((p, \G, \G_T,
e), g^{\vect{b}_{1,1}}, g^{\vect{b}_{1,2}}, g^{\vect{b}_2}, g^{\vect{b}_3})$
be the bilinear product group of basis vectors $\vect{b}_{1,1} = (1,0,a_1),
\vect{b}_{1,2} = (1,a_2,0), \vect{b}_2 = (a_2,-1,a_1 a_2 - a_3)$, and
$\vect{b}_3 = (a_1, a_3, -1)$. The Assumption \ref*{sec-sw-dhve}-4 is stated
as follows: given a challenge tuple
    \begin{align*}
    D = \big( (p, \G, \G_T, e),~
        g^{\vect{b}_{1,1}}, g^{\vect{b}_{1,2}}, g^{\vect{b}_2},
        g^{\vect{b}_3},
        (g^{\vect{b}_{1,1}})^{c_1}, (g^{\vect{b}_{1,1}})^{c_2},
        (g^{\vect{b}_{1,2}})^{c_1 c_2} (g^{\vect{b}_3})^{z_1},
        (g^{\vect{b}_{1,2}})^{c_1 c_2 c_3} (g^{\vect{b}_3})^{z_2}
    \big) \mbox{ and } T,
    \end{align*}
decides whether $T = T_0 = (g^{\vect{b}_{1,2}})^{c_3} (g^{\vect{b}_3})^{z_3}$
or $T = T_1 = (g^{\vect{b}_{1,2}})^d (g^{\vect{b}_3})^{z_3}$ with random
choices of $c_1, c_2, c_3, d \in \Z_p$, and $z_1, z_2, z_3 \in \Z_p$.

\begin{lemma} \label{lem-sw-dhve}
The above dHVE construction is selectively secure under the Assumptions
\ref*{sec-sw-dhve}-1, \ref*{sec-sw-dhve}-2, \ref*{sec-sw-dhve}-3, and
\ref*{sec-sw-dhve}-4.
\end{lemma}

\begin{proof}
The proof of this lemma is directly obtained from \cite{ShiW08} since the
Assumptions \ref*{sec-sw-dhve}-1, \ref*{sec-sw-dhve}-2, \ref*{sec-sw-dhve}-3,
and \ref*{sec-sw-dhve}-4 in prime order bilinear groups are correspond to the
Bilinear Diffie-Hellman (BDH), Bilinear Subgroup Decision (BSD), Composite
3-party Diffie-Hellman (C3DH), and Composite 3-party Diffie-Hellman (C3DH)
assumptions in composite order bilinear groups.
\end{proof}

\begin{lemma} \label{lem-sw-dhve-a1}
If the decisional BDH assumption holds, then the Assumption
\ref*{sec-sw-dhve}-1 also holds.
\end{lemma}

\begin{lemma} \label{lem-sw-dhve-a2}
The Assumption \ref*{sec-sw-dhve}-2 holds for all adversaries.
\end{lemma}

\begin{lemma} \label{lem-sw-dhve-a3}
If the decisional P3DH assumption holds, then the Assumption
\ref*{sec-sw-dhve}-3 also holds.
\end{lemma}

\noindent The Assumptions \ref*{sec-sw-dhve}-1, \ref*{sec-sw-dhve}-2, and
\ref*{sec-sw-dhve}-3 are the same as the Assumptions \ref*{sec-ll-hve}-1,
\ref*{sec-ll-hve}-2, and \ref*{sec-ll-hve}-3. Thus we omits the proof of
Lemma \ref*{lem-sw-dhve-a1}, \ref*{lem-sw-dhve-a2}, \ref*{lem-sw-dhve-a3}.

\begin{lemma} \label{lem-sw-dhve-a4}
If the decisional P3DH assumption holds, then the Assumption
\ref*{sec-sw-dhve}-4 also holds.
\end{lemma}

\begin{proof}
Suppose there exists an adversary $\mathcal{A}$ that breaks the Assumption
\ref*{sec-sw-dhve}-4 with a non-negligible advantage. An algorithm
$\mathcal{B}$ that solves the decisional P3DH assumption using $\mathcal{A}$
is given: a challenge tuple
    $D = ((p, \G, \G_T, e), (g,f),
    (g^{c_1}, f^{c_1}), \lb (g^{c_2}, f^{c_2}),
    (g^{c_1 c_2} f^{z_1}, g^{z_1}), (g^{c_1 c_2 c_3} f^{z_2}, g^{z_2}))$
and $T = T_{\gamma} = (T_{\gamma,1}, T_{\gamma,2})$ where $T = T_0 = (g^{c_3}
f^{z_3}, g^{z_3})$ or $T = T_1 = (g^d f^{z_3}, g^{z_3})$. $\mathcal{B}$ first
chooses random values $a_2, a_3 \in \Z_p$ and sets
    \begin{align*}
    &   g^{\vect{b}_{1,1}} = (g, 1, f),~
        g^{\vect{b}_{1,2}} = (g, g^{a_2}, 1),~
        g^{\vect{b}_2} = (g^{a_2}, g^{-1}, g^{a_3}),~
        g^{\vect{b}_3} = (f, f^{a_2} g^{-a_3}, g^{-1}),~ \\
    &   (g^{\vect{b}_{1,1}})^{c_1} = (g^{c_1}, 1, f^{c_1}),~
        (g^{\vect{b}_{1,1}})^{c_2} = (g^{c_2}, 1, f^{c_2}),~ \\
    &   (g^{\vect{b}_{1,2}})^{c_1 c_2} (g^{\vect{b}_3})^{z_1}
        = (g^{c_1 c_2} f^{z_1}, (g^{c_1 c_2} f^{z_1})^{a_2} (g^{z_1})^{-a_3},
            (g^{z_1})^{-1}),~ \\
    &   (g^{\vect{b}_{1,2}})^{c_1 c_2 c_3} (g^{\vect{b}_3})^{z_2}
        = (g^{c_1 c_2 c_3} f^{z_2}, (g^{c_1 c_2 c_3} f^{z_2})^{a_2} (g^{z_2})^{-a_3},
            (g^{z_2})^{-1}),~ \\
    &   T' = (T_{\gamma,1}, (T_{\gamma,1})^{a_2} (T_{\gamma,2})^{-a_3},
            (T_{\gamma,2})^{-1}).
    \end{align*}
Intuitively, it sets $a'_1 = \text{dlog}(f), a'_2 = a_2, a'_3 = a_1 a_2 -
a_3$ where $a'_1, a'_2, a'_3$ are elements of basis vectors for the
Assumption \ref*{sec-sw-dhve}-4. Next, it gives the tuple
    $D' = ((p, \G, \G_T, e),
    g^{\vect{b}_{1,1}}, g^{\vect{b}_{1,2}}, g^{\vect{b}_2}, g^{\vect{b}_3},
    (g^{\vect{b}_{1,1}})^{c_1}, (g^{\vect{b}_{1,1}})^{c_2}, \lb
    (g^{\vect{b}_{1,2}})^{c_1 c_2} (g^{\vect{b}_3})^{z_1},
    (g^{\vect{b}_{1,2}})^{c_1 c_2 c_3} (g^{\vect{b}_3})^{z_2} )$ and $T'$
to $\mathcal{A}$. Then $\mathcal{A}$ outputs a guess $\gamma'$. $\mathcal{B}$
also outputs $\gamma'$. If the advantage of $\mathcal{A}$ is $\epsilon$, then
the advantage of $\mathcal{B}$ is greater than $\epsilon$ since the
distribution of the challenge tuple to $\mathcal{A}$ is equal to the
Assumption \ref*{sec-sw-dhve}-4.
\end{proof}

%% file: chap-fully-secure-hve.tex
\section{Overview}

In this chapter, we propose a fully secure HVE scheme with short tokens. Our
construction based on composite order bilinear groups of products of four
primes and proved under four static assumptions.

The full security model is the right security model for predicate encryption.
However, it is not easy to provide full security model with reasonable
security reduction loss. Recently, Waters proposed a novel proof technique
called the dual system encryption \cite{Waters09}. In the dual system
encryption, the security proof consists of hybrid games that change the
original security game to a new game that can not be distinguishable from the
adversary's view.

The dual system encryption was very successful to prove the full security
model of hierarchical identity-based encryption, attribute-based encryption,
and public-key broadcast encryption. However, this technique does not work
well in predicate encryption. Main reason of this difficulty is that
predicate encryption should provide the attribute hiding property that
guarantees the anonymity of the ciphertexts and the adversary of predicate
encryption can query a predicate that satisfies with the challenge
ciphertext.

To overcome this problem, we restrict the adversary's capability as he can
only query predicates $f$ such that $f(x_0) = f(x_1) = 0$ where $x_0, x_1$
are the challenge vectors. That is, the adversary can not query a predicate
that satisfies with the challenge ciphertext.

\section{HVE in Composite Order Groups}

In this section, we construct an efficient HVE scheme in composite order
bilinear groups and prove its full model security under static assumptions.

\subsection{Construction}

Let $\Sigma = \Z_m$ for some integer $m$ and set $\Sigma_* = \Z_m \cup
\{*\}$. Our scheme is described as follows.

\begin{description}
\item[\normalfont{\textsf{Setup}($1^{\lambda}, l$)}:] The setup algorithm
    first generates the bilinear group $\G$ of composite order $n=p_1 p_2
    p_3 p_4$ where $p_1, p_2, p_3$ and $p_4$ are random primes of bit size
    $\Theta(\lambda)$. It chooses random elements $g, \{u_i, h_i\}_{i=1}^l
    \in \G_{p_1}, Z \in \G_{p_2}, Y \in \G_{p_3}$ and a random exponent
    $\alpha \in \Z_{p_1}$. It keeps $g, g^{\alpha}, \{u_i, h_i\}_{i=1}^l,
    Y$ as a master key $\textsf{MK}$. Next, it selects random elements
    $Z_v, \{Z_{u,i}, Z_{h,i}\}_{i=1}^l \in \G_{p_2}$ and publishes a public
    key \textsf{PK} as
    \begin{align*}
    \textsf{PK} = \Big(~
    &   V = g Z_v,~
        \big\{ U_i = u_i Z_{u,i},~ H_i = h_i Z_{h,i} \big\}_{i=1}^l,~
        Z,~ \Omega = e(g,g)^{\alpha} ~\Big).
    \end{align*}

\item[\normalfont{\textsf{GenToken}($\vect{\sigma}, \textsf{SK},
    \textsf{PK}$)}:] The token generation algorithm takes as input a vector
    $\vect{\sigma} = (\sigma_1, \ldots, \sigma_l) \in \Sigma_*^l$ and the
    secret key $\textsf{SK}$. Let $S$ be the set of indexes that are not
    wild cards in the vector $\vect{\sigma}$. It selects a random exponent
    $r \in \Z_n$ and random elements $Y_1, Y_2 \in \G_{p_3}$. Then it
    outputs a token as
    \begin{align*}
    \textsf{TK}_{\vect{\sigma}} = \Big(~
    &   K_1 = g^{\alpha} (\prod_{i\in S} u_i^{\sigma_i} h_i)^r Y_1,~
        K_2 = g^r Y_2
    ~\Big).
    \end{align*}

\item[\normalfont{\textsf{Encrypt}($\vect{x}, M, \textsf{PK}$)}:] The
    encryption algorithm takes as input a vector $\vect{x} = (x_1, \ldots,
    x_l) \in \Sigma^l$, a message $M \in \mathcal{M} \subseteq \G_T$, and
    the public key $\textsf{PK}$. It first chooses a random exponent $t \in
    \Z_n$ and random elements $Z_1, \{Z_{2,i}\}_{i=1}^l \in \G_{p_2}$. Then
    it outputs a ciphertext as
    \begin{align*}
    \textsf{CT} = \Big(~
    &   C_0 = \Omega^t M,~ C_1 = V^t Z_1,~
        \big\{ C_{2,i} = (U_i^{x_i} H_i)^t Z_{2,i} \big\}_{i=1}^l
    ~\Big).
    \end{align*}

\item[\normalfont{\textsf{Query}($\textsf{CT}, \textsf{TK}_{\vect{\sigma}},
    \textsf{PK}$)}:] The query algorithm takes as input a ciphertext
    $\textsf{CT}$ and a token $\textsf{TK}_{\vect{\sigma}}$ of a vector
    $\vect{\sigma}$. It first computes
    \begin{align*}
    M \leftarrow C_0 \cdot
        e(C_1, K_1)^{-1} \cdot e(\prod_{i\in S} C_{2,i}, K_2).
    \end{align*}
    If $M \notin \mathcal{M}$, it outputs $\perp$ indicating that the
    predicate $f_{\vect{\sigma}}$ is not satisfied. Otherwise, it outputs
    $M$ indicating that the predicate $f_{\vect{\sigma}}$ is satisfied.
\end{description}

\subsection{Correctness}

If $f_{\vect{\sigma}}(\vect{x}) = 1$, then the following calculations shows
that $\textsf{Query}(\textsf{CT}, \textsf{TK}_{\vect{\sigma}}, \textsf{PK}) =
M$ as
    \begin{align*}
    e(C_1, K_1)^{-1} \cdot e(\prod_{i\in S} C_{2,i}, K_2)
    =&  e(V^t Z_1, g^{\alpha} (\prod_{i\in S} u_i^{\sigma_i} h_i)^r Y_1)^{-1} \cdot
        e(\prod_{i\in S} (U_i^{x_i} H_i)^t Z_{2,i}, g^r Y_2) \\
    =&  e(g^t, g^{\alpha})^{-1} \cdot
        e((\prod_{i\in S} u_i^{(-\sigma_i + x_i)} h_i)^r, g^t)
    =   e(g,g)^{-\alpha t}.
    \end{align*}
Otherwise, that is $f_{\vect{\sigma}}(\vect{x}) = 0$, then the probability of
$\textsf{Query}(\textsf{CT},\textsf{TK}_{\vect{\sigma}},\textsf{PK}) \neq
\perp$ is negligible by limiting $|\mathcal{M}|$ to less than $|\G_T|^{1/4}$.

\subsection{Complexity Assumptions}

We introduce four static assumptions under composite order bilinear groups.

\vs \noindent \textbf{Assumption 1 (Subgroup Decision Assumption)} Let $(n,
\G, \G_T, e)$ be a description of the bilinear group of composite order
$n=p_1 p_2 p_3 p_4$. Let $g_{p_1}, g_{p_2}, g_{p_3}, g_{p_4}$ be generators
of subgroups of order $p_1, p_2, p_3, p_4$ of $\G$ respectively. The
Assumption 1 is stated as follows: given a challenge tuple
    $$D = ((n, \G, \G_T, e),~
    g_{p_1}, g_{p_2}, g_{p_3}) \mbox{ and } T,$$
decides whether $T = T_0 = Z_1 \in \G_{p_2}$ or $T = T_1 = Z_1 R_1 \in
\G_{p_2 p_4}$ with random choices of $Z_1 \in \G_{p_2}, R_1 \in \G_{p_4}$.

\vs \noindent \textbf{Assumption 2} Let $(n, \G, \G_T, e)$ be a description
of the bilinear group of composite order $n=p_1 p_2 p_3 p_4$. Let $g_{p_1},
g_{p_2}, g_{p_3}, g_{p_4}$ be generators of subgroups of order $p_1, p_2,
p_3, p_4$ of $\G$ respectively. The Assumption 2 is stated as follows: given
a challenge tuple
    $$D = ((n, \G, \G_T, e),~
    g_{p_1}, g_{p_2}, g_{p_3}, X_1 R_1, Y_1 R_2)
    \mbox{ and } T,$$
decides whether $T = T_0 = X_2 Y_2$ or $T = T_1 = X_2 Y_2 R_3$ with random
choices of $X_1, X_2 \in \G_{p_1}$, $Y_1, Y_2 \in \G_{p_3}$, $R_1, R_2, R_3
\in \G_{p_4}$.

\vs \noindent \textbf{Assumption 3} Let $(n, \G, \G_T, e)$ be a description
of the bilinear group of composite order $n=p_1 p_2 p_3 p_4$. Let $g_{p_1},
g_{p_2}, g_{p_3}, g_{p_4}$ be generators of subgroups of order $p_1, p_2,
p_3, p_4$ of $\G$ respectively. The Assumption 2 is stated as follows: given
a challenge tuple
    $$D = ((n, \G, \G_T, e),~
    g_{p_1}, g_{p_2}, g_{p_3}, g_{p_4},
    g_{p_1}^a Z_1, g_{p_1}^a Y_1 R_1, Y_2 R_1, g_{p_1}^b Z_2 R_2)
    \mbox{ and } T,$$
decides whether $T = T_0 = g_{p_1}^{ab} Z_3 R_3$ or $T = T_1 = g_{p_1}^c Z_3
R_3$ with random choices of $a, b, c \in \Z_{p_1}$, $Z_1, Z_2, Z_3 \in
\G_{p_2}$, $Y_1, Y_2 \in \G_{p_3}$, $R_1, R_2, R_3 \in \G_{p_4}$.

\vs \noindent \textbf{Assumption 4} Let $(n, \G, \G_T, e)$ be a description
of the bilinear group of composite order $n=p_1 p_2 p_3 p_4$. Let $g_{p_1},
g_{p_2}, g_{p_3}, g_{p_4}$ be generators of subgroups of order $p_1, p_2,
p_3, p_4$ of $\G$ respectively. The Assumption 3 is stated as follows: given
a challenge tuple
    $$D = ((n, \G, \G_T, e),~
    g_{p_1}, g_{p_2}, g_{p_3}, g_{p_4},
    g_{p_1}^a R_1, g_{p_1}^b R_2) \mbox{ and } T,$$
decides whether $T = T_0 = e(g_{p_1}, g_{p_1})^{ab}$ or $T = T_1 = e(g_{p_1},
g_{p_1})^c$ with random choices of $a, b, c, d \in \Z_{p_1}$, $R_1, R_2 \in
\G_{p_4}$.

\subsection{Security}

We describe a semi-functional ciphertext and a semi-functional token. They
are not used in a real system, but they are used in the proof of its
security. We let $g_{p_4}$ be a generator of the subgroup $\G_{p_4}$. Let
$(K'_1, K'_2)$ be a normal token and $y, z_k$ be random exponents in $\Z_n$.
Then the semi-functional token is generated as
    \begin{align*}
    semi\text{-}\textsf{SK}_{\vect{\sigma}} = \big(~
        K_1 = K'_1 \cdot g_{p_4}^{y z_k},~
        K_2 = K'_2 \cdot g_{p_4}^{y}
    ~\big).
    \end{align*}
Let $(C'_0, C'_1, \{ C'_{2,i} \})$ be a normal ciphertext and $x, z_{c,1},
\ldots, z_{c,l}$ be random exponents in $\Z_n$. Then the semi-functional
ciphertext is generated as
    \begin{align*}
    semi\text{-}\textsf{CT} = \big(~
        C_0 = C'_0,~ C_1 = C'_1 \cdot g_{p_4}^{x},~
        \{ C_{2,i} = C'_{2,i} \cdot g_{p_4}^{x z_{c,i}} \}_{i=1}^l
    ~\big).
    \end{align*}
Note that if a semi-functional token is used to decrypt a semi-functional
ciphertext, the decrypt algorithm will output the blinding factor multiplied
by the additional term $e(g_{p_4}, g_{p_4})^{xy(z_k - \sum_{i\in S}
z_{c,i})}$. If $z_k = \sum_{i\in S} z_{c,i}$, the the decrypt algorithm will
still work.

\begin{theorem} \label{thm-fhve-comp}
The above HVE construction is fully secure (match revealing) under the
Assumptions 1, 2, 3, and 4.
\end{theorem}

\begin{proof}
The proof uses a sequence of games. The first game will be the original
security game and the last one will be a game such that the adversary has no
advantage. We define the games as follows.

\vs \noindent $\textsf{Game}_0$. This game is the original full security
game. Note that the private keys and the challenge ciphertext are normal.

\svs \noindent $\textsf{Game}_1$. This game is almost identical to
$\textsf{Game}_0$ except that the challenge ciphertext of a challenge vector
$\vect{x}_{\gamma}$ is semi-functional.

\svs \noindent $\textsf{Game}_2$. This game is the same with the
$\textsf{Game}_1$ except that the tokens will be semi-functional. At this
moment, the tokens and the challenge ciphertexts are all semi-functional.

\svs \noindent $\textsf{Game}_3$. In this game we will replace the challenge
semi-functional ciphertext components $\{ C_{2,i} \}_{i=1}^l$ to random
elements in $\G_{p_1 p_2 p_4}$. In this case, the challenge ciphertext gives
no information about the challenge vector $\vect{x}_{\gamma}$.

\svs \noindent $\textsf{Game}_4$. We now define a new game $\textsf{Game}_4$.
This game differs from $\textsf{Game}_3$ in that the semi-functional
challenge ciphertext component $C_0$ is replaced by a random element in
$\G_{T,p_1}$. Note that in $\textsf{Game}_4$, the challenge ciphertext gives
no information about the vector $\vect{x}_{\gamma}$ and the encrypted message
$M_{\gamma}$. Therefore, the adversary can win this game with probability at
most $1/2$.

\vs Through the following four lemmas, we prove that it is hard to
distinguish $\textsf{Game}_{i-1}$ from $\textsf{Game}_{i}$ under the given
assumptions. Thus, the proof is easily obtained by the following four lemmas.
This completes our proof.
\end{proof}

\begin{lemma} \label{lem-fhve-comp-1}
If the Assumption 1 holds, then no polynomial-time adversary can distinguish
between $\textsf{Game}_0$ and $\textsf{Game}_1$ with a non-negligible
advantage.
\end{lemma}

\begin{proof}
Suppose there exists an adversary $\mathcal{A}$ that distinguishes between
$\textsf{Game}_{0}$ and $\textsf{Game}_{1}$ with a non-negligible advantage.
The simulator $\mathcal{B}$ that solves the Assumption 1 using $\mathcal{A}$
is given: a challenge tuple $D = ((n, \G, \G_T, e), g_{p_1}, g_{p_2},
g_{p_3})$ and $T$ where $T = Z_1 \in \G_{p_2}$ or $T = Z_1 R_1 \in \G_{p_2
p_4}$. Then $\mathcal{B}$ that interacts with $\mathcal{A}$ is described as
follows.

\begin{description}
\item [Setup:] $\mathcal{B}$ first chooses random elements $\{ u_i, h_i
    \}_{i=1}^l \in \G_{p_1}$ and a random exponent $\alpha \in \Z_n$. It
    selects random elements $Z_v, \{ Z_{u,i}, Z_{h,i} \}_{i=1}^l \in
    \G_{p_2}$ and publishes a public key as
    \begin{align*}
    V = g_{p_1} Z_v,~ \{ U_i = u_i Z_{u,i},~ H_i = h_i Z_{h,i} \}_{i=1}^l,~
    Z = g_{p_2},~ \Omega = e(g_{p_1}, g_{p_1})^{\alpha}.
    \end{align*}

\item [Query 1:] $\mathcal{A}$ adaptively requests a token query.
    $\mathcal{B}$ simply runs the token generation algorithm to create a
    normal token using the master key. Note that it can only create the
    normal tokens since it does not known $g_{p_4}$.

\item [Challenge:] $\mathcal{A}$ submits two vector $\vect{x}_0,
    \vect{x}_1$ and two messages $M_0, M_1$. $\mathcal{B}$ flips a random
    coin $\gamma$ internally, and it chooses random exponents $t, \{
    z_{c,i} \}_{i=1}^l \in \Z_n$. Then it outputs a ciphertext using random
    elements $ \{Z'_{2,i}\}_{i=1}^l \in \G_{p_2}$ as
    \begin{align*}
    C_0 = \Omega^{t} M_{\gamma},~ C_1 = V^t T,~
    \{ C_{2,i} = (U_i^{x_{\gamma,i}} H_i)^t T^{z_{c,i}} Z'_{2,i} \}_{i=1}^l.
    \end{align*}
    If $T = Z_1 \in \G_{p_2}$, then $\mathcal{B}$ is playing
    $\textsf{Game}_0$. Otherwise, it is playing $\textsf{Game}_1$. Note
    that it implicitly sets $g_{p_4}^x = R_1$.

\item [Query 2:] Same as Query Phase 1.

\item [Guess:] $\mathcal{A}$ outputs a guess $\gamma'$. If $\gamma =
    \gamma'$, it outputs $0$. Otherwise, it outputs $1$.
\end{description}

\noindent This completes our proof.
\end{proof}

\begin{lemma} \label{lem-fhve-comp-2}
If the Assumption 2 holds, then no polynomial-time adversary can distinguish
between $\textsf{Game}_1$ and $\textsf{Game}_2$ with a non-negligible
advantage.
\end{lemma}

\begin{proof}
Suppose that an adversary makes at most $q$ private key queries. We define a
sequence of games $\textsf{Game}_{1,0}, \textsf{Game}_{1,1}, \ldots,
\textsf{Game}_{1,q}$ where $\textsf{Game}_{1,0} = \textsf{Game}_1$. In
$\textsf{Game}_{1,i}$, for all $j$-th private key query such that $j > i$, a
normal private key is given to the adversary. However, for all $j$-th private
key query such that $j \leq i$, a semi-functional private key is given to the
adversary. It is obvious that $\textsf{Game}_{1,q}$ is equal with
$\textsf{Game}_2$.

\vs Suppose there exists an adversary $\mathcal{A}$ that distinguishes
between $\textsf{Game}_{1,k-1}$ and $\textsf{Game}_{1,k}$ with a
non-negligible advantage. A simulator $\mathcal{B}$ that solves the
Assumption 2 using $\mathcal{A}$ is given: a challenge tuple $D = ((n, \G,
\G_T, e), g_{p_1}, g_{p_2}, g_{p_3}, X_1 R_1, Y_1 R_2)$ and $T$ where $T =
X_2 Y_2$ or $T = X_2 Y_2 R_3$. Then $\mathcal{B}$ that interacts with
$\mathcal{A}$ is described as follows.

\begin{description}
\item [Setup:] $\mathcal{B}$ first chooses random exponents $\{u'_i, h'_i
    \}_{i=1}^l, \alpha \in \Z_n$ and sets $\{ u_i = g_{p_1}^{u'_i}, h_i =
    g_{p_1}^{h'_i} \}_{i=1}^l, Z = g_{p_2}, Y = g_{p_3}$. It selects random
    elements $Z_v, \{Z_{u,i}, Z_{h,i}\}_{i=1}^l \in \G_{p_2}$ and publishes
    a public key as
    \begin{align*}
    V = g_{p_1} Z_v,~
    \{ U_i = u_i Z_{u,i},~ H_i = h_i Z_{h,i} \}_{i=1}^l,~
    Z,~ \Omega = e(g_{p_1}, g_{p_1})^{\alpha}.
    \end{align*}

\item [Query 1:] $\mathcal{A}$ adaptively requests a token query for a
    vector $\vect{\sigma} = (\sigma_1, \ldots, \sigma_l) \in \Sigma_{*}^l$.
    If this is a $\rho$-th token query, then $\mathcal{B}$ handles this
    query as follows.
    \begin{description}
    \item [Case $\rho < k$ :] It selects random exponents $r, y, z_k \in
        \Z_n$. Then it chooses random elements $Y'_1, Y'_2 \in \G_{p_3}$
        and outputs a semi-functional token as
        \begin{align*}
        K_1 = g_{p_1}^{\alpha} (\prod_{i\in S} u_i^{\sigma_i} h_i)^r
              (Y_1 R_2)^{y z_k} Y'_1,~
        K_2 = g_{p_1}^r (Y_1 R_2)^{y} Y'_2.
        \end{align*}

    \item [Case $\rho = k$ :] It selects a random element $Y'_1 \in
        \G_{p_3}$ and outputs a token as
        \begin{align*}
        K_1 = g_{p_1}^{\alpha} T^{\sum_{i\in S} (u'_i \sigma_i + h'_i)} Y'_1,~
        K_2 = T.
        \end{align*}
    If $T = X_2 Y_2$, then $\mathcal{B}$ is playing $\textsf{Game}_{1,k-1}$.
    Otherwise, it is playing $\textsf{Game}_{1,k}$. Note that it implicitly
    sets $r = \text{dlog}(X_2), y = \text{dlog}(R_3)$, and $z_k = \sum_{i\in
    S} (u'_i \sigma_i + h'_i)$. It is obvious that the distribution of token
    is correct as follows
        \begin{align*}
        &   g_{p_1}^{\alpha} (\prod_{i\in S} u_i^{\sigma_i} h_i)^r Y_1 =
            g_{p_1}^{\alpha} g_{p_1}^{\sum_{i\in S} (u'_i \sigma_i + h'_i) r} Y_1 =
            g_{p_1}^{\alpha} (X_2 Y_2)^{\sum_{i\in S} (u'_i \sigma_i + h'_i)}
            Y'_1,~
            g_{p_1}^r Y_2 = X_2 Y_2.
        \end{align*}

    \item [Case $\rho > k$ :] It simply runs the token generation algorithm
        to create a normal token since it knows the master key.
\end{description}

\item [Challenge:] $\mathcal{A}$ submits two vectors $\vect{x}_0,
    \vect{x}_1$ and two messages $M_0, M_1$. $\mathcal{B}$ flips a random
    coin $\gamma$ internally and selects random elements $Z'_1,
    \{Z_{2,i}\}_{i=1}^l \in \G_{p_2}$. Then it outputs a semi-functional
    ciphertext as
    \begin{align*}
    C_0 = e(X_1 R_1, g_{p_1})^{\alpha} M_{\gamma},~
    C_1 = (X_1 R_1) Z'_1,~
    \{ C_{2,i} = (X_1 R_1)^{u'_i \sigma_{\gamma,i} + h'_i} Z'_{2,i} \}_{i=1}^l.
    \end{align*}
    Note that it by implicitly sets $t = dlog(X_1), x = dlog(R_1)$, and
    $z_{c,i} = u'_i \sigma_{\gamma,i} + h'_i$.

\item [Query 2:] Same as Query Phase 1.

\item [Guess:] $\mathcal{A}$ outputs a guess $\gamma'$. If $\gamma =
    \gamma'$, it outputs 0. Otherwise, it outputs 1.
\end{description}

\noindent The paradox of dual system encryption is solved since $\{ z_{c,i}
\}$ of the ciphertext with a vector $\vect{x}$ and $z_k$ of the $k$-th token
with a vector $\vect{\sigma}$ have the relation $z_k = \sum_{i\in S} z_{c,i}$
if $f_{\vect{\sigma}}(\vect{x}) = 1$. Additionally, the adversary cannot
detect any relationship between $\{z_{c,i}\}$ of the ciphertext and $z_k$ of
the $k$-th token since the function $u'_i \sigma_i + h'_i$ is a pairwise
independent function. This completes our proof.
\end{proof}

\begin{lemma} \label{lem-fhve-comp-3}
If the Assumption 3 holds, then no polynomial-time adversary can distinguish
between $\textsf{Game}_2$ and $\textsf{Game}_3$ with a non-negligible
advantage.
\end{lemma}

\begin{proof}
For the proof of this lemma, we define a sequence of games
$\textsf{Game}_{2,0}, \textsf{Game}_{2,1}, \lb \ldots, \textsf{Game}_{2,l}$
where $\textsf{Game}_{2,0} = \textsf{Game}_2$. In $\textsf{Game}_{2,i}$, the
semi-functional ciphertext components $\{C_{2,j}\}_{j=1}^i$ are replaced by
random elements in $\G_{p_1 p_2 p_4}$. It is obvious that
$\textsf{Game}_{2,l}$ is equal with $\textsf{Game}_3$.

\vs Suppose there exists an adversary $\mathcal{A}$ that distinguishes
between $\textsf{Game}_{2,k-1}$ and $\textsf{Game}_{2,k}$ with a
non-negligible advantage. A simulator $\mathcal{B}$ that solves the
Assumption 3 using $\mathcal{A}$ is given: a challenge tuple $D = ((n, \G,
\G_T, e), g_{p_1}, g_{p_2}, g_{p_3}, g_{p_4}, g_{p_1}^a Z_1, \lb g_{p_1}^a
Y_1 R_1, Y_2 R_1, g_{p_1}^b Z_2 R_2)$ and $T$ where $T = g_{p_1}^{ab} Z_3
R_3$ or $T = g_{p_1}^c Z_3 R_3$. Then $\mathcal{B}$ that interacts with
$\mathcal{A}$ is described as follows.

\begin{description}
\item [Setup:] $\mathcal{B}$ first chooses random exponents $\{u'_i,
    h'_i\}_{i=1}^l, \alpha \in \Z_n$. It selects $Z_v, \{Z_{u,i}, Z_{h,i}
    \}_{i=1}^l \lb \in \G_{p_2}$ and publishes a public key as
    \begin{align*}
    &   V = g_{p_1} Z_v,~
        \forall i\leq k : U_i = (g_{p_1}^a Z_1)^{u'_i} Z_{u,i},~
        \forall i > k   : U_i = g_{p_1}^{u'_i} Z_{u,i},~ \\
    &   \{ H_i = g_{p_1}^{h'_i} Z_{h,i} \}_{i=1}^l,~
        Z = g_{p_2},~ \Omega = e(g_{p_1}, g_{p_1})^{\alpha}.
    \end{align*}

\item [Query 1:] $\mathcal{A}$ adaptively requests a token query for a
    vector $\vect{\sigma} = (\sigma_1, \ldots, \sigma_l)$. $\mathcal{B}$
    selects random exponents $r, w \in \Z_n$, and random elements $Y'_1,
    Y'_2 \in \G_{p_3}$. Then it outputs a semi-functional token as
    \begin{align*}
    &   K_1 = g_{p_1}^{\alpha} (\prod_{i\in S_1} (g_{p_1}^a Y_1 R_1)^{u'_i \sigma_i} h_i)^r
        (\prod_{i\in S_2} g_{p_1}^{u'_i \sigma_i} h_i)^r (Y_2 R_1)^{rw}
        Y'_1,~
        K_2 = g_{p_1}^r (Y_2 R_1)^{r} Y'_2.
    \end{align*}
    Note that it implicitly sets $y = \text{dlog}(R_1) r$ and $z_k = \sum_{i\in
    S_1} u'_i \sigma_i + w$.

\item [Challenge:] $\mathcal{A}$ submits two vectors $\vect{x}_0,
    \vect{x}_1$ and two messages $M_0, M_1$. $\mathcal{B}$ flips a random
    coin $\gamma$ internally, and it chooses a random exponent $t \in
    \Z_n$, random elements $Z_1, Z_2 \in \G_{p_2}$. Then it outputs a
    semi-functional ciphertext as
    \begin{align*}
    &   C_0 = e(g_{p_1}^b Z_2 R_2, g_{p_1})^{\alpha} M_{\gamma},~
        C_1 = (g_{p_1}^b Z_2 R_2) Z'_1,~
        \forall i<k: C_{2,i} = P_i (Z_3 R_3)^{} Z'_{2,i},~ \\
    &   C_{2,k} = T^{u'_k \sigma_k + h'_k} Z'_{2,i},~
        \forall i>k: C_{2,i} = (g_{p_1}^b Z_2 R_2)^{u'_i \sigma_i + h'_i}
            Z'_{2,i}.
    \end{align*}
    If $T = g_{p_1}^{ab} Z_3 R_3$, then $\mathcal{B}$ is playing
    $\textsf{Game}_2$. Otherwise, it is playing $\textsf{Game}_3$. Note that it
    implicitly sets $t = b, x = \text{dlog}(R_2)$, and $z_{c,i} = u'_i \sigma_i
    + h'_i$.

\item [Query 2:] Same as Query Phase 1.

\item [Guess:] $\mathcal{A}$ outputs a guess $\gamma'$. If $\gamma =
    \gamma'$, it outputs 0. Otherwise, it outputs 1.
\end{description}

\noindent This completes our proof.
\end{proof}

\begin{lemma} \label{lem-fhve-comp-4}
If the Assumption 4 holds, then no polynomial-time adversary can distinguish
between $\textsf{Game}_3$ and $\textsf{Game}_4$ with a non-negligible
advantage.
\end{lemma}

\begin{proof}
Suppose there exists an adversary $\mathcal{A}$ that distinguishes between
$\textsf{Game}_2$ and $\textsf{Game}_3$ with a non-negligible advantage. A
simulator $\mathcal{B}$ that solves the Assumption 3 using $\mathcal{A}$ is
given: a challenge tuple $D = ((n, \G, \G_T, e), g_{p_1}, g_{p_2}, g_{p_3},
g_{p_4}, g_{p_1}^a R_1, g_{p_1}^b R_2)$ and $T$ where $T = e(g_{p_1},
g_{p_1})^{ab}$ or $T = e(g_{p_1}, g_{p_1})^c$. Then $\mathcal{B}$ that
interacts with $\mathcal{A}$ is described as follows.

\begin{description}
\item [Setup:] $\mathcal{B}$ first chooses random elements $\{u_i,
    h_i\}_{i=1}^l \in \G_{p_1}$. It implicitly sets $\alpha = a$ and
    publishes a public key using $Z_v, \{Z_{u,i}, Z_{h,i}\}_{i=1}^l \in
    \G_{p_2}$ as
    \begin{align*}
    &   V = g_{p_1} Z_v,~ \{ U_i = u_i Z_{u,i},~ H_i = h_i Z_{u,i} \},~
        Z = g_{p_2},~ \Omega = e(g_{p_1}, g_{p_1}^a R_1).
    \end{align*}

\item [Query 1:] $\mathcal{A}$ adaptively requests a token query for a
    vector $\vect{\sigma}$. $\mathcal{B}$ selects random exponents $r, z'_k
    \in \Z_n$ and a random element $R'_1 \in \G_{p_4}$. Then it outputs a
    semi-functional token using random elements $Y'_1, Y'_2 \in \G_{p_3}$
    as
    \begin{align*}
    K_1 = (g_{p_1}^a R_1) (\prod_{i\in S} u_i^{\sigma_i} h_i)^r (R'_1)^{z'_k} Y'_1,~
    K_2 = g_{p_1}^r (R'_1) Y'_2.
    \end{align*}
    Note that it implicitly sets $y = \text{dlog}(R'_1)$ and $z_k =
    \text{dlog}(R_1)/\text{dlog}(R'_1) + z'_k$.

\item [Challenge:] $\mathcal{A}$ submits two vectors $\vect{x}_0,
    \vect{x}_1$ and two messages $M_0, M_1$. $\mathcal{B}$ flips a random
    coin $\gamma$ internally, and it chooses a random exponent
    $\{w_i\}_{i=1}^l \in \Z_n$ and random elements $Z'_1,
    \{Z_{2,i}\}_{i=1}^l \in \G_{p_2}$. Then it outputs a semi-functional
    ciphertext with randomized $\{C_{2,i}\}$ components by implicitly
    setting $t = b$ as
    \begin{align*}
    &   C_0 = T M_{\gamma},~
        C_1 = (g_{p_1}^b R_2) Z'_1,~
        \{ C_{2,i} = (g_{p_1}^b R_2)^{w_i} Z'_{2,i} \}_{i=1}^l.
    \end{align*}
    If $T = e(g_{p_1}, g_{p_1})^{ab}$, then $\mathcal{B}$ is playing
    $\textsf{Game}_3$. Otherwise, it is playing $\textsf{Game}_4$.

\item [Query 2:] Same as Query Phase 1.

\item [Guess:] $\mathcal{A}$ outputs a guess $\gamma'$. If $\gamma =
    \gamma'$, it outputs 0. Otherwise, it outputs 1.
\end{description}

\noindent This completes our proof.
\end{proof}

%% file: chap-hve-applications.tex
In this chapter, we show that the HVE scheme supports conjunctive equality,
conjunctive comparison, conjunctive range, and conjunctive range queries on
encrypted data. The constructions of this chapter are based on
\cite{BonehW07}.

\section{Conjunctive Equality Queries}

It is trivial to construct a searchable encryption system that supports
conjunctive equality queries since the HVE scheme naturally supports
conjunctive equality queries. Therefore, we omits the construction.

\section{Conjunctive Comparison Queries}

Let $\Sigma_{01} = \{0,1\}$ and $\Sigma_{01*} = \{0,1,*\}$. Let
$(\textsf{Setup}_{HVE}, \textsf{GenToken}_{HVE}, \textsf{Encrypt}_{HVE}, \lb
\textsf{Query}_{HVE})$ be a secure HVE scheme over $\Sigma_{01}^{nw}$ where
$l=nw$. The searchable encryption for conjunctive comparison queries is
described as follows.

\begin{description}
\item[\normalfont{\textsf{Setup}($1^{\lambda}, n, w$)}:] The setup runs
    $\textsf{Setup}_{HVE}(1^{\lambda}, nw)$.

\item[\normalfont{\textsf{GenToken}($f_{\vect{a}}, \textsf{SK},
    \textsf{PK}$)}:] The token generation algorithm takes as input a
    predicate with a vector $\vect{a} = (a_1, \ldots, a_w) \lb \in \{1,
    \ldots, n\}^w$ and the secret key $\textsf{SK}$. It first defines
    $\sigma_*(\vect{a}) = (\sigma_{i,j}) \in \Sigma_{01*}^{nw}$ as follows:
    \begin{align*}
        \sigma_{i,j} = \left\{ \begin{array}{ll}
            1 & \mbox{ if } x_i = j, \\
            * & \mbox{ otherwise }
        \end{array} \right.
    \end{align*}
    It outputs $\textsf{GenToken}_{HVE}(\sigma_*(a), \textsf{SK}, \textsf{PK})$
    where the token size is $O(w)$.

\item[\normalfont{\textsf{Encrypt}($\vect{b}, M, \textsf{PK}$)}:] The
    encryption algorithm takes as input a vector $\vect{b} = (b_1, \ldots,
    b_w) \in \{1, \ldots, n\}^w$, a message $M \in \mathcal{M}$, and the
    public key $\textsf{PK}$. It first defines a vector $\vect{x}(\vect{b})
    = (x_{i,j}) \in \Sigma_{01}^{nw}$ as follows:
    \begin{align*}
        x_{i,j} = \left\{ \begin{array}{ll}
            1 & \mbox{ if } j \geq x_i, \\
            0 & \mbox{ otherwise }
        \end{array} \right.
    \end{align*}
    Then it outputs $\textsf{Encrypt}_{HVE}(\vect{\sigma}, M, \textsf{PK})$
    where the ciphertext size is $O(nw)$.

\item[\normalfont{\textsf{Query}($\textsf{CT}, \textsf{TK}_{\vect{a}},
    \textsf{PK}$)}:] The query algorithm outputs
    $\textsf{Query}_{HVE}(\textsf{CT}, \textsf{TK}_{\vect{a}})$.
\end{description}

\section{Conjunctive Range Queries}

In previous section, we constructed a searchable encryption system that
support comparison queries such that $x \leq a$ where the ciphertext contains
$x$ and the token contains $a$. It is easy to support comparison queries such
that $x \geq b$ by changing bit value of the ciphertext. Therefore, we can
construct a searchable encryption system that support range queries by
combining two comparison queries as $x \leq a \wedge x \geq b$ where the
ciphertext contains the pair $(x,x)$.

\section{Subset Queries}

Let $T$ be a set of size $n$. For a subset $A \subseteq T$, we define a
subset predicate as follows:
    \begin{align*}
        f_{A}(x) = \left\{ \begin{array}{ll}
            1 & \mbox{ if } x \in A, \\
            0 & \mbox{ otherwise }
        \end{array} \right.
    \end{align*}
The conjunctive subset predicates are naturally defined. Let $\Sigma_{01} =
\{0,1\}$ and $\Sigma_{01*} = \{0,1,*\}$. Let $(\textsf{Setup}_{HVE}, \lb
\textsf{GenToken}_{HVE}, \textsf{Encrypt}_{HVE}, \textsf{Query}_{HVE})$ be a
secure HVE scheme over $\Sigma_{01}^{nw}$ where $l=nw$. The searchable
encryption for conjunctive subset queries is described as follows.

\begin{description}
\item[\normalfont{\textsf{Setup}($1^{\lambda}, n, w$)}:] The setup runs
    $\textsf{Setup}_{HVE}(1^{\lambda}, nw)$.

\item[\normalfont{\textsf{GenToken}($f_{\vect{A}}, \textsf{SK},
    \textsf{PK}$)}:] The token generation algorithm takes as input a
    predicate with a vector $\vect{A} = (A_1, \ldots, A_w)$ and the secret
    key $\textsf{SK}$. It first defines $\sigma_*(\vect{A}) =
    (\sigma_{i,j}) \in \Sigma_{01*}^{nw}$ as follows:
    \begin{align*}
        \sigma_{i,j} = \left\{ \begin{array}{ll}
            0 & \mbox{ if } j \notin A_i, \\
            * & \mbox{ otherwise }
        \end{array} \right.
    \end{align*}
    It outputs $\textsf{GenToken}_{HVE}(\sigma_*(\vect{A}), \textsf{SK},
    \textsf{PK})$ where the token size is $O(nw)$.

\item[\normalfont{\textsf{Encrypt}($\vect{b}, M, \textsf{PK}$)}:] The
    encryption algorithm takes as input a vector $\vect{b} = (b_1, \ldots,
    b_w) \in T^w$, a message $M \in \mathcal{M}$, and the public key
    $\textsf{PK}$. It first defines a vector $\vect{x}(\vect{b}) =
    (x_{i,j}) \in \Sigma_{01}^{nw}$ as follows:
    \begin{align*}
        x_{i,j} = \left\{ \begin{array}{ll}
            1 & \mbox{ if } x_i = j, \\
            0 & \mbox{ otherwise }
        \end{array} \right.
    \end{align*}
    Then it outputs $\textsf{Encrypt}_{HVE}(\vect{x}, M, \textsf{PK})$ where
    the ciphertext size is $O(nw)$.

\item[\normalfont{\textsf{Query}($\textsf{CT}, \textsf{TK}_{\vect{A}},
    \textsf{PK}$)}:] The query algorithm outputs
    $\textsf{Query}_{HVE}(\textsf{CT}, \textsf{TK}_{\vect{A}})$.
\end{description}

%% file: chap-generic-group-model.tex
\section{Overview}

In this chapter, we prove that the new assumption of this thesis is secure
under the generic group model. The generic group model was introduced by
Shoup \cite{Shoup97}. The generic group model is a tool for analyzing generic
algorithms that work independently of the group representation. In the
generic group model, an adversary is given a random encoding of a group
element or an arbitrary index of a group element instead of an actual
representation of a group element. Thus, the adversary performs group
operations through oracles that provided by a simulator, and the adversary
only can check the equality of group elements. The detailed explanation of
the generic group model is given in \cite{BonehBG05,KatzSW08}.

The master theorems that can be used for the analysis of assumptions in
bilinear groups were presented in \cite{BonehBG05,KatzSW08,Freeman10}.
However, the new assumption of this paper can not be analyzed by the previous
master theorems. The reason of this difficulty is that the new assumption is
based on symmetric bilinear groups of prime order, the target group of our
assumption is $\mathbb{G}$ instead of $\mathbb{G}_T$, and the target consists
of many group elements instead of just one.

\section{Master Theorem}

To analyze the new assumption of this paper, we generalize the master theorem
of Katz et al. \cite{KatzSW08} to use prime order bilinear groups instead of
composite order bilinear groups and to use multiple groups elements in the
target instead of just one element.

Let $\mathbb{G}, \mathbb{G}_T$ be cyclic bilinear groups of order $p$ where
$p$ is a large prime. The bilinear map is defined as $e:\mathbb{G} \times
\mathbb{G} \rightarrow \mathbb{G}_T$. In the generic group model, a random
group element of $\mathbb{G}, \mathbb{G}_T$ is represented as a random
variable $P_i, R_i$ respectively where $P_i, R_i$ are chosen uniformly in
$\mathbb{Z}_p$. We say that a random variable has degree $t$ if the maximum
degree of any variable is $t$. Then we can naturally define the dependence
and independence of random variables as in Definition \ref{def-dep-indep}.

\begin{definition} \label{def-dep-indep}
Let $P = \{P_1, \ldots, P_u\},~ T_0 = \{T_{0,1}, \ldots, T_{0,m}\},~ T_1 =
\{T_{1,1}, \ldots, T_{1,m}\}$ be random variables over $\mathbb{G}$ where
$T_{0,i} \neq T_{1,i}$ for all $1\leq i\leq m$, and let $R = \{R_1, \ldots,
R_v\}$ be random variables over $\mathbb{G}_T$. We say that $T_b$ is
dependent on $A$ if there exists constants $\{\alpha_i\}, \{\beta_i\}$ such
that
    \begin{align*}
    \sum_i^m \alpha_i T_{b,i} = \sum_i^u \beta_i \cdot P_i
    \end{align*}
where $\alpha_i \neq 0$ for at least one $i$. We say that $T_b$ is
independent of $P$ if $T_b$ is not dependent on $P$.

Let $S_1 = \{(i,j) ~|~ e(T_{0,i}, T_{0,j}) \neq e(T_{1,i}, T_{1,j})\}$ and
$S_2 = \{(i,j) ~|~ e(T_{0,i}, P_j) \neq e(T_{1,i}, P_j)\}$. We say that
$\{e(T_{b,i}, T_{b,j})\}_{(i,j) \in S_1} \cup \{e(T_{b,i},P_j)\}_{(i,j) \in
S_2}$ is dependent on $P \cup R \cup \{e(T_{b,i}, T_{b,j})\}_{(i,j) \notin
S_1} \cup \{e(T_{b,i},P_j)\}_{(i,j) \notin S_2}$ if there exist constants
$\{\alpha_{i,j}\}, \{\alpha'_{i,j}\}, \lb \{\beta_{i,j}\}, \{\beta'_{i,j}\},
\{\gamma_{i,j}\}, \{\delta_i\}$ such that
    \begin{align*}
    &   \sum_{(i,j) \in S_1} \alpha_{i,j} \cdot e(T_{b,i}, T_{b,j}) +
        \sum_{(i,j) \notin S_1} \alpha'_{i,j} \cdot e(T_{b,i}, T_{b,j}) +
        \sum_{(i,j) \in S_2} \beta_{i,j} \cdot e(T_{b,i}, P_j) +
        \sum_{(i,j) \notin S_2} \beta'_{i,j} \cdot e(T_{b,i}, P_j) \\
    &   = \sum_i^u \sum_j^u \gamma_{i,j} \cdot e(P_i, P_j) +
        \sum_i^v \delta_i \cdot R_i.
    \end{align*}
where $\alpha_{i,j} \neq 0$ for at least one $(i,j) \in S_1$ or $\beta_{i,j}
\neq 0$ for at least one $(i,j) \in S_2$. We say that $\{e(T_{b,i},
T_{b,j})\}_{(i,j) \in S_1} \cup \{e(T_{b,i},P_j)\}_{(i,j) \in S_2}$ is
independent of $P \cup R \cup \{e(T_{b,i}, T_{b,j})\}_{(i,j) \notin S_1} \cup
\{e(T_{b,i},P_j)\}_{(i,j) \notin S_2}$ if $\{e(T_{b,i}, T_{b,j})\}_{(i,j) \in
S_1} \cup \{e(T_{b,i},P_j)\}_{(i,j) \in S_2}$ is not dependent on $P \cup R
\cup \{e(T_{b,i}, T_{b,j})\}_{(i,j) \notin S_1} \cup
\{e(T_{b,i},P_j)\}_{(i,j) \notin S_2}$.
\end{definition}

Using the above dependence and independence of random variables, we can
generalize the master theorem of Katz et al. as Theorem \ref{thm-master}.

\begin{theorem} \cite{KatzSW08} \label{thm-master}
Let $P = \{P_1, \ldots, P_u\},~ T_0 = \{T_{0,1}, \ldots, T_{0,m}\},~ T_1 =
\{T_{1,1}, \ldots, T_{1,m}\}$ be random variables over $\mathbb{G}$ where
$T_{0,i} \neq T_{1,i}$ for all $1\leq i\leq m$, and let $R = \{R_1, \ldots,
R_v\}$ be random variables over $\mathbb{G}_T$. Consider the following
experiment in the generic group model:
    \begin{quote}
    An algorithm is given $P = \{P_1, \ldots, P_u\}$ and $R = \{R_1,
    \ldots, R_v\}$. A random bit $b$ is chosen, and the adversary is given
    $T_b = \{T_{b,1}, \ldots, T_{b,m}\}$. The algorithm outputs a bit $b'$,
    and succeeds if $b'=b$. The algorithm's advantage is the absolute value
    of the difference between its success probability and $1/2$.
    \end{quote}
Let $S_1 = \{(i,j) ~|~ e(T_{0,i}, T_{0,j}) \neq e(T_{1,i}, T_{1,j})\}$ and
$S_2 = \{(i,j) ~|~ e(T_{0,i}, P_j) \neq e(T_{1,i}, P_j)\}$. If $T_b$ is
independent of $P$ for all $b \in \{0,1\}$, and $\{e(T_{b,i},
T_{b,j})\}_{(i,j) \in S_1} \cup \{e(T_{b,i},P_j)\}_{(i,j) \in S_2}$ is
independent of $P \cup R \cup \{e(T_{b,i}, T_{b,j})\}_{(i,j) \notin S_1} \cup
\{e(T_{b,i},P_j)\}_{(i,j) \notin S_2}$ for all $b \in \{0,1\}$, then any
algorithm $\mathcal{A}$ issuing at most $q$ instructions has an advantage at
most $O(q^2t/p)$.
\end{theorem}

The master theorem of Katz et al. still holds in prime order bilinear groups
since the dependent equation of an adversary can be used to distinguish the
target $T_b$ of the assumption. Additionally, it still holds when the target
consists of multiple group elements since the adversary can only make a
dependent equation in Definition \ref{def-dep-indep}.

\section{Analysis of Our Assumptions}

To prove that our assumption holds in the generic group model by applying the
master theorem of previous section, we only need to show the independence of
$T_0, T_1$ random variables.

\subsection{P3DH Assumption}

Using the notation of previous section, the decisional P3DH assumption can be
written as follows
    \begin{align*}
    &   P = \{ 1, X, A, XA, B, XB, AB + XZ_1, Z_1, C + XZ_2, Z_2 \},~
        R = \{ 1 \} \\
    &   T_0 = \{ ABC + XZ_3, Z_3 \},~ T_1 = \{ D + XZ_3, Z_3 \}.
    \end{align*}

The $T_1$ has a random variable $D$ that does not exists in $P$. Thus the
independence of $T_1$ is easily obtained. Therefore, we only need to consider
the independence of $T_0$. First, $T_0$ is independent of $P$ since $T_0$
contains $Z_3$ that does not exist in $P$. For the independence of
$\{e(T_{0,i},T_{0,j})\}_{(i,j) \in S_1} \cup \{e(T_{0,i},P_j)\}_{(i,j) \in
S_2}$, we should define two sets $S_1, S_2$. We obtain that $S_1 = \{(1,1),
(1,2), (2,1), (2,2)\}$. However, $e(T_{0,i},T_{0,j})$ contains $Z_3^2$
because of $Z_3$ in $T_0$, and $Z_3^2$ can not be obtained from the right
part of the equation in Definition \ref{def-dep-indep}. Thus, the constants
$\alpha_{i,j}$ should be zero for all $(i,j)$. From this, we obtain the
simple equations as follows
    \begin{align*}
    \sum_{(i,j) \in S_2} \beta_{i,j} \cdot e(T_{b,i}, P_j) +
    \sum_{(i,j) \notin S_2} \beta'_{i,j} \cdot e(T_{b,i}, P_j)
    = \sum_i^u \sum_j^u \gamma_{i,j} \cdot e(P_i, P_j) +
    \sum_i^v \delta_i \cdot R_i.
    \end{align*}

The set $S_2$ is defined as $\{(i,j) ~|~ \forall i,j\}$ because of $D$ in
$T_1$. However, $Z_3$ in $T_0$ should be removed to construct a dependent
equation since $Z_3$ does not exists in $P,R$. To remove $Z_3$ from the left
part of the above simple equation, two random variables $Y, XY$ should be
paired with $T_{0,i}$ for some $Y \in P$. If $Z_3$ is remove in the left part
of the above simple equation, then the left part has at least a degree $3$
and it contains $ABC$. To have a degree $3$ in the right part of the above
simple equation, $AB+XZ_1, Z_1$ should be used. However, the right part of
the above equation can not contain $ABC$ since $C, XC$ do not exist in $P$.
Therefore, the independence of $T_0$ is obtained.

%% file: chap-conclusion.tex
In this thesis, we proposed efficient HVE schemes with short tokens. We first
presented the efficient HVE schemes that have the constant size of tokens and
the constant cost of pairing computations in decryption. The scheme was based
on composite order bilinear groups where the order is a product of three
primes. Additionally, we constructed a scheme in asymmetric bilinear groups
where there are no efficiently computable isomorphisms between two groups.
Next, we presented a general framework that converts HVE schemes from
composite order bilinear groups to prime order bilinear groups. Using this
framework, we constructed HVE schemes that are secure under any kind of
pairing types. Finally, we proposed a fully secure HVE scheme with short
tokens in composite order bilinear groups by adapting the dual system
encryption technique.

There are many interesting problems that should be solved. The first one is to
construct a delegatable HVE scheme with short tokens. The delegation property
was achieved in hierarchical identity based encryption and attribute based
encryption. Though Shi and Waters constructed a delegatable HVE scheme, the
decryption const of their construction is proportional to the number of
attributes in tokens. The second one is to construct a HVE scheme with
constant size of ciphertexts. In HIBE, the scheme with constant size of
ciphertexts was proposed. In HVE, it is not easy because the scheme should
support wild-card in tokens. The third one is to construct a fully secure HVE
scheme without any restrictions on the capability of the adversary.